\documentclass[a4paper, amsfonts, amssymb, amsmath, reprint, showkeys, nofootinbib, twoside,superscriptaddress]{revtex4-1}
\usepackage[english]{babel}
\usepackage[utf8]{inputenc}
\usepackage[colorinlistoftodos, color=green!40, prependcaption]{todonotes}
\usepackage{amsthm}
\usepackage{mathtools}
\usepackage{physics}
\usepackage{xcolor}
\usepackage{graphicx}
\usepackage[left=23mm,right=13mm,top=35mm,columnsep=15pt]{geometry} 
\usepackage{adjustbox}
\usepackage{placeins}
\usepackage[T1]{fontenc}
\usepackage{lipsum}
\usepackage{csquotes}
\usepackage{mathrsfs,amsmath} 
\usepackage{amsfonts}
\usepackage{stmaryrd}
\usepackage[bbgreekl]{mathbbol}
\usepackage{tabularx}

\usepackage{amsthm}
\newcounter{theorem}
\newtheorem{lemma}[theorem]{Lemma}
\newenvironment{example}[1][]
{ 
\vspace{4mm}
\noindent\makebox[\linewidth]{\rule{\hsize}{1.5pt}}
\textbf{Example #1}\\
}

\DeclareSymbolFontAlphabet{\mathbbm}{bbold}
\DeclareSymbolFontAlphabet{\mathbb}{AMSb}%

\usepackage{caption}
\usepackage{bm}
\usepackage{adjustbox}

\usepackage{ dsfont }
\usepackage{multirow}
\usepackage{tabularx,ragged2e,booktabs}
\usepackage{hhline}
\usepackage{cases}
\usepackage{subcaption}
\usepackage{calrsfs}
\usepackage{verbatim}
\DeclareMathAlphabet{\pazocal}{OMS}{zplm}{m}{n}

\newcommand{\Cb}{\pazocal{C}}
\newcommand{\Db}{\pazocal{D}}

\newcommand{\Vb}{\pazocal{V}}

\usepackage[pdftex, pdftitle={Article}, pdfauthor={Author}]{hyperref} % For hyperlinks in the PDF

\allowdisplaybreaks
\begin{document}

\title{Multivariate Quadratic Hawkes Processes -- Part II: \\ Non-Parametric Empirical Calibration}

\author{Cecilia Aubrun}
\affiliation{Chair of Econophysics and Complex Systems, \'Ecole polytechnique, 91128 Palaiseau Cedex, France}
\affiliation{LadHyX UMR CNRS 7646, \'Ecole polytechnique, 91128 Palaiseau Cedex, France}
\affiliation{Capital Fund Management, 23 Rue de l’Universit\'e, 75007 Paris, France}

\author{Michael Benzaquen}
\email{michael.benzaquen@polytechnique.edu}
\affiliation{Chair of Econophysics and Complex Systems, \'Ecole polytechnique, 91128 Palaiseau Cedex, France}
\affiliation{LadHyX UMR CNRS 7646, \'Ecole polytechnique, 91128 Palaiseau Cedex, France}
\affiliation{Capital Fund Management, 23 Rue de l’Universit\'e, 75007 Paris, France}
\author{Jean-Philippe Bouchaud}
\affiliation{Chair of Econophysics and Complex Systems, \'Ecole polytechnique, 91128 Palaiseau Cedex, France}
\affiliation{Capital Fund Management, 23 Rue de l’Universit\'e, 75007 Paris, France}
\affiliation{Académie des Sciences, 23 Quai de Conti, 75006 Paris, France\smallskip}

\date{\today} % Leave empty to omit a date

\begin{abstract}
This is the second part of our work on Multivariate Quadratic Hawkes (MQHawkes) Processes, devoted to the calibration of the model defined and studied analytically in Aubrun, C., Benzaquen, M., \& Bouchaud, J. P., Quantitative Finance, 23(5), 741-758 (2023). We propose a non-parametric calibration method based on the general method of moments applied to a coarse-grained version of the MQHawkes model. This allows us to bypass challenges inherent to tick by tick data. Our main methodological innovation is a multi-step calibration procedure, first focusing on ``self'' feedback kernels, and then progressively including cross-effects. Indeed, while cross-effects are significant and interpretable, they are usually one order of magnitude smaller than self-effects, and must therefore be disentangled from noise with care. For numerical stability, we also restrict to pair interactions and only calibrate bi-variate QHawkes, neglecting higher-order interactions. Our main findings are: (a) While cross-Hawkes feedback effects have been empirically studied previously, cross-Zumbach effects are clearly identified here for the first time. The effect of recent trends of the {\sc e-mini} futures contract onto the volatility of other futures contracts is especially strong; (b) We have identified a new type of feedback that couples past realized {\it covariance} between two assets and future {\it volatility} of these two assets, with the pair {\sc e-mini}/{\sc tbond} as a case in point; (c) A cross-leverage effect, whereby the sign of the return of one asset impacts the volatility of another asset, is also clearly identified. The cross-leverage effect between the {\sc e-mini} and the residual volatility of single stocks is notable, and surprisingly universal across the universe of stocks that we considered.   
\end{abstract}

\keywords{Multivariate QHawkes, QGARCH, non-parametric calibration}

\maketitle
%\tableofcontents
% \input{sections/0intro.tex}

% \textbf{Context+Stylised facts}
\section{Introduction}\label{subsec:chap2bintro}
\addcontentsline{toc}{section}{\nameref{subsec:chap2bintro}}

The increasing amount of high-frequency data in financial markets has made it possible to study and calibrate microstructure models. An increasingly renowned family of such models has been invented by the late Alan G. Hawkes \cite{hawkes1971point, hawkes1971spectra}. Initially introduced to model seismic activity, the adaptability and interpretability of Hawkes processes have made them highly appealing for financial data \cite{hawkes2018hawkes, bacry2014hawkes}. They allow in particular to identify the feedback loop that underlies some of the most prominent stylized facts that characterize financial prices. 

In a previous paper \cite{aubrun2023multivariate}, we introduced a Multivariate version of the Quadratic Hawkes (QHawkes) process proposed by Blanc et al. \cite{blanc2017quadratic}. However, the calibration of MQHawkes processes on empirical data is far from trivial and poses specific difficulties that the present work aims at resolving.   

Many methods exist to calibrate Hawkes processes  on empirical data, using either events binned in a regular time grid, or successive time events. A good review is presented in Appendix C of \cite{bacry2015hawkes}. Methods fall into two categories: parametric and non-parametric, referring to the way the feedback ``kernel'' is specified. 

Among the  parametric approaches frequently employed, the maximum likelihood \cite{ogata1978estimators,ogata1998linear} and Expectation Maximisation (EM) method \cite{veen2008estimation} stand out as prominent examples. Non-parametric methods offer the advantage of agnostic kernel forms, as they do not impose specific shapes. The primary non-parametric approaches include the Expectation Maximisation-based method \cite{lewis2011nonparametric,marsan2008extending}, the minimisation of contrast function \cite{hansen2015lasso,reynaud2010adaptive}, and the method of moments or Wiener-Hopf-based method promoted by Bacry et al. \cite{bacry2016estimation,bacry2014hawkes,bacry2014second,fosset2021non,bacry2012non}. Additionally, the \textit{method of cumulants}, a non-parametric method \cite{hardiman2014branching,achab2017uncovering}, allows one to assess the average endogeneity of the process without having to fully estimate the feedback kernels.

In the present study we want to remain agnostic regarding the kernel shape and implement the method of moments on binned data for several pairs of stocks and futures (henceforth called ``futures''). This can later be used as an initial condition for maximum likelihood estimation. Our main methodological innovation is to perform the calibration of the MQHawkes in successive steps, first focusing on ``self'' (univariate) feedback kernels, and then progressively including cross-effects. Such a procedure is needed for the calibration to converge towards stable and meaningful results. Indeed, while cross-effects are significant and interpretable, they are usually one order of magnitude smaller than self-effects, and must therefore be disentangled from noise with care. We also separate quadratic effects (invariant when the sign of returns are flipped) from linear effects (that change sign when sign of returns are flipped), again to improve identifiability. 

For the sake of numerical robustness, we restrict to bi-variate calibrations and think of a full multivariate calibration of the MQHawkes model as resulting from the combination of all pairwise calibrations.   Along the way, we use the pair {\sc e-mini} vs. {\sc tbond} as a canonical test case for our method. We also simplify our model to deal with single stocks, by only considering interactions between the market factor (i.e. the {\sc e-mini} in our case) and the individual residual component of each stock separately, within the context of a one factor model.  

The outline of the paper is as follows. Section II recalls the definition of the MQHawkes model we aim to calibrate. Section III describes the data and the pre-processing methods we apply to it. Section IV presents the key ingredients for the calibration: a discrete-time (bin) approximation, the set of covariance functions  and the calibration method. Finally, the last sections are dedicated to our calibration results on empirical data (futures and single stocks) and their interpretation. Calibration on synthetic data, as a proof of concept, is given in Appendix~\ref{app:calib_sur_simu}. Many technical details can be found in other Appendices. 

The main takeaways of our study are the following:
\begin{itemize}
    \item The two well-studied ``self'' feedback effects, namely the Hawkes feedback (i.e. activity $\to$ activity \cite{bacry2015hawkes, hawkes2018hawkes}) and the Zumbach feedback (i.e. squared trend $\to$ activity \cite{blanc2017quadratic}) have ``cross'' counterparts. While the cross-Hawkes terms were empirically studied previously (see e.g. \cite{bormetti2015modelling, bacry2015hawkes}) cross-Zumbach effects are clearly identified here for the first time (see also \cite{aubrun2023multivariate}). The effect of recent trends of the {\sc e-mini} futures contract onto the volatility of other futures contracts is especially strong.
    \item We have identified a new type of feedback that couples past realized {\it covariance} between two assets and future {\it volatility} of these two assets, with the pair {\sc e-mini}--{\sc tbond} serving as a striking illustrative example.
    \item A cross-leverage effect, whereby the sign of the return of one asset impacts the volatility of another asset, is also clearly identified. Once again, the effect is particularly strong in the case of the {\sc e-mini} on other futures, although the sign of the returns of the {\sc tbond} also plays a significant role. The cross-leverage effect between the {\sc e-mini} and the residual volatility of single stocks is also noteworthy, and appears to be remarkably universal across the set of stocks we examined (see Fig. \ref{fig:Lx_average} below).
\end{itemize}

\section{MQHawkes \& MQARCH models}

In this section we recall essential ingredients of the original (point process) MQHawkes model presented in~\cite{aubrun2023multivariate}. We then formulate a corresponding multivariate quadratic ARCH model (MQARCH) that can be seen as a coarse-grained version of the MQHawkes model, see \cite{blanc2014fine} and below. Such a formulation is better adapted to the 1-minute binned data  at our disposal (see below) and could be used as a starting point for a more fine-grained calibration using tick by tick multivariate data. 

\subsection{MQHawkes}

Let $(N_t)_{t\geq0}$ be a Hawkes process, that is an inhomogenous %\footnote{Here, inhomogenous means that the intensity of the process is time dependent.} 
Poisson process with intensity $\lambda_t$ defined from the past realisations of the process itself, given by:\newline
\begin{equation}\label{eq:Hawkes_def}
    \lambda_t= \lambda_\infty + \int_{-\infty}^t\phi(t-u)\, {\rm d}N_u
\end{equation}
    with
\begin{equation}
    n = \int_\mathbb{R}\phi(u)\, {\rm d}u < 1.
\end{equation}
In this context, $\lambda_\infty$ denotes the baseline intensity, $\phi$ is the feedback kernel and $n$ represents the ``endogeneity ratio'', which must be strictly less than one to ensure that the process reaches a stationary state.

QHawkes processes, akin to Hawkes processes, are inhomogeneous Poisson processes reliant on past occurrences for their intensity calculations.
%Like Hawkes processes, QHawkes processes are also inhomogenous Poisson processes with an intensity defined from past events. 
%%=with the difference of now including past price changes ${\rm d}P_t$ (in the context of financial markets). 
In the context of financial markets, the event process $(N_t)_{t>0}$ represents the sequence of price changing events, where the price change (in tick units) is defined by ${\rm d}P_t = \epsilon_t {\rm d}N_t$~\cite{blanc2017quadratic} where $\epsilon_t = \pm 1$ is a random sign. The intensity self-exciting feedback loop is then defined on the return stream ${\rm d}P_t$ rather than on the events ${\rm d}N_t$. To wit, the sign of the returns influences the process intensity as \cite{blanc2017quadratic}: 
\begin{equation}\label{eq:Lambda_def_QH}
\begin{aligned}
    \lambda_t =& \lambda_{\infty} +  \int^t_{-\infty}L(t-s)\,{\rm d}P_s\\
    &+\iint^t_{-\infty} K(t-s,t-u)\, {\rm d}P_s{\rm d}P_u,
\end{aligned}
\end{equation}
where $\lambda_\infty$ is again the baseline intensity, $L(\cdot)$ is the so-called leverage kernel (capturing the increase of activity following a drop in prices) and $K(\cdot,\cdot)$ is the quadratic Hawkes kernel, with a time-diagonal part that boils down to the standard Hawkes feedback and a non-diagonal part that captures the path-dependency of the activity on past price changes \cite{blanc2014fine, guyon2023volatility, guyon2025discrete}.   

In a multivariate framework, that is encompassing $N$ financial assets with prices $(P^i_t)_{t\geq0,i=1, \dots,N}$, we associate such price processes with jump processes $(N^i_t)_{t\geq0}$, with: 
\begin{equation}
    {\rm d}P^i_t = \epsilon_i {\rm d}N^i_t, 
\end{equation} 
where each jump process $({\rm d}N^i_t)_{t\geq0}$ is a conditionally independent\footnote{This means that for given intensities $\lambda_{i,t}$, the inhomogeneous Poisson processes ${\rm d}N^i_t$ are independent.} Quadratic Hawkes process with intensity $(\lambda_{i,t})_{t\geq0}$  such that:
\begin{equation}\label{eq:Lambda_def_MQH}
\begin{aligned}
    \lambda_{i,t} =& \lambda_{i,\infty} + \sum_{j=1}^N  \int^t_{-\infty}L_j^i(t-s)\,{{\rm d}P^j_s} \\
    &+\sum_{j \leq k}^N  \iint^t_{-\infty} K_{jk}^i(t-s,t-u)\, {{\rm d}P^j_s}\, {\rm d}P^k_u.
\end{aligned}
\end{equation}
In this notation, the superscript in the kernels indicates which asset is affected by the feedback, while the subscripts identify the assets responsible for it.

We will generically refer to the kernels $K_{jk}^i$ with $j=k$ as the ``diagonal'' feedback terms. These components describe the quadratic feedback from two price changes of the same asset $j$ onto the activity of asset $i$. Similarly, $K_{jk}^i$ with $j < k$  outlines cross-effects from two different assets $j, k$ onto asset i's activity. As we argued in \cite{aubrun2023multivariate}, one can neglect the case where the three indices $i,j,k$ are all different, since this would correspond to scenario where the instantaneous return correlation between two assets impacts the activity on a third one. Although not impossible a priori, we will not consider this possibility, which opens the path to a complete calibration of an $N$-dimensional MQHawkes from $N(N-1)/2$ pairwise calibrations.   

Finally, we distinguish the ``time-diagonal'' of a kernel $K_{jk}(\tau_1,\tau_2)$, by which we mean $K_{jk}(\tau,\tau)$, from its ``diagonal'', which refers to the diagonal components in asset space $K_{ii}$.

 %\noindent \red{[M: je n'ai pas re-regarde le précédent papier mais dans cette section il faut veiller à ce qu'il n'y ait pas de phrases identiques. Si et quand c'est le cas, un petit ChatGPT "rephrase" devrait faire l'affaire.]}

 \subsection{MQARCH}

Here, we give the equations defining the multivariate, quadratic ARCH model that represents a coarse-grained version of the MQHawkes continuous point process. Specifically, relying on the work of Blanc \textit{et al.} \cite{blanc2017quadratic} on the univariate QHawkes, the QHawkes intensity $\lambda_t$ can be approximated by the squared volatility as follows:
\begin{equation}\label{eq:secMethod_approximation}
    \frac{{\left(\sigma_t^{(\Delta t)}\right)}^2}{\Delta t}  \xrightarrow[\Delta t \to 0^+]{}  \lambda_t,
\end{equation}
where $\sigma_t^{(\Delta t)}$ is a volatility estimate over the time bin of size $\Delta t$.

Thus, the Hawkes continuous point process ${\rm d}N_t$ can be approximated by the discrete volatility process $\sigma^{(\Delta t)}$ over a time bin of size $\Delta t$. This approximation critically depends on the choice of the time bin size $\Delta t$, which in turn significantly influences the calibration of the underlying QHawkes model.

Indeed if $\Delta t \gg 1/\bar{\lambda}$, the QHawkes relation $\Delta P_t = \pm {\Delta}N_t$, is no longer valid. Consequently, some of the terms in the Yule-Walker equations (relating kernels to covariances \cite{aubrun2023multivariate}) would need to be modified. In particular, the relation between the diagonal contribution $K(s,s) \Delta P_{t-s}^2$ and the Hawkes contribution $\phi(s) {\Delta}N_{t-s}$ is lost. 

%if we decompose the quadratic kernel into a diagonal component and a regular one ($\mathbb{K}_{\text{d}}(s,s)=\mathbb{K}_{\text{d,reg.}}(s,s) + \bbphi_{\text{d}}(s)$), the diagonal component would no longer replicate the linear Hawkes feedback, but includes a feedback loop on $({\rm d}P)^2$.
%Additionally, the choice of ${\rm d}t$ highly influences the estimation of the covariance values that determine the Yule-Walker equations. For a small ${\rm d}t$, say ${\rm d}t\ll 1/\bar{\lambda}$, consecutive events are rare potentially leading to an underestimation of correlations. Conversely, with a large ${\rm d}t$ (${\rm d}t \gg 1/\bar{\lambda}$), numerous consecutive events occur, potentially resulting in an overestimation of correlations. 
%Appendix~\ref{app:calib_sur_simu}-Section~\ref{app_sec:calib_sur_simu_1D} discusses this aspect in details. 
%Following \cite{blanc2017quadratic}, we alleviate these difficulties by approximating the MQHawkes model by a MQARCH model. 
Relying on the QARCH framework introduced in \cite{sentana1995quadratic,chicheportiche2014fine,blanc2014fine}, we define the MQARCH model in the bi-variate case, for the time dependent volatility $\sigma_{.,t}^2$ of two assets $i,j \in (1,2)$, as follows: 
\begin{equation}\label{eq:2D_QGARCH_def_sigma}
\begin{aligned}
    \sigma^2_{i,t} =& \sigma^2_{i,\infty} + \sum_{j=1}^2 \sum_{\tau=1}^{+\infty}L^i_j(\tau)r_{j,t-\tau} + \sum_{j=1}^2 \sum_{\tau=1}^{+\infty}\phi^i_j(\tau)r^2_{j,t-\tau}\\&+2\sum_{j=1}^2 \sum_{\tau_1=1}^{+\infty}\sum_{\tau_2=\tau_1+1}^{+\infty}K^i_j(\tau_1,\tau_2)
    r_{j,t-\tau_1}r_{j,t-\tau_2} \\&+\sum_{\tau=1}^{+\infty} \phi^i_{\times}(\tau)(r_{i,t-\tau}r_{\overline{\imath},t-\tau}-C)
    \\&+\sum_{\tau_1=1}^{+\infty}\sum_{\tau_2 \neq \tau_1}^{\infty} K^i_{\times}(\tau_1,\tau_2)r_{i,t-\tau_1}r_{\overline{\imath},t-\tau_2},
\end{aligned}
\end{equation}
where we define $\overline{\imath}=2$ if $i=1$ and vice-versa, and $C = \mathbb{E}[r_{1,t} r_{2,t}]$ is the equal time covariance between the returns of the two assets $1,2$, and where all returns are de-drifted ($\mathbb{E}[r_{i,t}]=0$). 

In this framework, the returns $r_{i,t}$ are no longer defined at microscale but are related -- as in all ARCH models -- to the volatility through
\begin{equation}
    r_{i,t} = \sigma_{i,t} \xi_{i,t},
\end{equation}
where $\xi_{i,t}$ are independent random variables of zero mean and variance unity, possibly non-Gaussian and non-identical for different assets.\footnote{We assume here that all non-equal time covariances are zero, but see Section~\ref{sec:martingalisation} below.} The MQARCH framework is more consistent with the data used in this paper (see Section~\ref{sec:calib_data}), which is based on binned 1-min returns.

Additionally, using the fact that $\mathbb{E}[\xi_{i,t}\xi_{j,t'}] = \delta_{ij} \delta_{tt'}$, the mean squared volatility is determined through the following relation:
\begin{equation}\label{eq:sigmabar_relation}
    \begin{pmatrix}\overline{\sigma_{1}^2}\\
    \overline{\sigma_{2}^2}\end{pmatrix}= \begin{pmatrix}
        1-n^1_1&n^1_2\\
        n^2_1&1-n^2_2
    \end{pmatrix}^{-1} \begin{pmatrix}{\sigma_{1,\infty}^2}\\
    {\sigma_{2,\infty}^2}\end{pmatrix},
\end{equation}
where
\begin{equation}
    n_j^i := \sum_{\tau=1}^{+\infty} \phi_j^i(\tau).
\end{equation}
Eq. \eqref{eq:sigmabar_relation} is of course only valid if the spectral radius of the feedback matrix $(n_j^i)$ is less than unity. Note that this relation does not involve the quadratic kernels $K$ nor the cross kernel $\phi_\times, K_\times$.
 
\section{Data}\label{sec:calib_data}

We consider a set of 6 futures contracts ({\sc e-mini},  {\sc e-mini}-3 (3 months futures), {\sc nasdaq}, {\sc dow jones}, {\sc crude oil} and {\sc tbond}) for every trading day from 2013 to 2023, and 317 single stocks belonging to the S\&P500 during the same period. 

\subsection{Data description}

For each studied asset $i$, we collect the opening (o), closing (c), highest (h) and lowest (l) prices for 1-minute intervals for every trading day. We only consider the bins between 10 am and 3 pm to exclude specially high volatility at market openings (as participants react to overnight news) and at market closings (as some participants close positions or complete the execution of their metaorders before the market session ends). 

From these price time series, we compute the two quantities needed for the calibration for each 1-minute bin, the log-returns of asset $i$ $r_{t,d}^i$ and the corresponding volatility $\sigma^i_{t,d}$, as follows:
\begin{equation}\label{eq:secData_returns}
r_{t,d}^i = \log\left(\frac{\text{c}^i_{t,d}}{\text{o}^i_{t,d}}\right)    
\end{equation}
\begin{equation}\label{eq:secData_vol}
    \sigma^i_{t,d} = \frac{1}{3}\frac{\text{h}^i_{t,d}-\text{l}^i_{t,d}}{\text{o}^i_{t,d}}+\frac{2}{3}\frac{|\text{c}^i_{t,d}-\text{o}^i_{t,d}|}{\text{o}^i_{t,d}}
\end{equation}
where $d$ accounts for the day and $t$ for the 1-minute bin within  day $d$. %For all pairs, we use price time series spanning from 2013 to 2023. 
For single stocks, a special treatment will be implemented, as we will decompose stock returns into factor and idiosyncratic components before extracting the volatility of each bin, see Section~\ref{sec:factorxQGarch_datapreprocessing}. 

\subsection{Data pre-processing}\label{sec:preprocessing}

After collecting the log-returns $r_{t,d}^i$ and volatilities $\sigma^i_{t,d}$ time series, we proceed with two steps of data pre-processing: normalisation and ``martingalisation''. The first step removes  any remaining intraday pattern of the volatility and standardises the volatility across the entire studied period. The ``martingalisation'' process allows us to remove any linear correlation between returns, imposing that $\mathbb{E}[r_tr_{t-\tau}]=0$. We now detail the two steps in turn.

\subsubsection{Normalisation}

The objectives of these normalization procedures are twofold: first, to make the time series stationary over time, and second, to standardise the activity throughout the day. 

\begin{figure}
    \centering
    \begin{subfigure}[b]{0.39\textwidth}
         \centering
         \includegraphics[width=\textwidth]{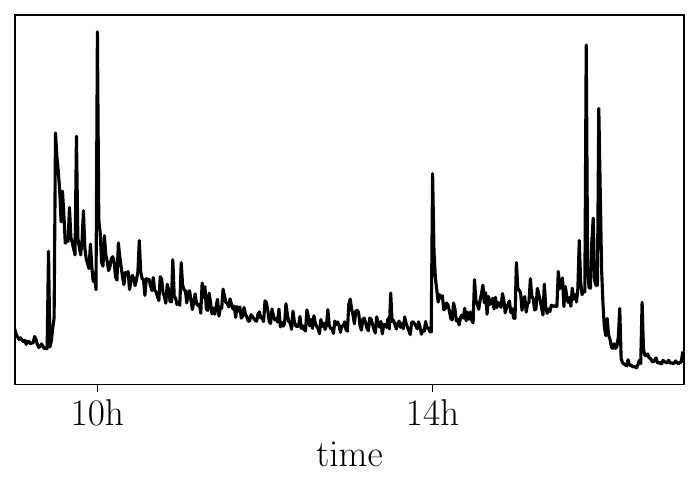}
         \caption{ {\sc e-mini} intraday squared volatility}
         \label{subfig:SPintra}
     \end{subfigure}
     \hfill
     \begin{subfigure}[b]{0.39\textwidth}
         \centering
         \includegraphics[width=\textwidth]{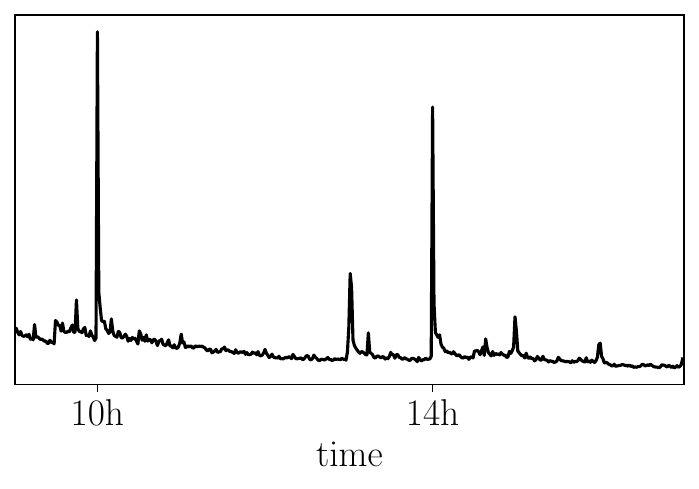}
         \caption{{\sc tbond} intraday squared volatility}
         \label{subfig:USintra}
     \end{subfigure}
    \caption{Average intraday profiles of the squared volatility Bachelier estimates, as defined by Equation~\eqref{eq:secData_vol}, for a period spanning from 2013 to 2023.}
    \label{fig:exampla_intraday_vol_profile}
\end{figure}

To achieve stationarity and remove intraday seasonalities, each 1-minute bin return is normalized by an estimate of the realized volatility over the past 100 days as described in Eqs.~\eqref{eq:norm1_dp} and \eqref{eq:norm1_sig}:
\begin{equation}\label{eq:norm1_dp}
    r_{t,d}^i \gets \frac{r_{t,d}^i}{\sqrt{\frac{1}{100}\sum_{k=d-100}^{d}(r_{t,k}^i)^2}},
\end{equation}
\begin{equation}\label{eq:norm1_sig}
    \sigma^i_{t,d} \gets \frac{\sigma^i_{t,d}}{\sqrt{\frac{1}{100}\sum_{k=d-100}^{d}(\sigma^i_{t,k})^2}}.
\end{equation}
Without this normalization, the daily activity profile exhibits marked a U-shape or even a J-shape, as it is the case for the  {\sc e-mini} and the {\sc tbond} as demonstrated in Figure~\ref{fig:exampla_intraday_vol_profile}, with low-frequency variations of the volatility due to the length of the time series. 

Even after such a normalisation, there is still a residual intraday profile that we suppress by the following operation:
\begin{equation}\label{eq:norm2_sig}
    \sigma^i_{t,d} \gets \frac{\sigma^i_{t,d}}{\sqrt{\langle (\sigma^i_{t,d})^2 \rangle_d}},
\end{equation}
\begin{equation}\label{eq:norm2_dp}
    r_{t,d}^i \gets \frac{r_{t,d}^i}{\sqrt{\langle (\sigma^i_{t,d})^2 \rangle_d}},
\end{equation}
where $\langle \dots \rangle_d$ means an average over all days in the time series.

\subsubsection{``Martingalisation''}\label{sec:martingalisation}

Upon examining empirical data, it appears that at the 1-minute time scale, returns show negligible auto-correlations, with a notable non-zero correlation between $r_{t}$ and $r_{t-1}$.
Although this correlation is weak and only significant over short time scales, it is essential to address this weakly non-martingale behavior because in the QHawkes framework presented in \cite{aubrun2023multivariate}, price returns are assumed to be martingales, and many properties rely on this assumption. Therefore, we perform by hand a ``martingalisation'' of the returns time series, meaning that the component of the returns at time $t$ that is predictable from time $t-1$ is removed to isolate the return's ``surprise'' component, as follows (see also \cite{fosset2021non} and Appendix \ref{app:surprisePrice} for more details): 
\begin{equation}
    r_{t,d}^i \gets r_{t,d}^i -  \mathbb{E}[r^i_{t,d} \vert \mathcal{F}_{t-1,d}].
\end{equation}
Such a procedure also removes any (small) average value of the returns. The data pre-processing is completed by normalising the returns one last time by their standard deviation. 

The subsequent section describes the calibration method to be applied on the returns and volatility time series.  

\section{Calibration method}\label{sec:calibration_met}

This section outlines the calibration method we implemented. For the sake of clarity, we present here the main elements and Appendices~\ref{app:YW_system_2DQGARCH} and~\ref{app:YW_calibration_matrix} provide more comprehensive details.

To stay agnostic on the shape of the kernels, we chose to implement a method of moments. This method relies on linear relationships between the kernels we want to recover and quantities that we can compute from data, that are covariance structures. Both were introduced in \cite{aubrun2023multivariate} for the MQHawkes model (see Equations (16), (17), (20) and (23) in Ref. \cite{aubrun2023multivariate}).

% These linear relationships, the so-called Yule-Walker equations, were introduced in Equations~\eqref{eq:C_YuleWalker_mQH} and~\eqref{eq:D_Yule_ind_mQH} for the MQHawkes model as defined in Chapter~\ref{chap:MQHawkesTheo}. 

The discretization of the data imposes these equations to be adapted; however, the same covariance structures are still a key ingredient to the calibration. 
% A second key ingredient is the set of covariance structures, indeed they compose the linear system required to recover the kernels. In Chapter~\ref{chap:MQHawkesTheo}, we introduced $\mathbb{C}$ and $\Db$ in Equations~\eqref{eq:defOfC} and~\eqref{eq:Ddef}. As for the Yule-Walker system of equations, the covariance structures need to be adapted to account for the discretization of data. 
We thus start by discussing the data binning and the modifications that follow.

\subsection{Approximation for discrete data -- toward the MQARCH model}\label{sec:approximationBinned}

As we just mentioned, the advantage of the method of moments is that it is non-parametric, unlike the Maximum Likelihood method for instance. The method of moments relies on the computation of covariance structures, such as those introduced in \cite{aubrun2023multivariate}, in terms of the different feedback kernels, without having to specify their shape. 

The estimation of these covariance structures can be done directly using the times of events, i.e. the times of price changes in the QHawkes model, which necessitates working with high frequency data. However, high frequency data is sometimes difficult to obtain and/or to process, due to the sheer quantity of data. Such data can also be heavily impacted by microstructure noise \cite{bacry2013modelling}. In particular, high frequency, particularly tick-by-tick, price changes reflect strong bid-ask bounce effects (strong mean reversion), which is not of interest in this study.
%\red{[pas sûr de ce dernier argument, on nous dira qu'il suffit de prendre le mid. Par ailleurs tu as parlé de martingalisation plus haut, qui aurait déjà naturellement supprimé le bounce.]}. 
Therefore, we choose to work with aggregated data, to which we apply the martingalisation process of Section~\ref{sec:martingalisation} to effectively eliminate any residual bid-ask bounce effect.

The subsequent section introduces the ``adapted'' covariance structures needed to characterise the kernels of Equation~\eqref{eq:2D_QGARCH_def_sigma}, that are $(L^i_j)_{i,j\in\{1,2\}}$, $(\phi^i_j)_{i,j\in\{1,2\}}$, $(K^i_{j})_{i,j\in\{1,2\}}$, $(\phi^i_\times)_{i\in\{1,2\}}$ and $(K^i_\times)_{i\in\{1,2\}}$ and baseline values $(\sigma_{i,\infty})_{i\in\{1,2\}}$ of Equation~\eqref{eq:2D_QGARCH_def_sigma} which cannot be directly observed in data.

\subsection{Covariance structures} \label{sec:cov}

\subsubsection{Definitions}

The method of moments relies on the fact that kernels can be deduced from covariance functions, which can easily be estimated from empirical data. Given the MQGARCH specification of the model delineated in Equation~\eqref{eq:2D_QGARCH_def_sigma}, we must consider adapting the covariance structures and the corresponding Yule-Walker system, derived for the event-based, point process MQHawkes model. 

Let us detail the number of unknowns to be determined. Considering a time grid up to lag $q$, 
\begin{itemize}
    \item $(L^i_j(\tau))_{i,j\in\{1,2\},1\leq\tau\leq q}$ and $(\phi^i_j(\tau))_{i,j\in\{1,2\},1\leq\tau\leq q}$ bring each $q\times 2\times2$ unknowns
    \item $(K^i_{j}(\tau_1,\tau_2))_{i,j\in\{1,2\}, 1\leq \tau_1<\tau_2\leq q}$ brings  $\frac{q(q-1)}{2}\times 2\times2$ unknowns. Indeed, since $(K^i_{j})_{i,j\in\{1,2\}}$ are time-symmetric, it is sufficient to only determine the upper triangle entries.
    \item $(\phi^i_\times(\tau))_{i\in\{1,2\},1\leq\tau\leq q}$ bring $q \times 2$ unknowns and 
 and $(K^i_\times(\tau_1,\tau_2))_{i\in\{1,2\},1\leq\tau_1,\tau_2\leq q, \tau_1\neq\tau_2}$ bring $q\times (q -1) \times 2$ unknowns.
\end{itemize}
All this adds up to $4q^2 + 6q$ unknowns. 

Now, given the form of the model, we need covariance structures containing information on both the time-diagonal of the kernels ($\tau_1 = \tau_2)$ and their off-time-diagonal $\tau_1\neq \tau_2$.\newline
The first quantity of interest is the covariance between volatilities and realized squared returns, for all $\tau>0$, for $i,j\in \{1,2\}$:
\begin{equation}\label{eq_MM:Cdef}
    \Cb_{ij}(\tau) := \mathbb{E}(\sigma^2_{i,t}r^2_{j,t-\tau}) - \mathbb{E}({\sigma^2_{i}})\;\mathbb{E}({\sigma^2_{j}})
\end{equation}
where we have noted that $\mathbb{E}({r^2_{j}}) \equiv \mathbb{E}({\sigma^2_{j}})$.

As we shall see, computing $(\Cb_{ij}(\tau))_{i,j\in\{1,2\},1\leq\tau\leq q}$ allows to set $q\times (2\times 2)$ equations and predominantly shape the kernels $(\phi^j_{i})_{i,j\in\{1,2\}}$. 

We also define two relevant three-point correlation structures, for $0<\tau_1<\tau_2$,\footnote{Note that $ \Db_{ij}(\tau,\tau) \equiv \Cb_{ij}(\tau)$.}
\begin{equation}\label{eq:D_def}
    \Db_{ij}(\tau_1,\tau_2) := \mathbb{E}\left(\left(\sigma^2_{i,t}-\langle\sigma^2_{i,t}\rangle_t\right)r_{j,t-\tau_1}r_{j,t-\tau_2}\right)
\end{equation}
and, for $0<\tau_1,\tau_2$,
\begin{equation}\label{eq:Dx_def}
    \Db_{\times j}(\tau_1,\tau_2):=\mathbb{E}\left(\left(\sigma^2_{j,t}-\langle\sigma^2_{j,t}\rangle_t\right)r_{j,t-\tau_1}r_{{\overline {\jmath}},t-\tau_2}\right).
\end{equation}
$\Db_{ij}$ is time-symmetric and allows to fix ${q(q-1)}/{2}\times (2\times2)$ equations. It mainly shapes the kernels $(K^i_{j})_{i,j\in\{1,2\}}$.  $\Db_{\times j}$ fixes $q\times q \times (2\times 1)$ equations, and mainly determines $(\phi^i_\times)_{i\in\{1,2\}}$ and $(K^i_\times)_{i\in\{1,2\}}$.

The above covariance structures fix $6q^2 + 2q$ values and enable one to characterize the kernels $(\phi^i_j)_{i,j\in\{1,2\}}$, $(K^i_{j})_{i,j\in\{1,2\}}$, $(\phi^i_\times)_{i\in\{1,2\}}$ and $(K^i_\times)_{i\in\{1,2\}}$ up to a lag $q$, after which the covariance structures and kernels are considered negligible. Moreover, Eq.~\eqref{eq:sigmabar_relation} enables to retrieve the baseline values from $(\phi^i_j)_{i,j\in\{1,2\}}$.

Finally, to recover the leverage feedback that measures the influence of past signed returns on future volatility and determine the missing $4q$ quantities, we define an additional two-point correlation: 
\begin{equation}\label{eq:V_def}
    \begin{aligned}
        \Vb_{ij}(\tau) := 
    \mathop{\mathbb{E}}\left(\sigma^2_{i,t}r_{j,t-\tau}\right).
    \end{aligned}
\end{equation}
$(\Vb_{ij}(\tau))_{i,j\in\{1,2\},1\leq\tau\leq q}$ allows to fix $q\times (2\times 2)$ equations and predominantly shapes the kernels $(L^j_{i})_{i,j\in\{1,2\}}$.\newline

Let us note that some of the aforementioned covariance structures involve fourth moments of financial returns which, due to their fat-tailed distribution, might be undefined and could lead to numerical instability. An interesting alternative would be to use mixed moments as used in Chicheportiche \textit{et al.} (2014) \cite{chicheportiche2014fine}. To avoid obscuring our message, the corresponding Yule-Walker equations linking the feedback kernels to these modified covariance structures are established in Appendix~\ref{app:YW_system_2DQGARCH}.\newline

\subsubsection{Fits and noise reduction}

\begin{figure}[h]
    \centering
    \includegraphics[width=\linewidth]
    {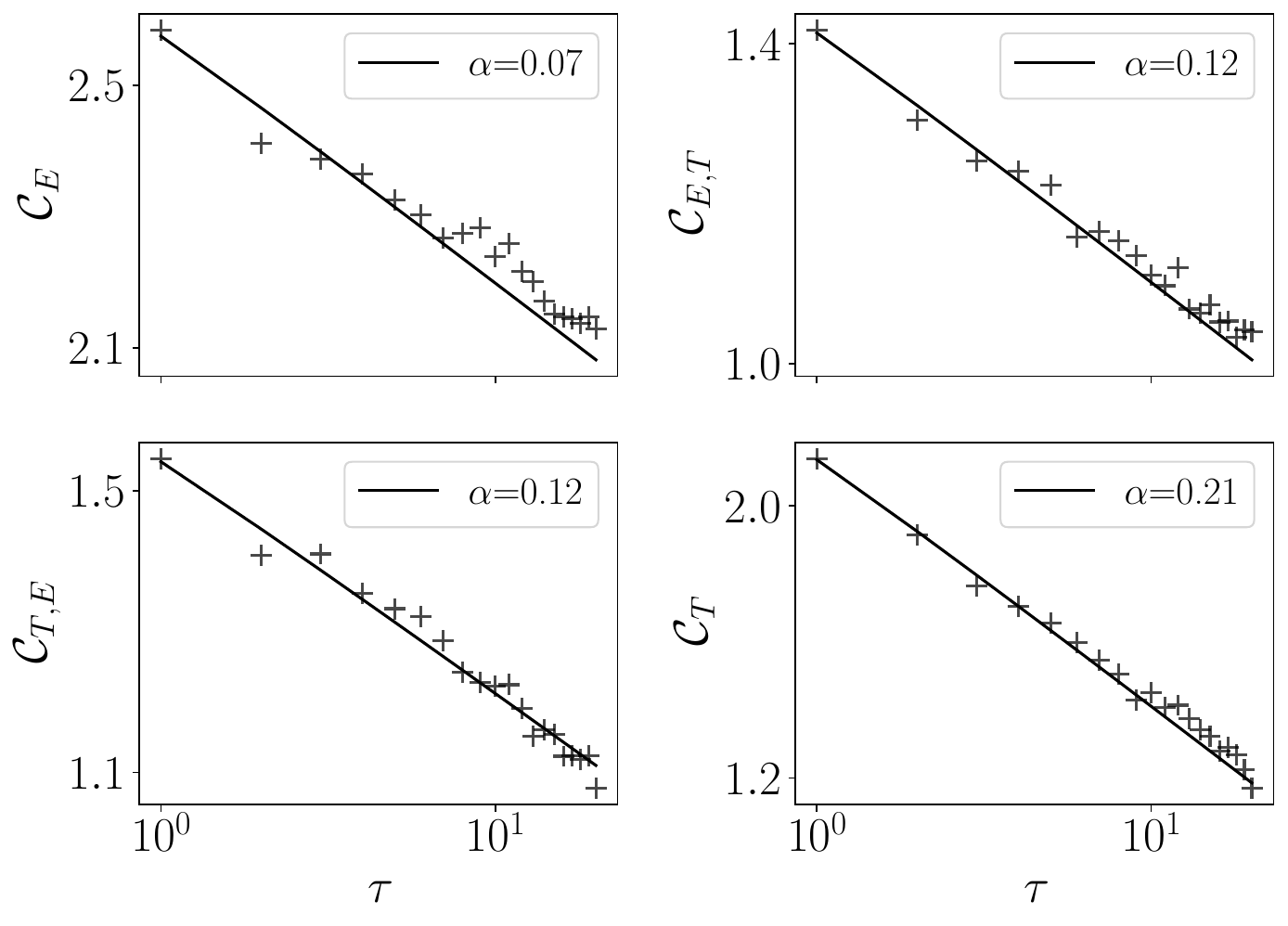}
    % {images/chap3_Csigret_SPUS.pdf}
    \caption{Two-points correlation $\Cb_{ij}(\tau)$, as defined in Equation~\eqref{eq_MM:Cdef}, calculated for the pair  {\sc e-mini} vs {\sc tbond}. The scatter crosses represent the empirical data, while the solid black line represents the fit used for the calibration, given by $\tau \xrightarrow{} {n\exp(-\beta \tau)}{(1+\gamma \tau)^{-\alpha}}$. The parameter $\alpha$
 is specified in each subplot's labels and the other parameters, $(\beta, \gamma, n)$ are given in Table~\ref{table:Cb_fit} in Appendix F. Note that the subplots are presented on a log-log scale.}% \red{[indiquer la pente ?]}}
    \label{fig:C_SPUS}
\end{figure}

\begin{figure}[h]
    \centering
    \includegraphics[width=\linewidth]{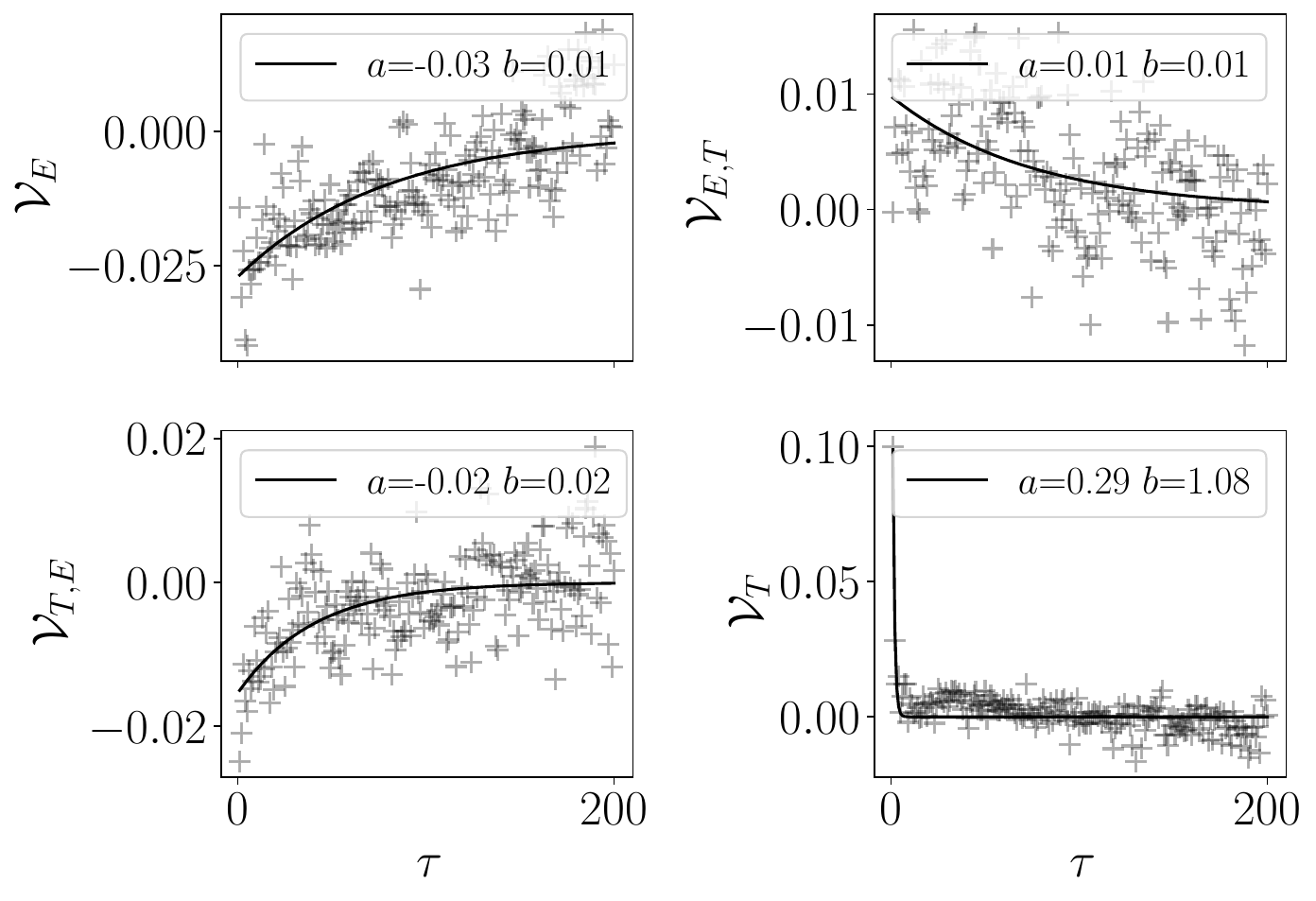}
    \caption{Two-points correlation $(\Vb_{ij})_{i,j\in\{1,2\}}$, as defined in Equation~\eqref{eq:V_def}, calculated for the pair  {\sc e-mini} vs {\sc tbond}. The scatter crosses represent the empirical data, whereas the solid black line represents the fit used for the calibration with the corresponding values of $a,b$ given in the caption.}%Table~\ref{table:Vb_fit}.}
    \label{fig:V_SPUS}
\end{figure}

In order to reduce the numerical noise, the four covariance functions $\Cb,\Db,\Db_\times,\Vb$ computed on empirical data are approximated (``smoothed'') in the following way: 

\begin{itemize}
    \item $(\Cb_{ij})_{i,j\in\{1,2\}}$ and the time-diagonal of $(\Db_{\times j})_{j\in\{1,2\}}$ are smoothed using their corresponding best fits  
     \[\tau \xrightarrow{} \frac{n\exp(-\beta \tau)}{(1+\gamma \tau)^\alpha}.\] 
    Figure~\ref{fig:C_SPUS} and Figure~\ref{fig:Dx_SPUS} illustrate the approximation of $(\Cb_{ij})_{i,j\in\{1,2\}}(\tau)$ and $(\Db_{\times j}(\tau,\tau))_{j\in\{1,2\}}$, respectively, for the pair  {\sc e-mini} vs. {\sc tbond}. 
    These figures suggest that $(\Cb_{ij})_{i,j\in\{1,2\}}$ and the time-diagonal of $(\Db_{\times j})_{j\in\{1,2\}}$ both exhibit long-range correlations (note the log-log scale in Figure \ref{fig:C_SPUS}). This observation justifies the use of a power law function to smooth the empirical values; it also means that volatility movements that happened early in the day still have influence on the end-of-day volatility.
   
\begin{figure}[h]
    \centering
    \includegraphics[width=\linewidth]{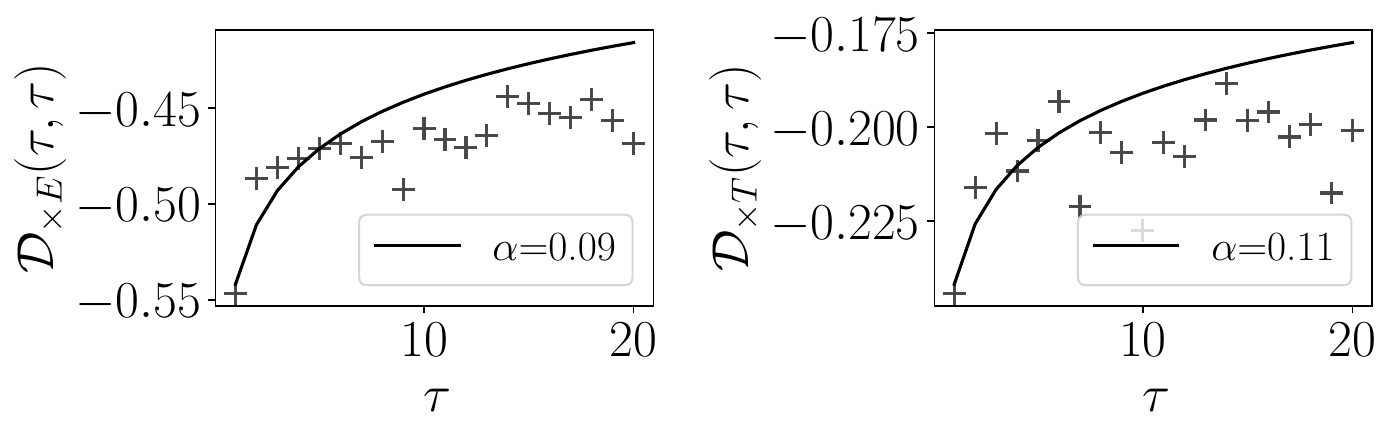}
    \caption{Three-points correlation $(\Db_{\times j})$ for the pair  {\sc e-mini} vs {\sc tbond}, as defined by Equation~\eqref{eq:Dx_def}. The scatter crosses represent the empirical data, whereas the solid black line represents the fit used for the calibration, described by $\tau \xrightarrow{} {n\exp(-\beta \tau)}{(1+\gamma \tau)^{-\alpha}}$. The parameter $\alpha$
 is specified in each subplot's labels and the other parameters, $(\beta, \gamma, n)$ are given in Table~\ref{table:Dx_fit} in Appendix F. }
    \label{fig:Dx_SPUS}
\end{figure}

% \newpage
    \item We observe that the largest eigenvalue the off-diagonal part of $(\Db_{ij}(\tau_1,\tau_2))_{1\leq \tau_1,\tau_2\leq q}$ for all $i,j \in \{1,2\}$ is a lot more significant that the subsequent ones. This suggests using a rank-one approximation for these four matrices, where we furthermore smooth the first eigenvector, which is replaced by its best fit of 
    \[\tau \xrightarrow{} a \exp(-b \tau).
    \]
    For instance, Figure~\ref{fig:Deig_SPUS} presents the first eigenvector and its corresponding fit of the 3-points correlation $\Db_{ij}$ for the pair  {\sc e-mini} vs {\sc tbond}. 
    
    Let us note that, the ``martingalisation'' process described in Section~\ref{sec:martingalisation} facilitates this exponential fit. Indeed, without applying the ``martingalisation'' process, the first eigenvector of the 3-point correlation matrix $\Db_{ij}$ would reflect the strong mean-reversion observed at this scale, characterized by the first two points having opposite signs. 
    
    We also observe from the eigenvalues of the off-diagonal part of $(\Db_{\times j})_{j\in \{1,2\}}$ that such coefficients are mostly noise hovering around zero, giving us an hint on the off-diagonal of the kernels $(K^j_\times)$.
    \begin{figure}[h]
    \centering
    \includegraphics[width=\linewidth]{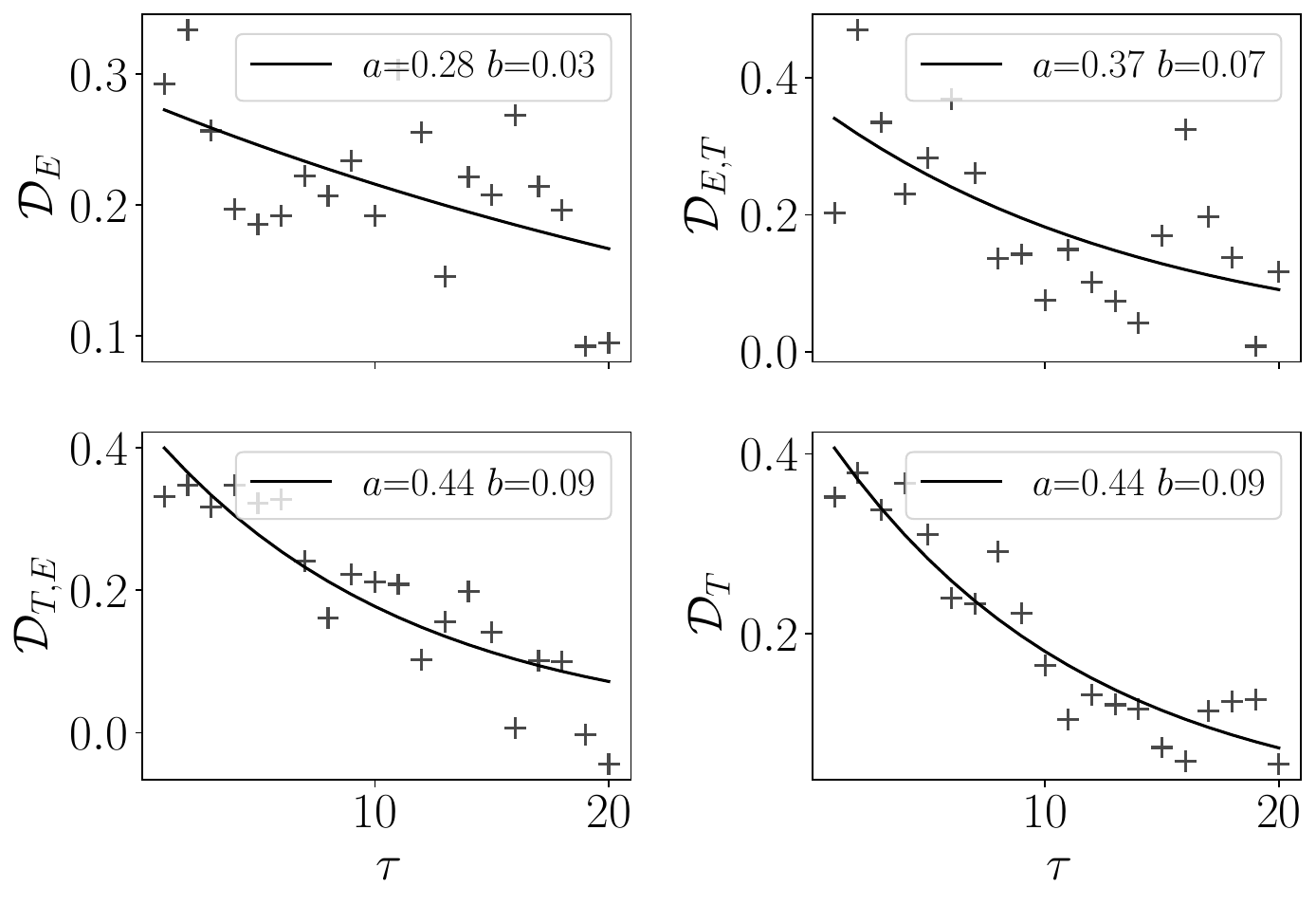}
    \caption{First eigenvector of the 3-points correlation $\Db_{ij}$, as defined in Equation~\eqref{eq:D_def}, for the pair  {\sc e-mini} vs {\sc tbond}. The scatter plots are the empirical values and the solid black line is the exponential fit $\tau \xrightarrow{} a \exp(-b \tau)$ with the corresponding values of $a,b$ given in the caption.}%Table~\ref{table:Deig_fit}***.}
    \label{fig:Deig_SPUS}
\end{figure}

    % \newpage

    \item Finally, $(\Vb_{ij})_{i,j\in\{1,2\}}$ is approximated by its best fit using $\tau \xrightarrow{} a \exp(-b \tau)$.  Figure~\ref{fig:V_SPUS} presents the approximation of all $(\Vb_{ij})_{i,j\in\{1,2\}}$ for the pair {\sc e-mini} vs. {\sc tbond}, with their corresponding fit.  

% \newpage
\end{itemize}

\subsubsection{Empirical results: discussion}

Let us comment on the empirical results obtained above. 

\begin{enumerate}
    \item First of all, the standard ``ARCH'' correlations $(\Cb_{ii})_{i\in\{1,2\}}$ between past square returns and future volatilities are strong and decay as power-laws, as well documented in the literature. We find that the corresponding cross-correlations between {\sc e-mini} and {\sc tbond} are of the same order of magnitude and also decay as power-laws. 
    \item Turning to the time-off-diagonal part of $(\Db_{ij}(\tau_1,\tau_2))$, we have seen that it is well reproduced by a rank-one matrix, with a corresponding vector that is well fitted by an exponential decay in $\tau$. Following Ref. \cite{blanc2017quadratic}, this is interpreted as self- and cross-Zumbach effects: the exponential moving average of the past returns of both {\sc e-mini} and {\sc tbond} (i.e., the ``trend'') is correlated with higher future volatilities.
    \item Moreover, Figure~\ref{fig:Dx_SPUS} demonstrates that the diagonals of $(\Db_{\times,j})$ for ${j\in \{\text{{\sc e},{\sc t}}}\}$ are negative and decay slowly, where here and henceforth we abbreviate subscripts {\sc e-mini} and {\sc tbond} by {\sc e}, {\sc t} respectively. Since the returns of the {\sc e-mini} and the {\sc tbond} are negatively correlated during the studied period (correlation coefficient is $C_{\text{{\sc et}}} \approx -0.15$), $(\Db_{\times,j}) < 0$ means that when past realized covariance is more negative than its average value $-0.15$, then future volatility tends to be higher than usual, and conversely, when the returns of the  {\sc e-mini} and the {\sc tbond} are less negatively correlated, then future volatility tends to be higher than usual. However, the time-diagonal of $(\Db_{\times j})$ also captures pure volatility effects which could explain such negative values as we know that volatility begets volatility. Only the determination of the influence kernels $K_\times^j$ can ascertain the impact of realized correlations in the determination of volatility.   

    \item Finally, Figure~\ref{fig:V_SPUS} recovers the well known self-leverage effect on the {\sc e-mini}: negative past returns are correlated to higher future volatility. But one also clearly sees that negative returns of the {\sc e-mini} increases the volatility of the {\sc tbond} as well (left panels). The upper right panel suggests that the cross-leverage correlation of the {\sc tbond} on the {\sc e-mini} is positive, although more noisy. This is however consistent with the negative correlation between the {\sc e-mini} and {\sc tbond} returns, and could be a spillover effect. Again, the determination of the influence kernels $L_i^j$ will resolve this issue. It is also notable that the {\sc tbond} self-leverage effect (lower right panel) is one order of magnitude larger but decays very fast with lag $\tau$.
\end{enumerate}

\subsection{Calibration using Yule-Walker equations}

This section aims at delineating the calibration method, and illustrating in on the example of the pair {\sc e-mini}--{\sc tbond}. Appendices~\ref{app:YW_system_2DQGARCH} and \ref{app:YW_calibration_matrix} provide further details.\newline

\begin{figure*}
    \centering
     \begin{subfigure}[t]{0.45\textwidth}
         \centering
         \includegraphics[width=\textwidth]{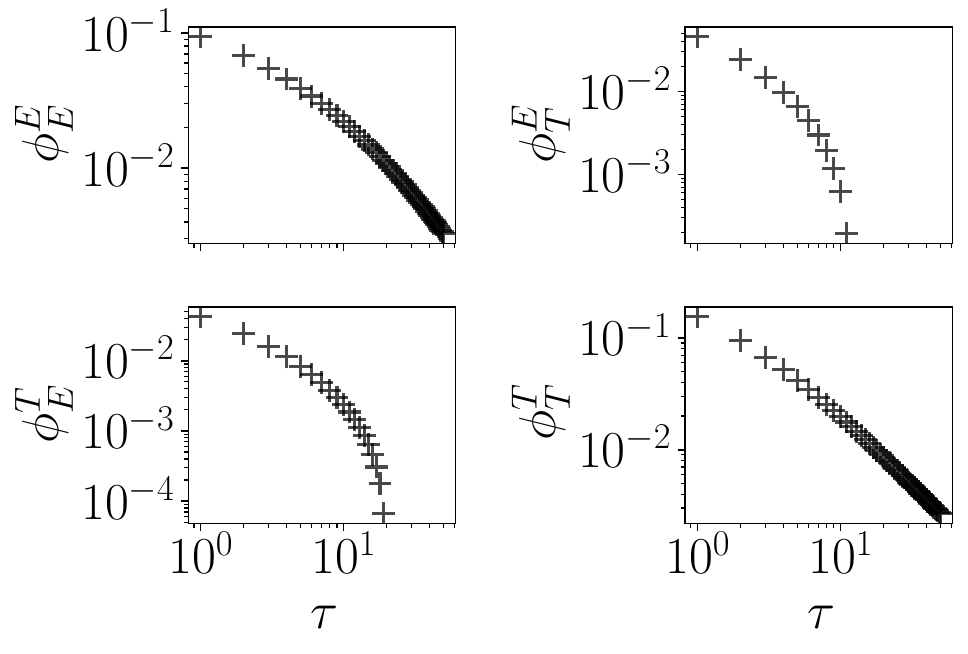}
         \caption{Linear kernels $(\phi^j_i)_{i,j \in {\text{{\sc e},{\sc t}}}}$ up to a lag $q=50$, in log-log scale.}
         \label{subfig:SPUS_h}
     \end{subfigure}
     \hfill
     \begin{subfigure}[t]{0.45\textwidth}
         \centering
         \includegraphics[width=\textwidth]{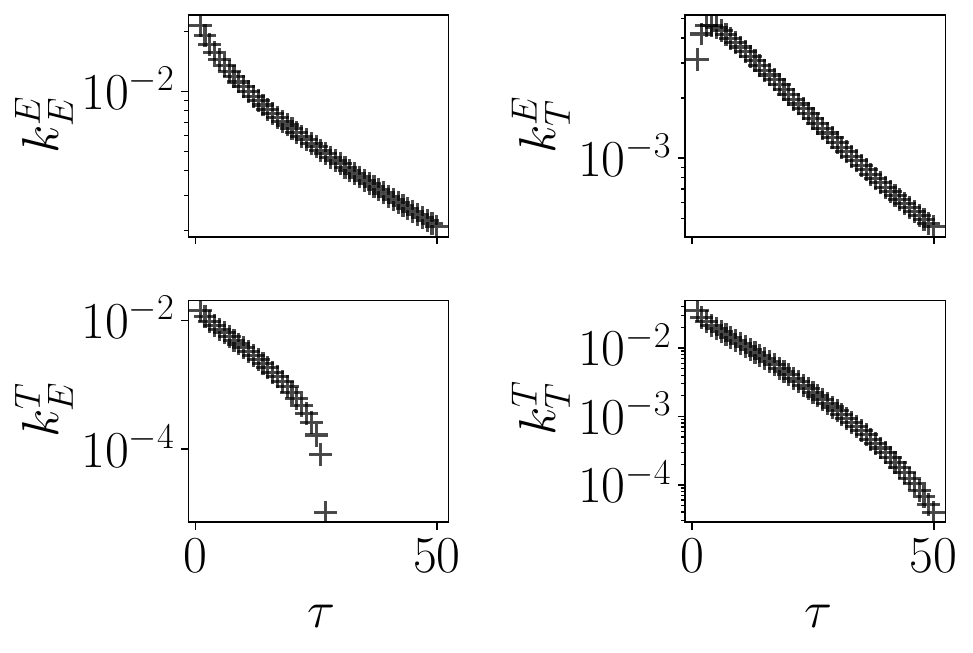}
         \caption{Rank one approximation of the off-time-diagonal of the quadratic kernel $(K^j_i)_{i,j \in {\text{{\sc e},{\sc t}}}}$ up to a lag $q=50$. The y-axis is in log-scale.}
         \label{subfig:SPUS_k}
     \end{subfigure}
     \hfill
     \begin{subfigure}[t]{0.45\textwidth}
         \centering
         \includegraphics[width=\textwidth]{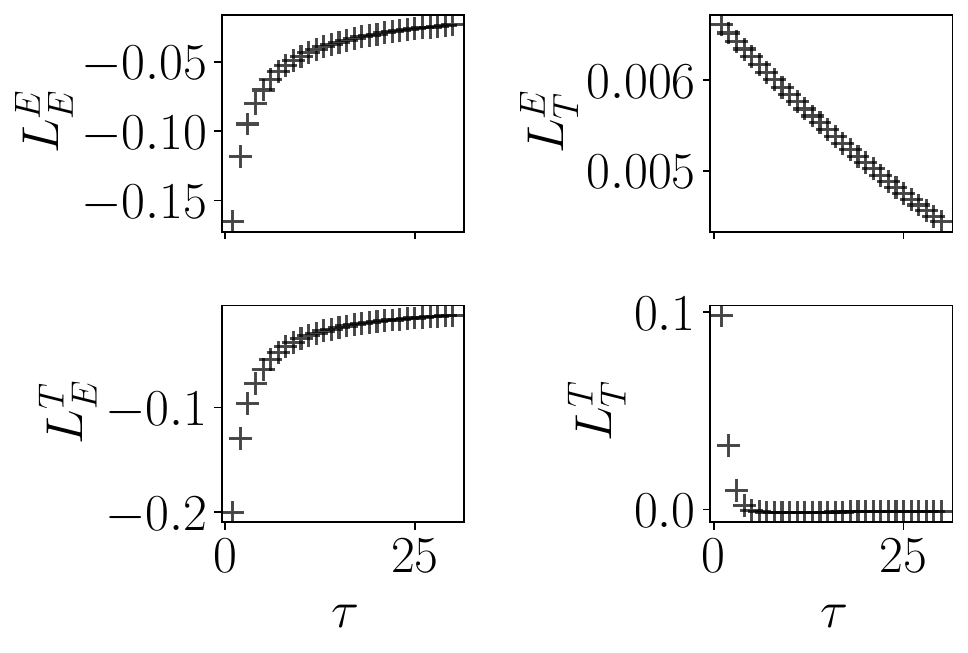}
         \caption{\underline{Left panels}: Leverage kernels of the  {\sc e-mini} feedback $(L^j_{\text{ {\sc e-mini}}})_{j \in {\text{{\sc e},{\sc t}}}}$ up to a lag $q=30$. \underline{Right panels}: Self-leverage kernel of the {\sc tbond} feedback $L^\text{\sc tbond}_{\text{{\sc tbond}}}$ up to a lag $q=30$.}
         \label{subfig:SP_US_L}
     \end{subfigure}
     \hfill
     \begin{subfigure}[t]{0.45\textwidth}
         \centering
         \includegraphics[width=0.45\textwidth]{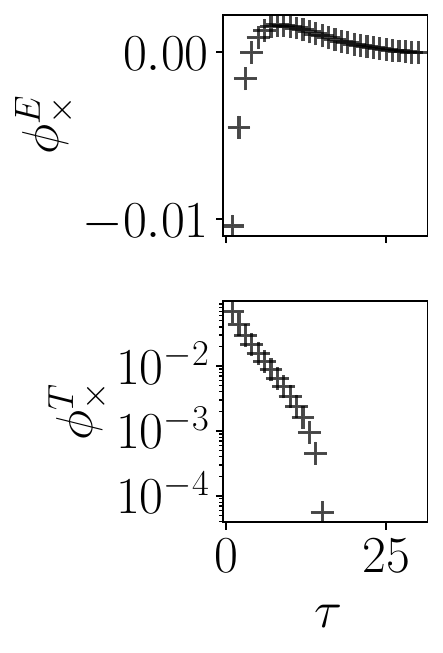}
         \caption{Returns past covariance kernels $(\phi^j_\times)_{j \in {\text{{\sc e}, {\sc t}}}}$ up to a log $q=30$. For the lower panel, the y-axis is in log-scale.}
         \label{subfig:SP_US_phix}
     \end{subfigure}
     %  \begin{subfigure}[b]{0.45\textwidth}
     %     \centering
     %     \includegraphics[width=\textwidth]{images/chap3_L_phix_SPUS.pdf}
     %     \caption{\underline{left panels}: Leverage kernels of the  {\sc e-mini} feedback$(L^j_{\text{Emini}})_{j \in {\text{{\sc e}, {\sc t}}}}$ up to a lag $q=30$. \underline{right panels}: cross kernels $(\phi^j_\times)_{j \in {\text{{\sc e}, {\sc t}}}}$ up to a log $q=30$. The y-axis is in log-scale.}
     %     \label{subfig:SP_US_L_phix}
     % \end{subfigure}
    \caption{Results of the 2D QGARCH calibration on the pair  {\sc e-mini} vs. {\sc tbond}.}
    \label{fig:results_calib_SPUS}
\end{figure*}

\subsubsection{A 4-step procedure}

As explained above, most cross-asset effects can be captured by pair-wise interactions only, so our calibration strategy focuses on the two-dimensional (2D) version of the MQHawkes/MQGARCH model which accounts for leverage, linear and quadratic effects. In order to maintain numerical stability, the calibration process is broken down in successive steps, which take advantage of the fact that the different terms in the MQGARCH model have different orders of magnitude and different symmetries: self-feedback effects ($K_i^i$) are much larger than cross-feedback effects  ($K_{\overline{\imath}}^i$), which are themselves larger than realized covariance effects  ($K^i_\times$). 
\newline 

All steps rely on linear relationships between the covariance and the kernels, the so-called Yule-Walker system of equations. (Appendix~\ref{app:YW_system_2DQGARCH} details those equations system for the model specified in Equation~\eqref{eq:2D_QGARCH_def_sigma}). Solving such linear system of equations is achieved by considering that after a lag $q$ the kernels and the covariance structures are negligible (see  Appendix~\ref{app:YW_calibration_matrix} for details on the implementation of the Yule-Walker systems for the 2D-QGARCH). We now outline the main steps of the calibration.\newline

To accurately distinguish leverage feedback (which break the $r \to -r$ symmetry) from quadratic and cross-feedback effects, we initially estimate the linear, quadratic, and cross kernels using symmetrized $r \to -r$ data. Subsequently, we estimate the leverage kernels on the original dataset, excluding from $\Vb$ the contributions accounted for by the linear feedback. \newline

Specifically, to completely eliminate the leverage effect, we duplicate our datasets as follows. For each time series $r_t, \sigma_t$, we generate a mirror time series $-r_t,\sigma_t$ that we append to compute all correlation functions, with averages noted $\mathbb{E}^\prime$. Hence leverage correlations are identically zero for this expanded data set, while all other correlations structures are strictly identical. Indeed, 
\[
\mathbb{E}^\prime(\sigma_t r_{t-\tau})=\frac12 [\mathbb{E}(\sigma_t r_{t-\tau})-\mathbb{E}(\sigma_t r_{t-\tau})]=0, \] while, for example,
\[
\mathbb{E}^\prime(\sigma_t r^2_{t-\tau})=\frac12 [\mathbb{E}(\sigma_t r^2_{t-\tau})+\mathbb{E}(\sigma_t r^2_{t-\tau})] \equiv \mathbb{E}(\sigma_t r^2_{t-\tau}). \] 
The linear, quadratic and cross kernels are then estimated from in the following three-step procedure, with a final step to retrieve the leverage kernels. \newline

\begin{enumerate}
    \item Since it is numerically stable, as shown in Appendix~\ref{app:calib_sur_simu}, the univariate QHawkes feedback kernels of each asset are estimated up to a certain lag $q$. From this first step, we obtain $(\phi^i_{i}(\tau))_{i\in\{1,2\}, 1\leq \tau\leq q}$ and $(K^i_{i}(\tau_1,\tau_2))_{i\in\{1,2\}, 1\leq \tau_1<\tau_2\leq q}$.
    \item Considering the cross contributions null in a first pass ($\phi_\times=k_\times=0$),  we estimate the cross linear and quadratic contributions characterised by the values $(\phi^i_{j}(\tau))_{i,j\in\{1,2\}, i\neq j,1\leq \tau\leq q}$ \\and $(K^i_{j}(\tau_1,\tau_2))_{i,j\in\{1,2\},i\neq j,1\leq \tau_1<\tau_2\leq q}$. After this second step, the estimation of the baseline activity $\sigma^2_{1,\infty}$ and $\sigma^2_{2,\infty}$ can be achieved using Equation~\eqref{eq:sigmabar_relation}.
    \item Accounting for the contributions of the kernels obtained in steps 1 and 2, we estimate the cross feedback components up to a certain lag $q$, i.e., $(\phi^i_\times(\tau))_{i\in\{1,2\}, 1\leq \tau\leq q}$ and \\$(K^i_{\times}(\tau_1,\tau_2))_{i\in\{1,2\}, 1\leq \tau_1,\tau_2\leq q,\tau_1\neq\tau_2}$.
    \item Once these three steps are completed, the leverage kernels, $(L^j_{i}(\tau))_{i,j\in\{1,2\}, 1\leq \tau\leq q}$, can then be estimated up to a lag $q$ on the original dataset $(r,\sigma)$ while removing the contributions of the quadratic kernels (we refer to Section~\ref{app_sec:leverage} of Appendix~\ref{app:YW_calibration_matrix} for further details).
\end{enumerate}

Note that the calibration using Maximum Likelihood estimation, whose implementation is detailed in Appendix~\ref{app:calib_ML}, has also been implemented and provides similar results as the method of moments. Notably, the Maximum Likelihood estimation is particularly efficient when the GMM outcome is used as an initial guess.
% \end{remark}

% It is worth noticing that only 4 functions are required to describe all the terms of Yule Walker equations (Appendix~\ref{app:YW_calibration_matrix}). Thus, it is enough to define 4 interactions bloc matrices to be combined in order to characterise the linear relationships between the kernels and the covariance structures.

\subsubsection{Some general empirical observations}

First, as already mentioned above, we observe that the empirical cross-covariance structure $\Db_{\times j}(\tau_1,\tau_2)$ is close to zero for $\tau_1\neq \tau_2$. This means that the product $r_{i,t-\tau_1}r_{\imath,t-\tau_2}$ is uncorrelated with $\sigma_{i,t}^2$ unless $\tau_1=\tau_2$: the realized correlation between the returns of two assets only affects future volatility when these returns are simultaneous. This observation leads to $K^j_\times(\tau_1,\tau_2)=0$ when $\tau_1\neq\tau_2$. Consequently, only the time-diagonal $\phi_\times^i (\tau):= K^i_\times(\tau,\tau) $ will be considered hereafter.\newline

Second, consistent with the empirical findings of \cite{blanc2017quadratic}, the calibration of the off-time-diagonal of $K^j_i$ reveals that these kernels can then be well approximated by a rank-one matrix, such that $K^j_i(\tau_1,\tau_2)  \approx k^j_i(\tau_1)k^j_i(\tau_2)$. Indeed, the first eigenvalue of the matrices $K^j_i(\tau_1,\tau_2)$ with a diagonal of zeros, is subtantially larger than the subsequent ones. Besides, the time-diagonal of these kernels is directly represented by $\phi^j_i(\tau)$.\newline

% Appendix~\ref{app:calib_sur_simu} provides a proof of concept of the calibration employing the method of moments on synthetic data.
% \begin{remark}
\section{Calibration of the 2D-QGARCH on futures pairs}

\subsection{Results for the pair  {\sc e-mini}--{\sc tbond}}

Implementing the above procedure on the time series of the  {\sc e-mini} and the {\sc tbond} resulted in the kernels presented in 
Figure~\ref{fig:results_calib_SPUS}. Figure~\ref{subfig:SPUS_k} displays the rank-one approximation $k^j_i(\tau)$ of the off-time-diagonal of the quadratic kernels $(K^j_i)_{i,j\in\{\text{{\sc e},{\sc t}}\}}$. Let us comment on the results.\newline

As already mentioned above, it appears that the covariance structures forms, $\Cb$, $\Db$, $\Db_\times$ and $\Vb$ (see Figures~\ref{fig:C_SPUS}, \ref{fig:Deig_SPUS}, \ref{fig:Dx_SPUS} and \ref{fig:V_SPUS}, respectively), shape the resulting kernels. 

Notably, the long-range nature of $(\Cb_{ii})_{i\in\{\text{{\sc e},{\sc t}}\}}$ is reflected in the kernels $(\phi^i_i)_{i\in\{\text{{\sc e},{\sc t}}\}}$ shown in Figure~\ref{subfig:SPUS_h}. The cross-influence Hawkes kernels $\phi_{\overline{\imath}}^i$ have a smaller amplitude and decay faster but still appear to be very significant. %Indeed, one can also notice that, since $\Cb_{\text{\sc tbond}\text{  {\sc e-mini}}}$ presents the shorter range of correlation, the kernel $\phi^{\text{\sc tbond}}_{\text{ {\sc e-mini}}}$ also has the shorter range.\newline

Figure~\ref{subfig:SPUS_k} exhibits the shape of the ``Zumbach'' kernels $k_i^j$, i.e. the filters that determine the trends on the {\sc e-mini} and the {\sc tbond} which feedback on future volatility of both contracts. Of note: (i) the amplitude and range of the self-kernels $k_i^i$ are similar for {\sc e-mini} and the {\sc tbond} and (ii) the cross-kernels $k_{\overline{\imath}}^i$ are similar magnitude as the self-kernels, although -- perhaps surprisingly -- the decay of $k_{\sc{e}}^{\sc{t}}$ is faster than that of $k_{\sc{t}}^{\sc{e}}$: trends on the {\sc tbond} has a longer lasting effect on the volatility of the 
{\sc e-mini} than the other way round. 

Turning now to Figure~\ref{subfig:SP_US_L}, we see that the negative exponential shape of the influence of the {\sc e-mini} past returns on both $\sigma^2_{\text{\sc e}}$ and $\sigma^2_{\text{\sc t}}$ is well-captured by the kernels $(L^i_{\text{{\sc e}}})_{i\in\{\text{{\sc e},{\sc t}}\}}$. Additionally, the leverage kernel capturing the feedback of {\sc tbond} returns on its own volatility, depicted in the lower right panel of Figure~\ref{subfig:SP_US_L}, exhibits a positive shape that rapidly decays to zero. This observation is consistent with the shape of $\Vb_{\text{\sc tbond}}$ in Figure~\ref{fig:V_SPUS}, and with the negative correlation between the {\sc e-mini} and the {\sc tbond} returns. Consistently with the observation of Figure~\ref{fig:V_SPUS}, the leverage kernel representing the feedback of the {\sc tbond} returns on the  {\sc e-mini} volatility is weakly positive, again consistent with the  negative correlation between the {\sc e-mini} and the {\sc tbond} returns. Note that stronger leverage signals are expected for larger time scales, as the leverage effect predominantly manifests at the daily scale, as demonstrated by the first figure of \cite{aubrun2023multivariate}.

Finally, Figure~\ref{subfig:SP_US_phix} represents the perhaps less trivial of all feedback kernels, $(\phi^j_\times)_{j\in\{\text{{\sc e},{\sc t}}\}}$, that describe the influence of the realized covariance between the returns of the {\sc e-mini} and the {\sc tbond} on future volatility of both assets. These graphs indicate that when the realized covariance is more negative than usual (a.k.a. a flight to quality mode), it tends to decrease the {\sc tbond} future volatility while it increases the {\sc e-mini} volatility for short time horizon before perhaps reverting sign for longer time horizons. 
%as already discussed in Figure~\ref{fig:Dx_SPUS}
% \blue{check $\phi_\times$ - 
Note that this is not exactly the effect that was previously discussed for $(\Db_{\times j})_{j\in \{\text{{\sc e},{\sc t}}\}}$ in Figure~\ref{fig:Dx_SPUS}. In fact, the kernels $(\phi^j_\times)_{j\in\{\text{{\sc e},{\sc t}}\}}$ reveal solely the effect of past returns covariance, while $(\Db_{\times j})_{j\in \{\text{{\sc e},{\sc t}}\}}$ conflate feedback effects arising from the Hawkes kernels. \newline  %}.

\subsection{Extension to other pairs of futures contracts}

We now extend the 2D-QGARCH calibration over all pairs composed by the following six futures: four on very correlated indices:  {\sc e-mini},  {\sc e-mini}-3 (3 months futures), {\sc nasdaq}, and {\sc dow jones}, one on commodities ({\sc crude oil}) and finally {\sc tbond}. 

The temporal structure of all kernels are found to be very similar to those obtained for the pair {\sc e-mini}--{\sc tbond}, so we chose to only display here the L1-norm of the corresponding kernels in Figure~\ref{fig:futures_norm}; see Appendix~\ref{app:future_on_indices_profiles} for their full profiles.\newline

\begin{figure*}
    \centering
     \begin{subfigure}[t]{0.45\textwidth}
         \centering
         \includegraphics[width=\textwidth]{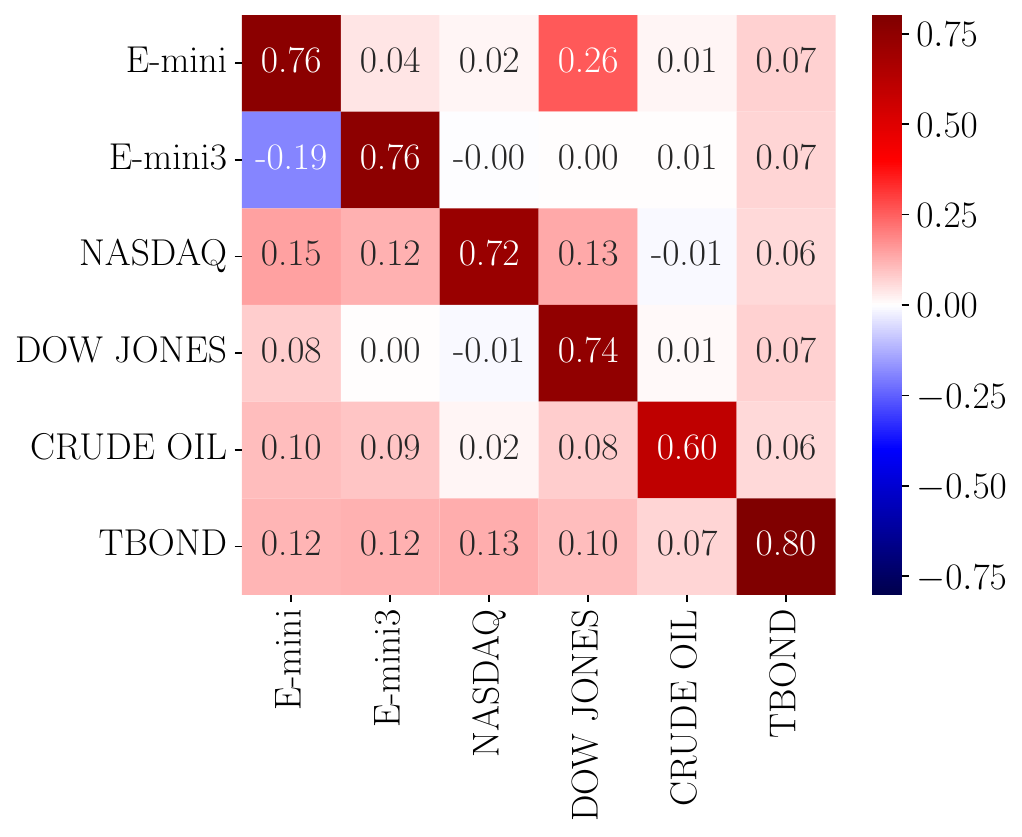}
         \caption{Norm of the linear kernels $||\phi^i_j||_1$, with $i$  the row index and $j$ the column index.}
         \label{subfig:norm_phi}
     \end{subfigure}
     \hfill
      \begin{subfigure}[t]{0.45\textwidth}
         \centering
         \includegraphics[width=\textwidth]{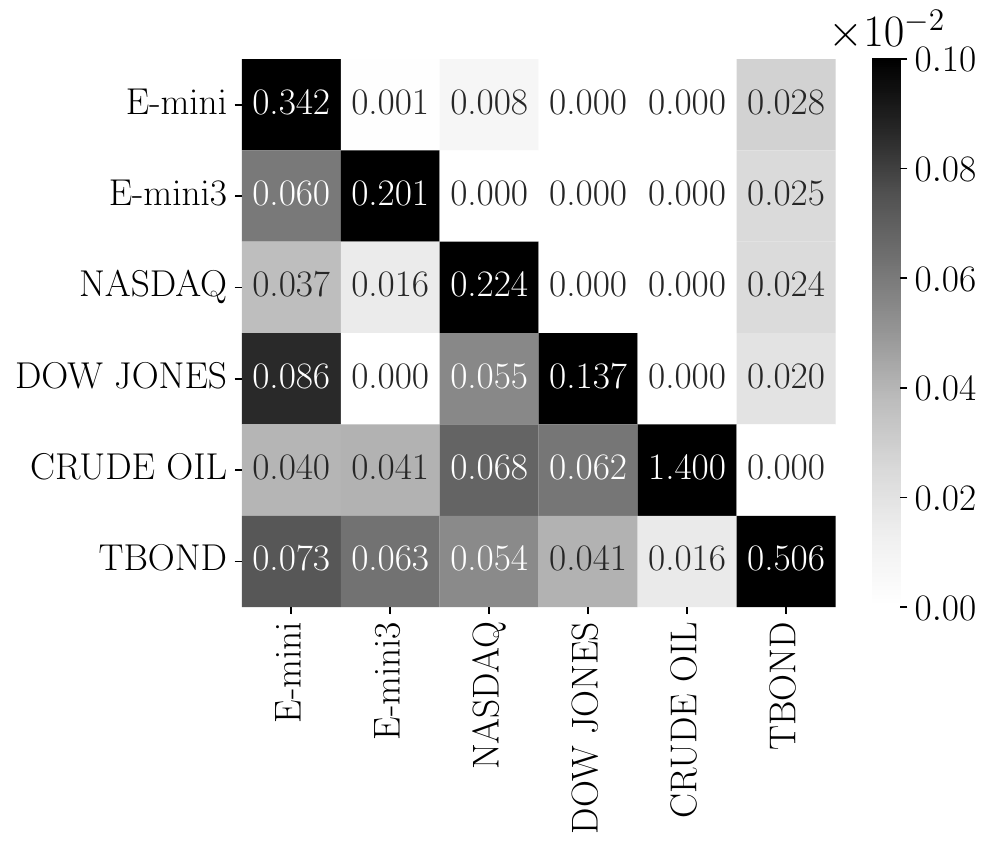}
         \caption{Norm of the rank-one approximation of the off-diagonal of the quadratic kernels $||(k^i_j)^2||_1$, with $i$  the row index and $j$ the column index.}
         \label{subfig:ksq_norm}
     \end{subfigure}
     \hfill
      \begin{subfigure}[t]{0.45\textwidth}
         \centering
         \includegraphics[width=\textwidth]{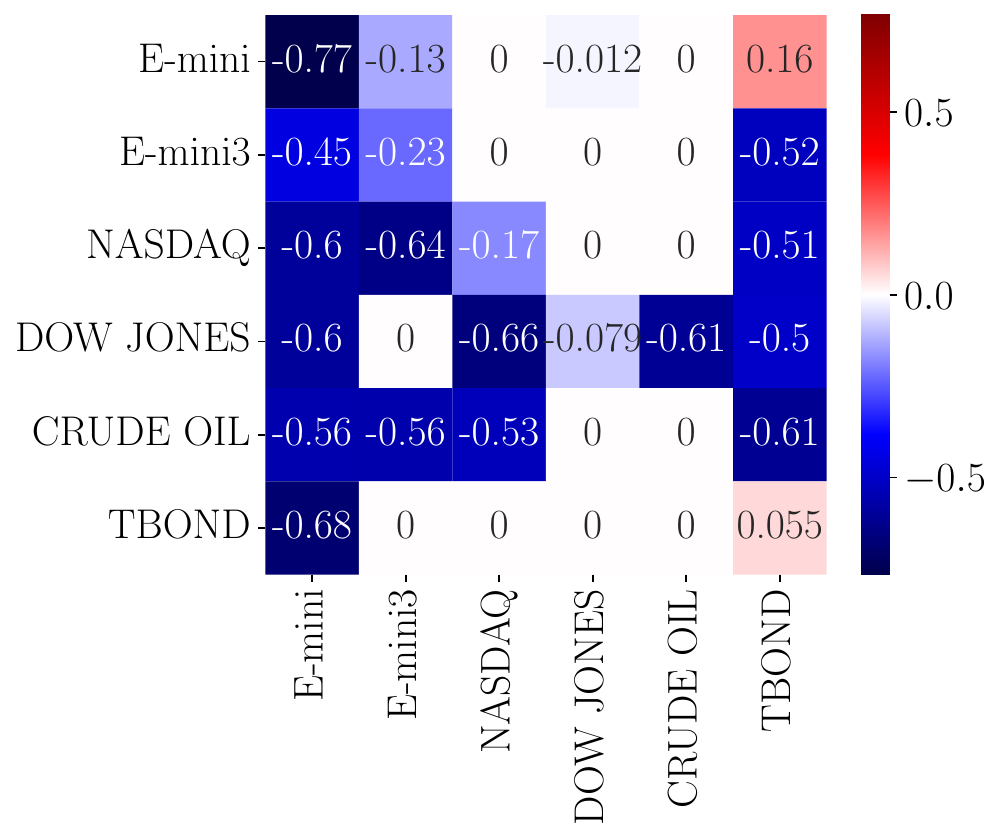}
         \caption{Norm of the leverage kernels $||L^i_j||_1$, with $i$  the row index and $j$ the column index.}
         \label{subfig:L_norm}
     \end{subfigure}
     \hfill
      \begin{subfigure}[t]{0.45\textwidth}
         \centering
         \includegraphics[width=\textwidth]{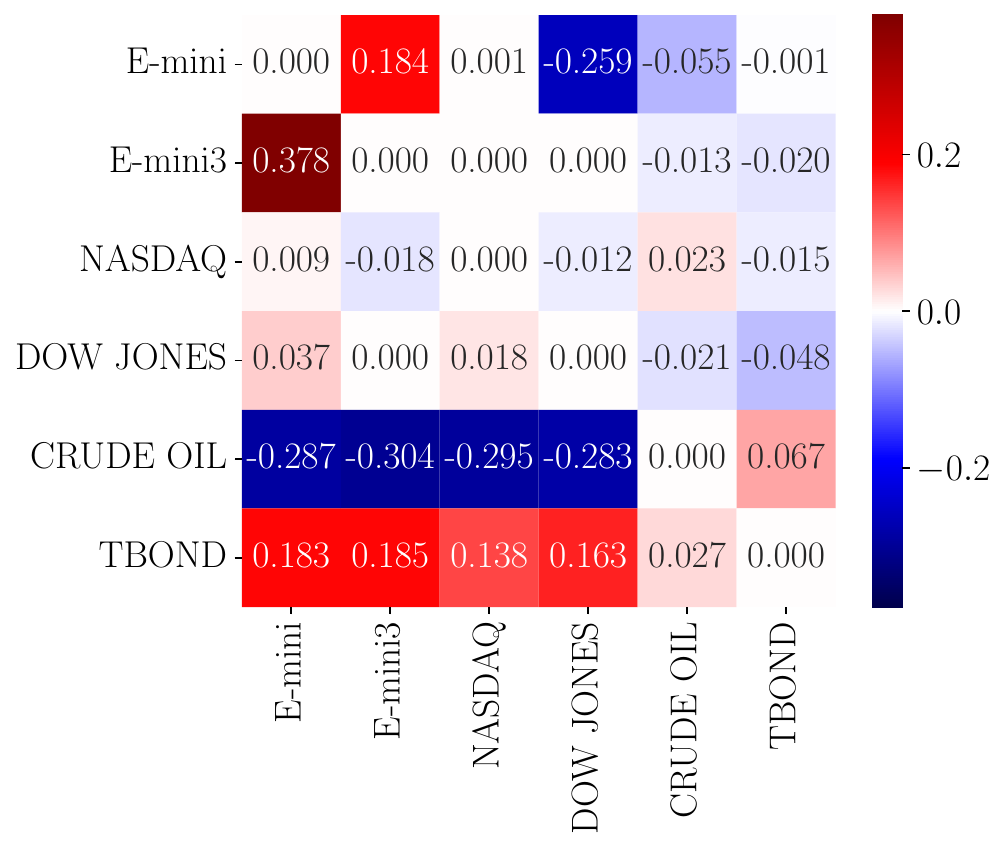}
         \caption{Norm of the cross kernels $||\phi^i_\times||_1$. The row index $i$ corresponds to the target, and the column index $j$ the other member of the pair defining $\times$. }
         \label{subfig:phi_times_norm}
     \end{subfigure}
     \hfill
    \caption{Norms of the kernels from the calibration of 2D-QGARCHs on pairs of futures on indices. For the four sub-figures, the x-labels determine the index $i$ corresponding to the ``source'' and the y-labels determine the index $j$ corresponding to the ``target'' of the feedback.}
    \label{fig:futures_norm}
\end{figure*}

\subsubsection{Volatility feedback (Hawkes) kernels $\phi^i_j$}

The primary observation from Figure~\ref{fig:futures_norm} concerns the kernels characterising the influence of past volatility on future volatility, $\phi^i_j$, shown in Figure~\ref{subfig:norm_phi}. The strong diagonal of Figure~\ref{subfig:norm_phi} indicates that most of the system endogeneity originates from the Hawkes feedback of the asset's own volatility on itself. Additionally, Figure~\ref{subfig:norm_phi} demonstrates that the system operates near criticality, as the diagonal values are all above 0.6 and the largest eigenvalues of the pairwise matrices $||\phi^i_j||_1)$ are between 0.73 and 0.90.
Of note is the rather strong column corresponding to the {\sc e-mini} as a trigger of volatility for all other markets, whereas the column corresponding to {\sc nasdaq} is surprisingly weak.
%Subsidiary observations regarding the other kernels are worth noting.\newline

\subsubsection{Zumbach trend feedback kernels $k^i_j$}

The norms of the rank-one approximations $k^i_j$ of the off-time-diagonal of the the quadratic kernels $K_j^i$, in Figure~\ref{subfig:ksq_norm}, are consistent with their profiles in Appendix~\ref{app:future_on_indices_profiles}. The strong diagonal of Figure~\ref{subfig:ksq_norm} shows that the self-Zumbach feedback is quite strong compared to the cross-Zumbach effects. Furthermore, some futures present particularly trend quadratic feedback, like {\sc crude oil}. 

As for the Hawkes feedback, the first column of Figure~\ref{subfig:ksq_norm} 
demonstrates that the cross-Zumbach effect arising from the  {\sc e-mini} on futures on other markets is quite strong compared to the other cross-Zumbach kernels, meaning that the S\&P500 index past trends predominantly increase the volatility of the other futures on indices. Similarly, the last column of Figure~\ref{subfig:ksq_norm} shows that the Zumbach feedback of the {\sc tbond} on stock indices is significant and consistent across contracts.
%meaning that the trend of the futures on the US-10Y-Treasury-bond increases uniformly the volatility of the futures on stock-indices. It is also interesting to note that futures on the {\sc crude oil} and on the 10Y-Treasury bond appear to be the most impacted by these trend quadratic feedback, as demonstrated by the two last rows of Figure~\ref{subfig:ksq_norm}.\newline \newline

\subsubsection{Leverage feedback kernels $L^i_j$}

% It appears that leverage kernels are not always relevant, as evidenced by the many zeros in  Figure~\ref{subfig:L_norm} and exemplified by the pair  {\sc e-mini} {\sc vs tbond}. 

It appears that leverage kernels are not always relevant at the intraday level, as evidenced by the many zeros in  Figure~\ref{subfig:L_norm}. In fact, certain leverage kernels derived from our calibration process exhibit several constraints inherent to our framework, prompting the exclusion of the corresponding norms and profiles depicted in Figures~\ref{subfig:L_norm} and~\ref{fig:L_profile_futures}. 
The primary limitation is the finite dimensionality of the calibration, which is further emphasized by our strategy to remove the influence of overnight effects by averaging correlations solely within individual days. This approach is particularly significant in the context of leverage kernels, as the leverage effect predominantly appears at the daily scale. 
Additionally, leverage kernels are computed only during the fourth and last step of our method, following the removal of the linear kernels', $\phi$, contribution from the correlation $\Vb$. Consequently, the final correlation used to derive $L$ is notably influenced by the outcomes of the initial correlation step. 

When present, except in the case of the {\sc tbond} as discussed above, the leverage feedback is negative, consistent with the idea that large negative returns trigger increased volatility. Notably, the {\sc dow jones} futures\footnote{US index.} seems to be particularly affected by the leverage effect from both itself and other futures on indices. We also point out again that the model is calibrated intraday, whereas the leverage effect predominantly appears on a daily scale.\newline

\subsubsection{Covariance feedback kernels $\phi^i_\times$}

Figure~\ref{subfig:phi_times_norm} reports on the influence of past realized covariance of the returns of the two assets involved in each pair on the volatility of both assets in the pair. 

The pair {\sc e-mini}--{\sc e-mini}-3 exhibits a particularly strong returns covariance feedback compared to the other pairs (see the upper left 2$\times$2 matrix of Figure~\ref{subfig:phi_times_norm}), which means that an increased past covariance between the returns of the {\sc e-mini} and those of the {\sc e-mini}-3 tends to increase the volatility of both futures. 

Additionally, the volatility of the ({\sc tbond}) and of the ({\sc crude oil}) appear to be significantly impacted by the past covariance of their returns with the returns of US stock-indices, as demonstrated by the last two rows of Figure~\ref{subfig:phi_times_norm}. Specifically, for all the pairs combining the {\sc crude oil} with stock-indices, the returns covariance feedback kernels $\phi_\times$ exhibit a negative norm, while the pairs combining a stock-index with the {\sc tbond}, display a positive returns covariance feedback. 

During the studied period, {\sc crude oil} and stocks were positively correlated while stocks and bonds were negatively correlated.  Therefore, the sign of the norm of the returns covariance feedback kernels $||\phi^i_\times||$ for pairs combining stock-indices with the {\sc tbond} and for pairs combining stock-indices with the {\sc crude oil} conveys the same meaning in both cases: when the absolute value of the past returns covariance increases, it tends to decrease future volatility of both {\sc tbond} and {\sc crude oil}. The opposite effect on the {\sc e-mini} is much weaker and has the opposite sign with the {\sc tbond}, as already noted above.
%Conversely, when the two involved assets, generally in phase (or in opposite phase in the case of the {\sc tbond}), decouple, it tends to increase future volatility.\newline
%While this effect appears to be symmetric, albeit slightly down, for pairs combining future on stock-index with the {\sc crude oil} futures, meaning that an increase past covariance between the {\sc crude oil} returns and those of a stock-index tends to decrease the stock-index volatility. It is, nevertheless, different for pairs combining future on stock-index with the {\sc tbond}. Notably, an increase, in absolute value, covariance between the {\sc tbond} returns and a stock-index returns tends to increase the stock-index volatility.
As for the pair {\sc crude oil}--{\sc tbond}, an increased, in absolute value, of the past returns covariance tends to decrease {\sc tbond} volatility while it increases {\sc crude oil} volatility.  

% See profiles in Appendix~\ref{app:future_on_indices_profiles}
% \newpage

\section{A MQARCH factor-model}

To calibrate on a large set of stocks, while accommodating for the specific  feedback from the index to which they belong, or from more general factors we define and calibrate a 2D-QGARCH factor model. The framework is outlined below.

\subsection{Model framework}

The starting point is a 1-factor model. Specifically, the returns of stock $i$ are decomposed into a sum of their market exposure, represented by the factor component $f_0$, with an exposure coefficient $\beta_i$, and a residual component $e_{i,t}$, which is idiosyncratic to the stock dynamics:
\begin{equation}\label{eq:factor_model_def}
    \begin{aligned}
        r^i_t = \beta_i f_{0,t} + e_{i,t}.
    \end{aligned}
\end{equation}
Empirically, $f_0$ represents the market mode and will be directly sourced from the  {\sc e-mini} returns studied in the previous section, while $\beta_i$ is estimated from the correlations between the returns of stock $i$ and $f_0$. 

The particularity of such framework, and our main interest, lies in the non-linear dynamics of $f_0$ and $e_i$. Specifically, we posit here that the volatility of the market factor $f_0$ is unaffected by the dynamics of residuals $e_i$ and given by a 1D-QGARCH, while the $e_{i}$s themselves follow a 2D-QGARCH model with (i) self-feedback and (ii) feedback from the market factor. This will account for the empirical observation that the volatility of the residuals $e_i$ depends on that of the market mode $f_0$ \cite{cizeau2001correlation,lillo2000symmetry,allez2011individual,chicheportiche2015nested}. 

Correspondingly, the volatility of $f_0$ is written as
\begin{equation*}
    \begin{aligned}
         \sigma^{2}_{0,t}=& \sigma^2_{0,\infty} + 
         \sum^{+\infty}_{k=1} L^0_0(k)f_{0,t-k} +  \sum^{+\infty}_{k=1} \phi^0_0(k)f^2_{0,t-k}\\&+2 \sum^{+\infty}_{k_1=1} \sum^{+\infty}_{k_2=k_1+1} K^0_0(k_1,k_2)f_{0,t-k_1}f_{0,t-k_2},
    \end{aligned}
\end{equation*}
while the volatility of $e_i$ is specified as
\begin{equation}
    \begin{aligned}
        \sigma^{2}_{i,t}=& \sigma^2_{i,\infty} + \sum^{+\infty}_{k=1} L^i_i(k)e_{i,t-k}+\sum^{+\infty}_{k=1} L^i_0(k)f_{0,t-k}\\&
        +\sum^{+\infty}_{k_1=1} \sum^{+\infty}_{k_2=1} K^i_i(k_1,k_2)e_{i,t-k_1}e_{i,t-k_2}\\&+\sum^{+\infty}_{k_1=1} \sum^{+\infty}_{k_2=1} K^i_0(k_1,k_2)f_{0,t-k_1}f_{0,t-k_2} \\&+ \sum^{+\infty}_{k_1,k_2=1} K^i_\times(k_1,k_2)e_{i,t-k_1}f_{0,t-k_2}
    \end{aligned}
\end{equation}
Scrutinizing this last expression, we see that the idiosyncratic volatility, $\sigma_i^2$, is influenced by the past returns of both the factor and the residual, characterised by (i) the leverage kernels $L_\times$ and $L_i$ respectively, (ii) by the past trends of both the factor and the residual, described by the quadratic kernels $K_0^i$ and $K_i^i$ respectively and (iii) a covariance feedback effect described by $K^i_\times$. Note that since the correlation between factor $f_0$ and residuals $e_i$ is zero by construction, we do not need to subtract the average $C_{0i}$ from $e_{i,t-k}f_{0,t-k}$ in the last term.

In fact, empirical data suggests that the covariance feedback effect between $e_i$ and $f_0$ is negligible on intraday timescales but may become relevant at lower frequencies (at variance with the case of the {\sc e-mini} -- {\sc tbond} pair). We leave this question open for future investigations and will drop the $K^i_\times$ terms henceforth. {As before, we can decompose the quadratic kernels to highlight the diagonal feedback (Hawkes feedback of the activity onto itself), which primarily characterizes the endogeneity of the system, and the off-diagonal feedback (Zumbach feedback of the trends) as follows:} 
\begin{equation}
    \begin{aligned}
        \sigma^{2}_{i,t}=& \sigma^2_{i,\infty} + \sum^{+\infty}_{k=1} L^i_i(k)e_{i,t-k}+\sum^{+\infty}_{k=1} L^i_0(k)f_{0,t-k}\\&
        +\sum^{+\infty}_{k=1} \phi^i_i(k)e^2_{i,t-k}+\sum^{+\infty}_{k=1} \phi^i_0(k)f^2_{0,t-k}\\&
        +2\sum^{+\infty}_{k_1=1} \sum^{+\infty}_{k_2=k_1+1} K^i_i(k_1,k_2)e_{i,t-k_1}e_{i,t-k_2}\\&+2\sum^{+\infty}_{k_1=1} \sum^{+\infty}_{k_2=k_1+1} K^i_0(k_1,k_2)f_{0,t-k_1}f_{0,t-k_2}
    \end{aligned}
\end{equation}

The goal of this section is to investigate the values of $\sigma_{i,\infty}^2$ and the features of the various feedback kernels determining the residual future volatility for US stocks, in particular the role of the index factor $f_0$ on this volatility since we know from the work of Refs.~\cite{cizeau2001correlation,lillo2000symmetry,allez2011individual,chicheportiche2015nested} that the equal time correlation between residual volatilities and the market factor is non-trivial. Of particular interest are the Zumbach self-trend and cross-trend effects.  

The subsequent section is dedicated to presenting the data used in this section, followed by a discussion of the results obtained.

\subsection{Data processing}\label{sec:factorxQGarch_datapreprocessing}

Our initial dataset contains 1-minute returns $(r^i)_{i=1,...,317}$ of 317 US stocks spanning from 2013 to 2023 and the 1-minute  {\sc e-mini} returns, $f_0$, for the same period. We consider only stocks which belonged to the S\&P500 index for the whole studied period. 

This initial dataset is then used to deduce the time series $(e_i)_i$ according to Equation~\eqref{eq:factor_model_def} and taking, for all $i$, $\beta_i$ as the covariance coefficients between the returns of stock $i$ and the returns of the  {\sc e-mini}. The squared volatility time series, $\sigma_{i,t}^2$ and $\sigma_{f_0,t}^2$, are then approximated by the squared 1-minute returns, $e_{i,t}^2$ and $f_{0,t}^2$ respectively. 

At this stage, we thus end up with two time series for the  {\sc e-mini} and for each stock of our 317 US stocks sample: one characterising the returns and the second one representing the squared volatility. These times series are then processed as before following the steps described in Section~\ref{sec:preprocessing}.\newline

The method of moments is then used to calibrate the 1D-QGARCH on the  {\sc e-mini}, resulting in the values of $\sigma^2_{0,\infty}$, $(L_0(k))_{1\leq k\leq q}$ and $(K_0(k_1,k_2))_{1\leq k_1,k_2\leq q}$, where $q$ is the lag beyond which feedback effects are considered negligible.

Finally, the calibration of the 2D-QGARCH on the 317 stocks residuals, adapting the method described in Section~\ref{sec:calibration_met} and Appendix~\ref{app:YW_calibration_matrix}, is performed, with results presented in the next section.

\subsection{MQARCH 1-factor model: Calibration Results}

\begin{figure*}
    \centering
    \includegraphics[width=\linewidth]{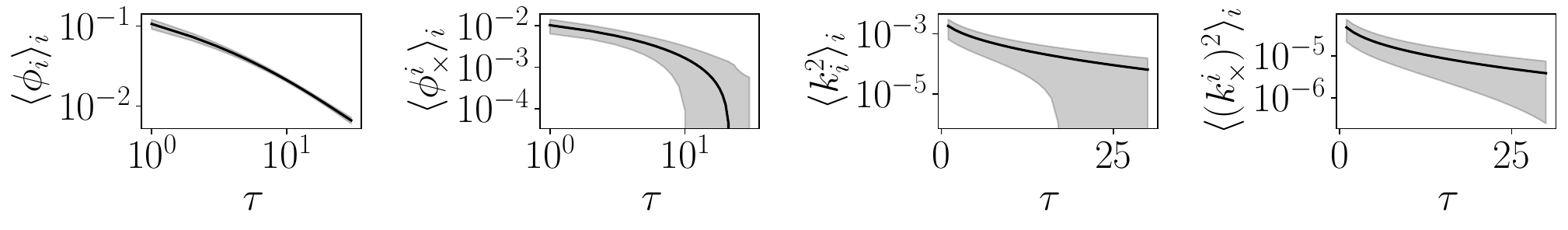}
    \caption{Average profiles of kernels for the dynamics of the residuals $e_i$. The solid black line represents the average profile and the grey area indicates the range of one standard deviation across the stocks.}
\label{fig:average_profiles_qhawkesxfactors}
\end{figure*}

\begin{figure}[b]
  % \begin{center}
    \centering
    \includegraphics[width=0.5\linewidth]{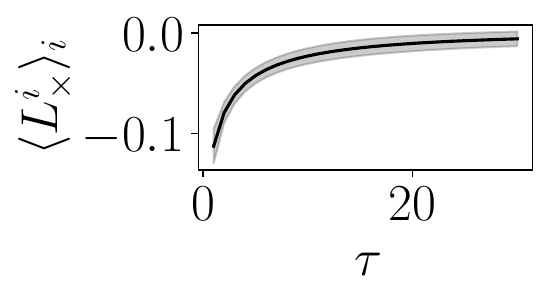}
  % \end{center}
  \caption{Average profile of the leverage kernel from the  {\sc e-mini} on the residual $L^i_\times$.}
  \label{fig:Lx_average}
  
\end{figure}

The kernels $L_0^0$, $\phi_0^0$ and $k_0^0$,  where $\phi_0^0$ and $k_0$ are respectively the time-diagonal of $K_0^0$ and the rank-one approximation of the off-time-diagonal of $K_0$, result from the calibration of a 1D-QGARCH on the  {\sc e-mini}, and are therefore identical to those already computed in the previous section and shown in Figure~\ref{fig:results_calib_SPUS}. 

Figure~\ref{fig:average_profiles_qhawkesxfactors} presents the {\it cross-sectionally averaged} profiles of the quadratic kernels shaping the volatility of the stock idiosyncratic dynamics, i.e. the volatility feedback kernels $\langle \phi_i^i \rangle_i$ and $\langle \phi^i_0 \rangle_i$, and the rank-one approximation of the off-time-diagonal of the feedback kernels, $\langle k_i^i \rangle_i$ and $\langle k_0^i \rangle_i$.

% \blue{TO DISCUSS}Interestingly, both the diagonal and the off-diagonal of the quadratic kernels, $(K_i)_i$ and $(K^i_\times)_i$, seem to exhibit a long-range decay scale even though the smoothing is still made out of an exponential decay for the three points correlations, as illustrated in Figure~\ref{fig:Deig_SPUS}. Consequently, both residual and index trends, as well as volatility events that occurred early in the day, continue to influence the residual volatility at the end of the day.\newline

%second version
% A first observation can be made concerning the long-range scale of both linear kernels characterising the residual volatility feedback, $(\phi_i)_i$,  and that of the factor $(\phi^i_\times)$. Indeed, the log-log scale of the two left panels of Figure~\ref{fig:average_profiles_qhawkesxfactors} illustrates that residual or index volatility events that occurred early in the day, continue to influence the residual volatility at the end of the day. 

%Additionally the right panels of Figure~\ref{fig:average_profiles_qhawkesxfactors}, even if only the y-axis is in log-scale, show that the rank-one approximation of the off-time-diagonal of the quadratic kernels  scale and influence the residual volatility for a long time  even though the smoothing is still made out of an exponential decay for the three points correlations, as illustrated in Figure~\ref{fig:Deig_SPUS}, the rank-one 

Interestingly, on average the norm of the residual volatility feedback kernel is close to one ($\langle||\phi^i_i||\rangle_i = 0.94$), supporting the aforementioned results suggesting that endogeneity arises chiefly from the self-volatility feedback. Additionally, the kernels characterising the residual volatility feedback on itself exhibit a long-range, power-law behaviour, as demonstrated by the log-log scale of the left panel of Figure~\ref{fig:average_profiles_qhawkesxfactors}. This means that the idiosyncratic activity that occurred early in the day influences the realized idiosyncratic volatility throughout the entire day (recall that we have removed any intraday seasonal pattern in the data). The thin grey area on the left panel of Figure~\ref{fig:average_profiles_qhawkesxfactors} indicates the cross-sectional dispersion of individual kernels around the sample mean. We thus conclude that the Hawkes kernels $\phi_i^i(\tau)$ are quite uniform across stocks, which is by no means a trivial result.

Moreover, the average endogeneity arising from the market contribution is not negligible, reaching up to $||\phi^i_0||_{\text{max}}=0.12$, with values hovering around $\langle||\phi^i_0||\rangle_i = 0.02$. Past volatility of the market factor appears to predominantly influence the residual volatility of stocks in the Financials, Real Estate, and Industrials sectors. Additionally, the log-log scale of the second panel from the left of Figure~\ref{fig:average_profiles_qhawkesxfactors} can be fitted by a power-law function truncated by an exponential.

As previously found, the off-time-diagonal of the trend feedback kernels, as characterised by their rank-one approximation, have lower values than the diagonal feedback kernels with a norm averaging at $\langle||k^2_i||\rangle_i = 0.012$. This is consistent with the results of the calibration of the QHawkes on stocks in \cite{blanc2017quadratic}. The average profile of $k_i$s does not reflect strong differences between stocks for small lags, but such differences tend to increase with lag as demonstrated by the large grey area in the third panel from the left of Figure~\ref{fig:average_profiles_qhawkesxfactors}. This might however be due to numerical noise, which becomes more important for large lags. Notably, the higher the residual activity, the longer the range of the residual Zumbach feedback. Additionally, the Real Estate and Utilities sectors, which show the highest average levels of idiosyncratic activity, consistently present trend feedback kernels with the highest norms and longer temporal spans on average, specifically,  $\langle||(k^i_i)^2||\rangle_{i\in \text{real estate}} = 0.017$ and $\langle||(k^i_i)^2||\rangle_{i\in \text{utilities}} = 0.018$. 

The kernels characterising the market past trends contribution to future residual volatility exhibit much smaller but still clearly discernible values,  with a norm averaging $\langle||(k^i_0)^2||\rangle_i=4.7\;\; 10^{-4}$, as shown by the profile in the right panel of Figure~\ref{fig:average_profiles_qhawkesxfactors}. Interestingly, the self-Zumbach effect of the index increases the activity of the index, which, in turn, increases the activity of the stock through $\phi_0^i$,
%the cross-Zumbach effect between the index and individual stocks is also mediated by the self-Zumbach effect on the index itself (i.e. $k_0^0 \to \phi_0^i$), 
with a net effect of the same order of magnitude as the direct coupling $k_0^i$. Once again, the Real Estate and Utilities sectors stand out as having the most significant direct trend index feedback.
%\blue{The log-scale on the y-axis of the two right panels of Figure~\ref{fig:average_profiles_qhawkesxfactors} seem to highlight that these rank-one approximation's shape is that of an  exponential, probably since the smoothing is still made out of an exponential decay for the three points correlations, as illustrated in Figure~\ref{fig:Deig_SPUS}}

Surprisingly, we found highly disparate results across our sample of stocks for the self-leverage kernels $L_i^i$. In fact, some stocks exhibit positive residual leverage feedback. Note however that the leverage effect predominantly manifests itself at the daily scale.\newline 

Nevertheless, the leverage feedback $L_0^i$ of the index on the volatility of the residuals is quite stable across stocks and its average profile is represented in Figure~\ref{fig:Lx_average}, together with a thin grey zone indicating a very small amount of cross-sectional dispersion. The negative exponential shape of $L_0^i$ indicates that the {\sc e-mini} negative past returns directly increase the residual volatility of single stocks. This sheds a complementary light to the results presented in \cite{reigneron2011principal}.

Finally, the endogeneity ratios of these 2D-QGARCH-1 factor systems\footnote{The endogeneity ratio of the 2D-QGARCH-1 factor model for stock $i$ is the largest eigenvalue of $\begin{pmatrix}
    ||\phi^0_0||&0\\||\phi_0^i||&||\phi_i^0||
\end{pmatrix}$, which is thus directly given by $\max(||\phi_0^0||,||\phi_i^0||)$.} are very close to 1, with all values exceeding 0.955 and an average of 0.96, indicating that the systems operate near criticality. Correspondingly the baseline fraction of the stock idiosyncratic activity averages at $\langle\sigma^2_{i,\infty}\rangle_i=0.04$. 

To recap, the calibration of our model on single stocks yields two main results. First, most of the system endogeneity stems from the long-range feedback of the idiosyncratic volatility, consistent with previous findings on the calibration of futures on indices. Second, past volatility and negative returns of the index both significantly impact the residual volatility of stocks, supporting empirical findings that the idiosyncratic volatility of stocks is strongly coupled to that of the market as a whole \cite{lillo2000symmetry,cizeau2001correlation,allez2011individual, chicheportiche2015nested}. %Finally, index and residual trend kernels seem to exhibit long-range decay, indicating that start-of-day trend events concerning both idiosyncratic and index dynamics, influence the stock residual volatility throughout the entire day.

A natural extension of this framework would be to investigate a factor model with more than one factor, thereby characterising the influence of the index not only on the stock's idiosyncratic component but also on the various industrial sectors, whose dynamics would be accounted for by the additional factors.

%%%%%%%%%%%%%%%%%%%%%%%%%%%%%%%%%%%%%%%%%%

\section{Conclusion}\label{subsec:chap2bconclu}
\addcontentsline{toc}{section}{\nameref{subsec:chap2bconclu}}

Let us summarise our work. 
Building upon the work of Blanc \textit{et al.} \cite{blanc2017quadratic} and our theoretical work presented in \cite{aubrun2023multivariate}, we presented a non-parametric calibration method to characterise the feedback mechanisms that influence the dynamics of future volatility. To overcome the challenges associated with estimating an underlying MQHawkes process from aggregated data, the micro-scale MQHawkes framework is approximated by a discrete MQGARCH framework. The key to this transformation is assimilating the Hawkes intensity as the squared volatility, up to a multiplicative factor. 

We introduced the key ingredients and steps required to calibrate the 2D-MQARCH framework using the method of moments, employing the pair  {\sc e-mini} vs. {\sc tbond} as a pedagogical example. The primary idea of the calibration is to build a linear system of equations using the covariance structures, which are directly observable in real data, to deduce the kernel functions. The difficulty we had to overcome to get stable and meaningful results is that these covariance structures have quite different orders of magnitude and relate to different symmetries of the feedback terms. This has led us to propose a robust 4-step calibration scheme, where dominant effects are first accounted for, before including weaker effects.   

Using this calibration method on pairs of futures revealed that the strongest feedback effect on future volatility originates from the past volatility itself and the sign of past returns (``leverage''), followed by the self-Zumbach effect. Cross effects are smaller, but clearly significant and interpretable. In particular, our analysis demonstrated that the past trends and returns of the {\sc e-mini} contribute significantly to the activity of other futures, compared to other cross-feedback effects. Interestingly, we have discovered a new stylized fact: realized {\it correlation} between the {\sc e-mini} and other futures contracts does influence the volatility of these contracts.  

To investigate the feedback mechanisms on stock volatility while distinguishing market dynamics from stock idiosyncratic dynamics, we combined the 2D-QGARCH framework with a 1-factor model, which allowed us to disentangle the self-influence from the influence of the market factor on the idiosyncratic activity of single stocks. Calibrating the 2D-QGARCH 1-factor model on 317 US stocks proved to be quite stable across stocks. Notably, while most of the system endogeneity stems from the influence of past residual activity on itself, the past returns and past volatility of the market factor still significantly impact the future volatility of the stock idiosyncratic component. In our analysis, we observed some disparities among industrial sectors. It would be interesting to further develop a MQARCH multi-factor framework to investigate the feedback loops between the market, the industrial sectors and the stock idiosyncratic component. We leave such a study for future research. 

Finally, it would be interesting to undertake a direct calibration of the MQHawkes process using synchronized, tick by tick data. However, some of the feedback terms that we have discussed seem to arise at lower frequencies, so for many purposes extending our MQARCH calibration method to daily data would also be insightful. \newline

\section*{Acknowledgements}
We would like to thank Jérôme Garnier-Brun and
Salma Elomari, for their insights on data analysis and
calibration.
We are also grateful to Natascha Hey and Jutta Kurth
for their research internships which gave a lot of insights
on the calibration on tick-by-tick data and on multivariate processes.
This research was conducted within the Econophysics
\& Complex Systems Research Chair, under the aegis of
the Fondation du Risque, the Fondation de l’Ecole polytechnique, the Ecole polytechnique and Capital Fund
Management.

\bibliographystyle{vancouver}
\bibliography{biblio}

\clearpage

\appendix

\small 
\onecolumngrid

\section{Calibration - Proof of concept}\label{app:calib_sur_simu}

This appendix provides a proof of concept for the proposed calibration method, demonstrating its results using synthetic data. Notably, we discuss the challenges associated with accurately retrieving the (Q)Hawkes process, an event-by-event point process, from aggregated data. These difficulties justify our decision to approximate the QHawkes process with a QGARCH model, as described by Equation~\eqref{app_eq_MM:sigma^2_def}.

This appendix is organised as follows.  It begins by comparing the calibration of the univariate (Q)Hawkes model with that of the univariate (Q)GARCH model. This comparison shows that, in certain regimes, the calibration of the Hawkes becomes biased due to data binning whereas the QGARCH remains stable across a wider range of parameters. 

The second part focuses on calibrating the multivariate (Q)GARCH using the method of moments, providing evidence that the calibration method is reliable.  

To compare the calibration of the (Q)Hawkes with the calibration on the (Q)GARCH, we generate datasets in two ways: 
\begin{itemize}
    \item The thinning algorithm \cite{ogata1981lewis} provide time and mark of events for the (Q)Hawkes, enabling the construction of price time series, according to the definition of the QHawkes, i.e. ${\rm d}P=\pm {\rm d}N$. These time series are then aggregated into 1 minute bins to create the volatility time series $\sigma^\mathcal{B}$ and the 1-minute returns time series ${\rm d}P$. These synthetic time series are used to test the calibration of the (Q)Hawkes model.
    \item The simulation of the (Q)GARCH, as described in Equation~\eqref{app_eq_MM:sigma^2_def}, directly results in two time series at 1 minute timescale: the volatility $\sigma^2$ and the returns $r$. These time series are used to test the calibration of the QGARCH.
\end{itemize}
In both cases, we use exponential feedback kernels with the general form: 
\begin{align*}
    \phi(t)=n_H\beta\exp(-\beta t)\quad\&\quad
    k(t)=\sqrt{2n_Z\omega}\exp(-\omega t)
\end{align*}

\subsection{Univariate cases: comparison between the (Q)Hawkes and the (Q)GARCH models}\label{app_sec:calib_sur_simu_1D}

This first part is dedicated to show and discuss the calibration results of the univariate (Q)Hawkes and of the univariate (Q)GARCH. It begins with the linear case and then deals with the quadratic extension.

\subsubsection{Univariate linear case}

\begin{figure}[h]
     \centering
     \begin{subfigure}[t]{0.35\textwidth}
         \centering
         \includegraphics[width=\textwidth]{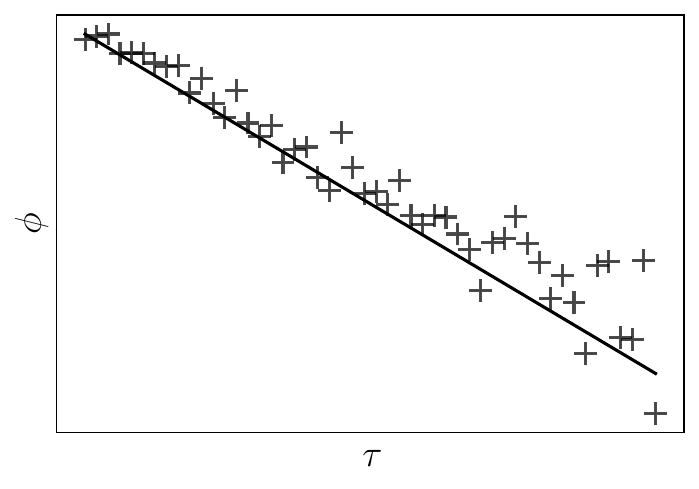}
         \caption{In the case $\bar{\lambda}=0.03 \text{ min}^{-1}$ ($\lambda_\infty = 0.01 \text{ min}^{-1}$).}
         \label{subfig:linHawkes_baseline001}
     \end{subfigure}
     \hfill
     \begin{subfigure}[t]{0.35\textwidth}
         \centering
         \includegraphics[width=\textwidth]{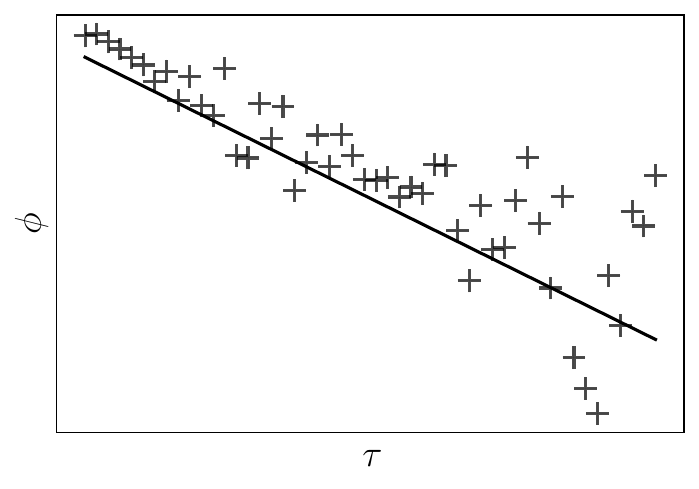}
         \caption{In the case $\bar{\lambda}=1.7 \text{ min}^{-1}$ ($\lambda_\infty = 0.5 \text{ min}^{-1}$).}
         \label{subfig:linHawkes_baseline05}
     \end{subfigure}
     \caption{Calibration of a 1D linear Hawkes with exponential kernel, for two different sets of parameters.
     In both cases, $n_H=0.7$ and $\beta=0.04 \text{ min}^{-1}$ and only the baseline varies. The scatter plot is the kernel obtained by the method of moments and the plain line is the kernel used to generate the synthetic data and that we want to recover.}
     \label{app_fig:lin_Hawkes_1D}
\end{figure}

As a reminder, the linear Hawkes process is a point process $N$, whose intensity is defined as follows: 

\begin{equation*}
    \lambda_t = \lambda_\infty+\int_{-\infty}^t \phi(t-s){\rm d}N_s \quad \text{ with returns defined as }\quad {\rm d}P_t=\pm {\rm d}N_t.
\end{equation*}

In the case of a linear GARCH process, the squared volatility $\sigma^2$ is driven by 

\begin{equation*}
    \sigma^2_t = \sigma^2_\infty+\sum_{k=1}^{+\infty}\phi(k)r^2_{t-k}\quad \text{ with returns defined as } r_t=\sigma_t \xi_t, \quad \text{with}\quad \xi_t\sim\mathcal{N}(0,1).
\end{equation*}

The connection between the two models lies in the relationship between the GARCH squared volatility and the Hawkes intensity: $\frac{\sigma^2}{\Delta} \equiv \lambda_t$, where $\Delta$ is the time bin used to compute $\sigma$ \cite{blanc2017quadratic}. It is interesting to note that, by definition, the Hawkes model is a continuous process describing each event of price change whereas the GARCH model is inherently discretized in time. 

The objective of the calibration is to retrieve the kernel $\phi$. Specifically, in the case of a exponential kernel,  $\phi=n_H\beta\exp(-\beta t)$, we aim to determine the parameters $n_H$, $\beta$ and $\lambda_\infty$ (or $\sigma^2_\infty$). When the shape of the kernel is unknown, as it is often the case with empirical data, adopting a non-parametric method such as the general method of moments (GMM) allows us to remain agnostic about the kernel's shape. In this appendix, we implement the GMM on synthetic data to test its validity.\newline

\begin{figure}[h!]
     \centering
     \begin{subfigure}[b]{0.35\textwidth}
         \centering
         \includegraphics[width=\textwidth]{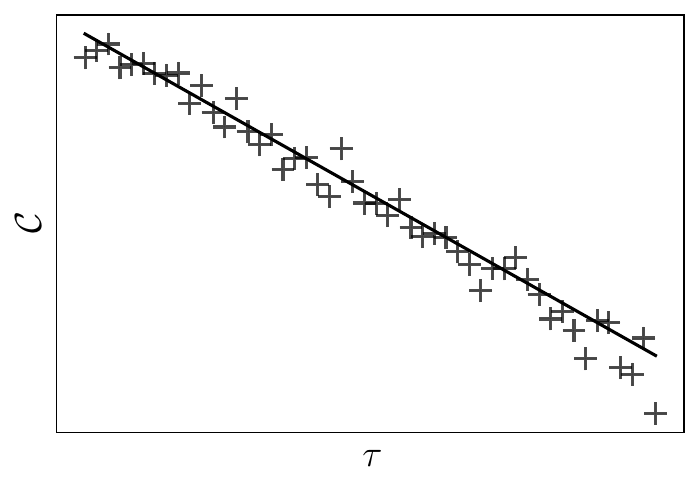}
         \caption{In the case $\bar{\lambda}=0.03 \text{ min}^{-1}$ ($\lambda_\infty = 0.01 \text{ min}^{-1}$).}
         \label{app_subfig:Cb001}
     \end{subfigure}
     \hfill
     \begin{subfigure}[b]{0.35\textwidth}
         \centering
         \includegraphics[width=\textwidth]{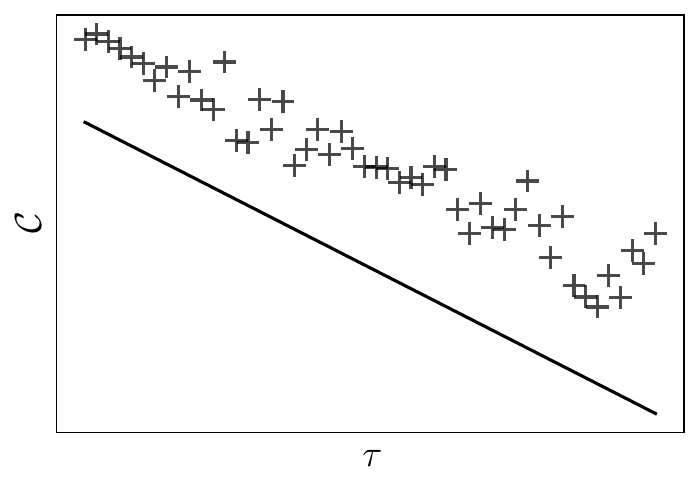}
         \caption{In the case $\bar{\lambda}=0.1.7 \text{ min}^{-1}$ ($\lambda_\infty = 0.5 \text{ min}^{-1}$).}
         \label{app_subfig:Cb05}
     \end{subfigure}
     \caption{Illustration of the estimation of the covariance structure $\Cb(\tau) = \frac{1}{\bar{\lambda}}\left(\mathbb{E}(\frac{{\rm d}N_t}{{\rm d}t}\frac{{\rm d}N_{t-\tau}}{{\rm d}t})-\bar{\lambda}^2\right)$ from the generation of a 1D linear Hawkes with exponential kernel, for 2 different sets of parameters. In both cases, $n_H=0.7$ and $\beta=0.04 \text{ min}^{-1}$ and only the baseline varies. The scatter plot is the obtained covariance $\Cb$, while the plain line is the theoretical covariance $\Cb$ \cite{bouchaud2018trades}.}
    \label{app_fig:Cb_linearHawkes} 
\end{figure}
% \newpage

Two key elements about the method of moments should be kept in mind. First, it relies on autocovariance structure, and in particular in the linear case, on the autocovariance of the process's activity $\Cb(\tau) = \frac{1}{\bar{\lambda}}\left(\mathbb{E}(\frac{{\rm d}N_t}{{\rm d}t}\frac{{\rm d}N_{t-\tau}}{{\rm d}t})-\bar{\lambda}^2\right)$ (for the GARCH, an equivalent expression is $\Cb(\tau) = \mathbb{E}(\sigma^2_tr^2_{t-\tau})-\overline{\sigma^2r^2}$). These values form the linear system of equations required to recover $\phi$ (for more comprehensive details see \cite{blanc2017quadratic} and Appendix~\ref{app:YW_calibration_matrix}, in particular Section~\ref{app_sec:A1_structure}).
 Secondly, the method of moments yields the values $(\phi(\tau))_{1\leq \tau\leq q}$, where $q$ is the lag after which the covariance and kernels are considered null.

We now proceed to the presentation and discussion of the calibration results of the Hawkes process.

Figure~\ref{app_fig:lin_Hawkes_1D} shows the estimations of the linear kernels $\phi$ calibrated on synthetic data generated by linear Hawkes processes for two sets of parameters. Specifically, we compare the results obtained from a simulation with $\bar{\lambda}{\rm d}t = 0.03$ to those from a second simulation with $\bar{\lambda}{\rm d}t = 1.7$. The observations are manifolds. Firstly, although the calibration is noisy, the method of moments successfully retrieves the exponential shape of the input kernel $\phi$. Secondly, while the calibration method remains consistent across both parameter sets, it appears to yield better results when $\bar{\lambda}{\rm d}t\ll1$. In this regime, the data binning to obtain $\sigma^\mathcal{B}$ from $N$ and $P$ ensures that there is at most one event per bin, and so $\sigma^\mathcal{B}\approx {\rm d}N$. However, when $\bar{\lambda}{\rm d}t\geq1$, there are more events within each bin, leading to a overestimation of the activity.

For illustration, Figure~\ref{app_fig:Cb_linearHawkes} compares the autocovariance, $\Cb$, of the synthetic activity with its theoretical values \cite{bouchaud2018trades}. As shown in Figure~\ref{app_subfig:Cb001}, when $\bar{\lambda}{\rm d}t\ll1$, the estimated autocovariance matched its theoretical values. In contrast, Figure~\ref{app_subfig:Cb05} demonstrates that $\Cb$ is overestimated when $\bar{\lambda}{\rm d}t=1.7$. These estimation errors in $\Cb$ result in a biased estimation of the kernel $\phi$ as the Yule-Walker system to find $\phi$ relies solely on $\Cb$.

For real data, due to the high activity of financial markets, we are situated in a regime where $\bar{\lambda}{\rm d}t\geq 6$. Hence, if the underlying process is a linear Hawkes, we expect the autocovariance $\Cb$ to be significantly overestimated. For the linear Hawkes process, it is possible to determine a correction to apply to $\Cb$ to retrieve the correct theoretical value and thus to correct the bias in the resulting kernel $\phi$. However, extending this correction to the quadratic case is challenging.

% \newpage

\begin{figure}[h]
    \centering
    \includegraphics[width=0.38\linewidth]{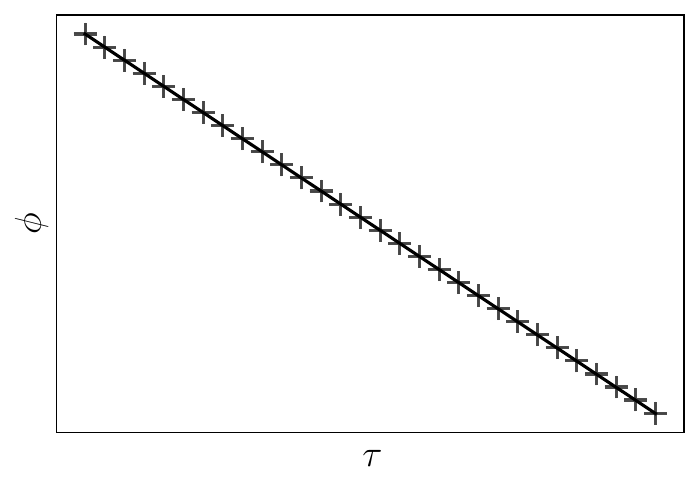}
    \caption{Proof of concept - Calibration of a 1D linear GARCH with exponential kernel. The scatter plot is the kernel obtained by the method of moments and the plain line is the kernel used to generate the synthetic data and that we want to recover. For this simulation $n_H=0.7$, $\beta=0.1\text{ min}^{-1}$ and $\sigma^2_\infty=0.1$}
    \label{app_fig:linearGarch1D}
\end{figure}

In comparison, as shown in Figure~\ref{app_fig:linearGarch1D}, for instance, the calibration of the univariate linear GARCH remains stable across a wide range of parameters, provided that an appropriate number of lags $q$ is chosen (see Appendix~\ref{app:YW_calibration_matrix}). A consistent choice is to take $q>\frac{3}{\beta}$.

Note also that the calibration is much less noisy for the GARCH model compared to the linear Hawkes model.

We now turn our attention to the quadratic case.

\newpage

\subsubsection{Univariate quadratic case}\label{app_sec:univariate_quadratic_calib_sur_simu}

The quadratic case, for both the QHawkes and the QGARCH, is a lot more delicate. 

Introduced in \cite{blanc2017quadratic}, the intensity of a QHawkes process is defined by

\begin{equation*}
    \lambda_t = \lambda_\infty+\int_{-\infty}^t\int_{-\infty}^t K(t-s,t-u){\rm d}P_s{\rm d}P_s  \quad \text{ with returns defined as }\quad {\rm d}P_t=\pm {\rm d}N_t.
\end{equation*}

In the case of a QGARCH process, as defined in \cite{chicheportiche2014fine}, the square volatility, $\sigma^2$, is driven by:   

\begin{equation*}
    \sigma^2_t = \sigma^2_\infty+\sum_{i=1}^{+\infty}\sum_{j=1}^{+\infty}K(i,j)r_{t-i}r_{t-j}\quad \text{ with returns defined as } r_t=\sigma_t \xi, \quad \text{with}\quad \xi\sim\mathcal{N}(0,1).
\end{equation*}

The connection between the two models remains the same as in the linear case, that is $\frac{\sigma^2}{\Delta} \equiv \lambda_t$.

In the quadratic cases, the kernel $K$ can be understood as a matrix such that $\mathbb{K}_{ij}=K(i,j)$. The objective of the calibration is to estimate the entries of the matrix $\mathbb{K}$ up to a certain lag $q$; these entries correspond to the values $(K(i,j))_{1\leq i,j\leq q}$. Moreover, since $K$ is symmetric, it is only necessary to determine the upper triangle of $\mathbb{K}$, specifically the values $(K(i,j))_{1\leq i\leq j\leq q}$.

\begin{figure}[h]
    \centering
    \includegraphics[width=0.7\linewidth]{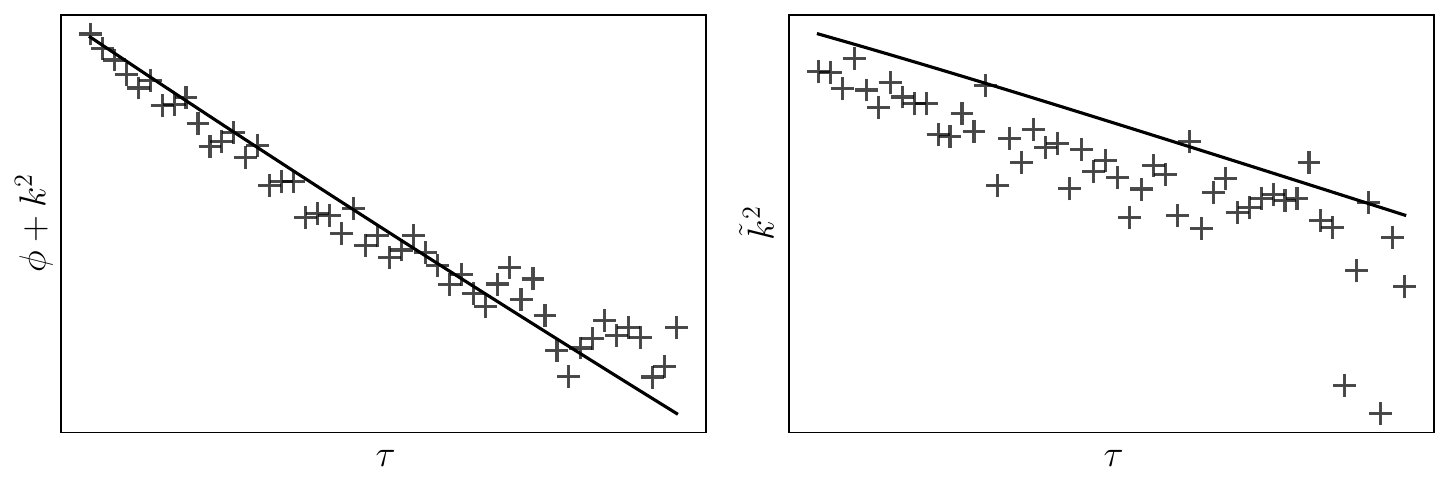}
    \caption{Calibration of a 1D QHawkes with method of moments from a simulation with parameters $n_H=0.6$, $\beta=0.04\text{ min}^{-1}$, $n_Z=0.2$, $\omega=0.03\text{ min}^{-1}$, $\lambda_\infty=0.003\text{ min}^{-1}$, hence $\bar{\lambda}{\rm d}t=0.015$. The left panel presents the diagonal of the kernel $\mathbb{K}(\tau,\tau)=\phi(\tau)+k^2(\tau)$ up to a lag $q=50$. The right panel presents the rank-one decomposition of the off-diagonal of $\mathbb{K}$. In both panels the scatter plot is the component from the method of moments, while the plain line is the theoretical component we want to recover.}
    \label{app_fig:calibQHawkes}
\end{figure}

Following the empirical observation of Blanc \textit{et al.} \cite{blanc2017quadratic}, we adopt the ZHawkes framework which expresses the kernels $K$ as the sum of a diagonal contribution and a rank-one component, meaning that  $K(\tau_1,\tau_2) = \phi(\tau_1)\delta_{\tau_1-\tau_2} + k(\tau_1)k(\tau_2)$. For exponential kernels, $\phi$ and $k$ take the form defined above. Note that the ZHawkes framework also greatly simplifies the simulation and limit the number of parameters to determine. Consistently with the explanation above, the method of moments, as detailed in Appendix~\ref{app:YW_calibration_matrix}, returns the estimated diagonal of $\mathbb{K}$ up to a certain lag $q$, that is $K(\tau,\tau)_{\tau \in \llbracket 1, q\rrbracket}$, and its upper triangle values, $K(\tau_1,\tau_2)_{1\leq\tau_1<\tau_2\leq q}$. Therefore, within the ZHawkes framework, the diagonal of $\mathbb{K}$ ($K(\tau,\tau)_{\tau \in \llbracket 1, q\rrbracket}$) provides an an estimation of $\phi + k^2$, and the upper triangle values of $\mathbb{K}$ ($K(\tau_1,\tau_2)_{1\leq\tau_1<\tau_2\leq q}$) allow to rebuild the off-diagonal of $\mathbb{K}$ (since $\mathbb{K}$ is symmetric) and approximate it with a rank-one representation, so called hereafter $\Tilde{k}$. Figure~\ref{app_fig:calibQHawkes} shows the result of such calibration on synthetic data generated under the condition $\Bar{\lambda}{\rm d}t\ll 1$. Similar to the linear Hawkes process, this regime ensures that there are very few events per bin, thereby maintaining the relationship ${\rm d}P=\pm {\rm d}N$ and approximating $\sigma^{\mathcal{B}}\approx{\rm d}N$.
 
\begin{figure}
    \centering
    \includegraphics[width=0.7\linewidth]{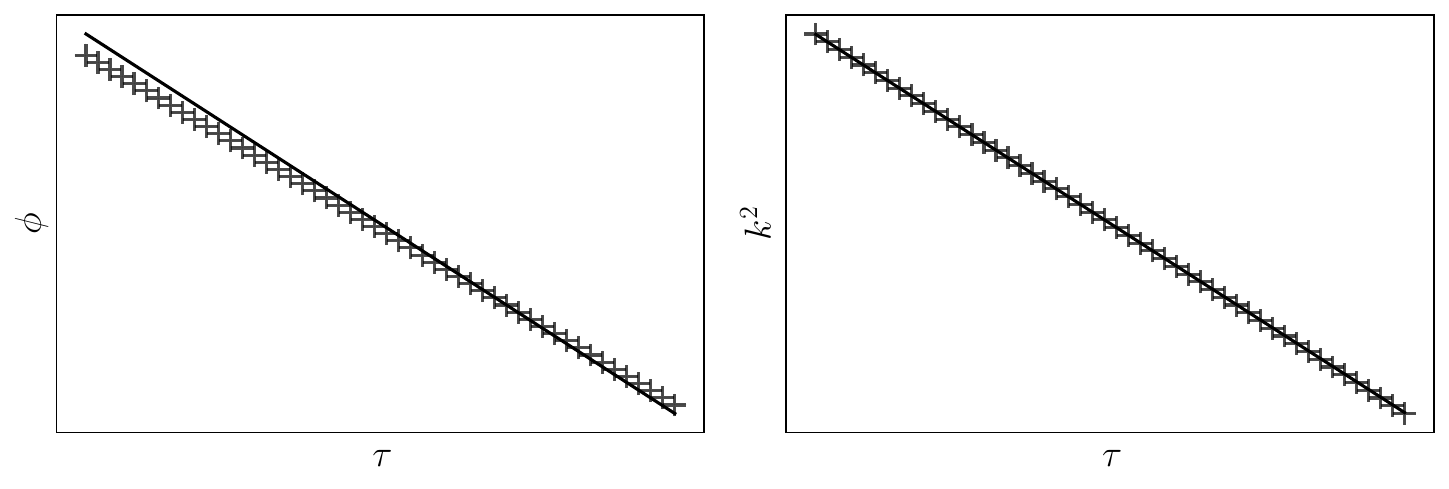}
    \caption{Calibration of a 1D QHawkes with method of moments combined with a parametric minimisation from a simulation with parameters $n_H=0.6$, $\beta=0.04\text{ min}^{-1}$, $n_Z=0.2$, $\omega=0.03\text{ min}^{-1}$, $\lambda_\infty=0.003\text{ min}^{-1}$, hence $\bar{\lambda}{\rm d}t=0.015$. The left panel presents the kernel $\phi$. The right panel present the kernel $k$. In both panels the scatter plot is the optimised kernel obtained from the method of moments, while the plain line is the theoretical kernel with input parameters, we want to determine.}
    \label{app_fig:calibQHawkes_opt}
\end{figure}

From the method of moments estimation, one can reconstruct the entire kernel matrix $\mathbb{K}$ up to a lag $q$ by combining the diagonal and the upper triangular entries, considering that $\mathbb{K}$ is a symmetric $q\times q$ matrix. Subsequently, the parameters $n_H$, $n_Z$, $\beta$ and $\omega$ can be retrieved by minimizing $[\hat{K}(\tau_1, \tau_2) - \phi(\tau_1)\delta_{\tau_1-\tau_2} - k(\tau_1)k(\tau_2)]^2$, where $\hat{K}$ is the estimated matrix from the method of moments and $\phi$ and $k$ are the parametric kernels defined above. The results of such minimisation are illustrated in Figure~\ref{app_fig:calibQHawkes_opt}. Note that the obtained values of $n_H$, $n_Z$, $\beta$ and $\omega$ can then serve as initial guesses to perform an estimation by maximum likelihood as described in Appendix~\ref{app:calib_ML}.

\begin{figure}[h]
     \centering
     \begin{subfigure}[b]{\textwidth}
         \centering
         \includegraphics[width=0.7\textwidth]{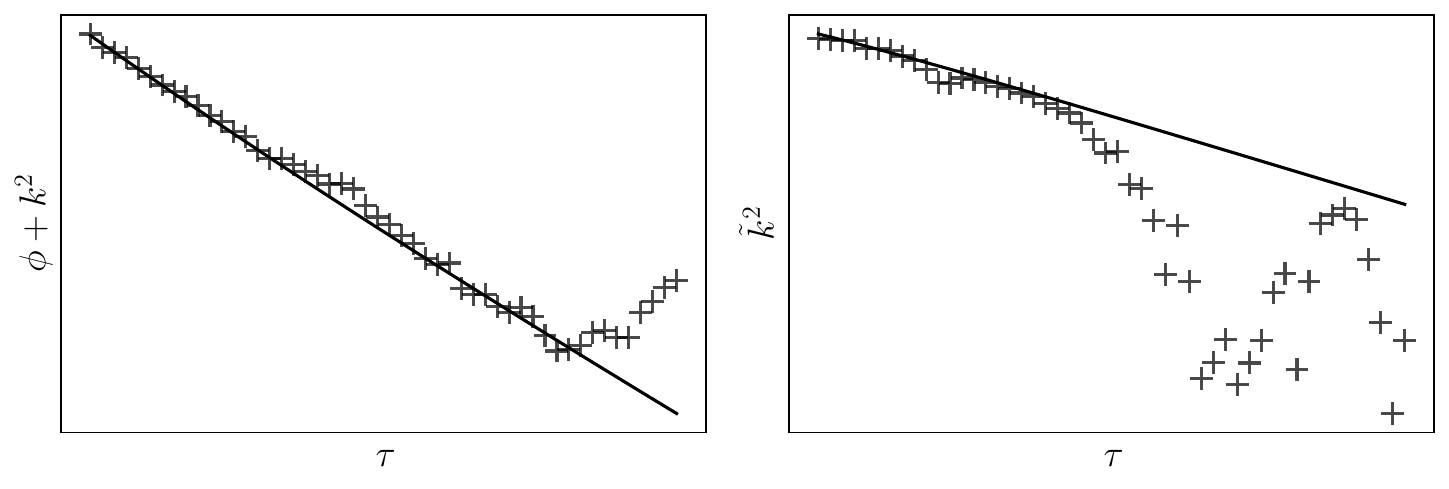}
         \caption{Calibration of a QGARCH for $\bar{\lambda}=0.7$, obtained with $\sigma^2_\infty=0.1$}
         \label{app_subfig:QGARCH07}
     \end{subfigure}
     \hfill
     \begin{subfigure}[b]{\textwidth}
         \centering
         \includegraphics[width=0.7\textwidth]{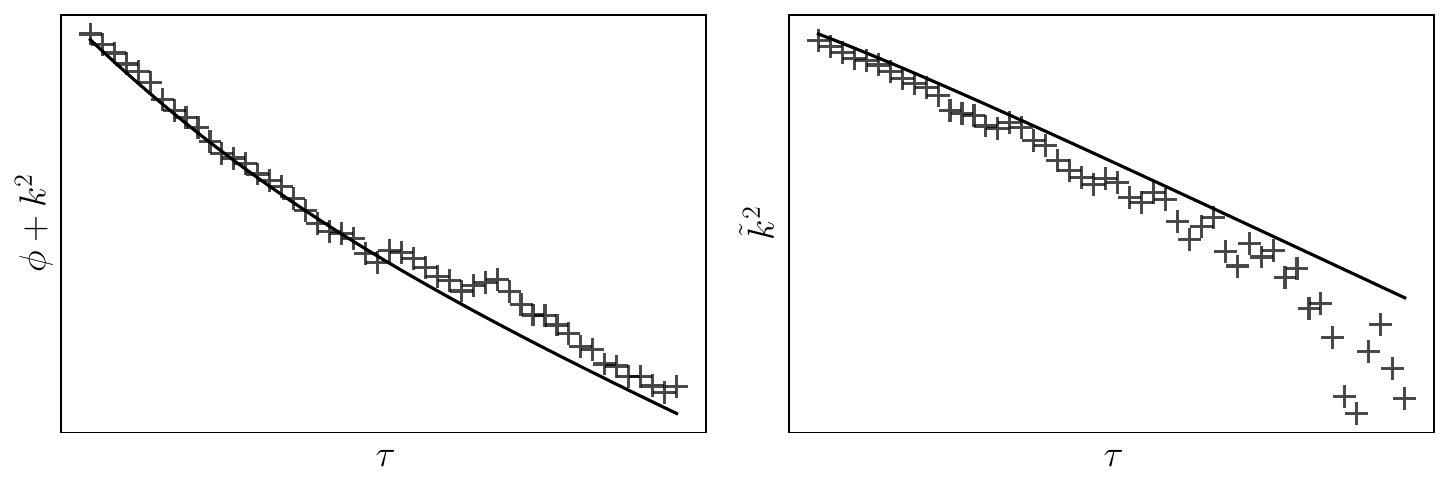}
         \caption{Calibration of a QGARCH for $\bar{\lambda}=12$, obtained with $\sigma^2_\infty=1.5$}
         \label{app_subfig:QGARCH12}
     \end{subfigure}
     \caption{Proof of concept - Calibration of a 1D quadratic GARCH with exponential kernels. The scatter plot are the components obtained by the method of moments and the plain line are the theoretical components used to generate the synthetic data and that we want to recover. The left panel presents the diagonal of the kernel $K=\phi+k^2$ up to a lag $q=50$. The right panel presents the rank one decomposition of $K$ considering its diagonal has only zeros. For this simulation, we set $n_H=0.7$, $n_Z=0.2$, $\beta=0.06$, $\omega=0.05$ and $\sigma^2_\infty$ varies between the two sub-panels.}
    \label{app_fig:QGARCH_1D} 
\end{figure}

 For the quadratic case, it is essential to remain in the regime where $\Bar{\lambda}{\rm d}t\ll 1$; otherwise, the calibration results deviate significantly from the input values, and the method of moments, even when combined with maximum likelihood estimation, does not yield reliable parameters. Since, for real data at the 1-minute scale, we observe that $\Bar{\lambda}{\rm d}t\gg 1$, properly estimating the QHawkes model as it stands is very challenging. Therefore, following the comparison and the approximation made in \cite{blanc2017quadratic}, we have turned to the QGARCH model introduced in \cite{chicheportiche2014fine}. Indeed, the method of moments to calibrate the QGARCH behaves properly for a large set of parameters as shown by Figure~\ref{app_fig:QGARCH_1D}. 

% \begin{figure}[h]
%     \centering
%     \includegraphics[width=\linewidth]{images/app_1D_quadra_GARCH_07.pdf}
%     \caption{$\bar{\lambda}=0.7$}
%     \label{fig:enter-label}
% \end{figure}

% \begin{figure}[h]
%     \centering
%     \includegraphics[width=\linewidth]{images/app_1D_quadra_GARCH_12.pdf}
%     \caption{$\bar{\lambda}=12$}
%     \label{fig:enter-label}
% \end{figure}

% \newpage
\subsection{Calibration of the Multivariate QGARCH}

Since retrieving the univariate QHawkes from aggregated data is already challenging, for the multivariate case, we present only the calibration results on synthetic data for the MQARCH model as a proof of concept. We first examine the 2D linear GARCH, followed by the 2D quadratic GARCH model.   

\subsubsection{Calibration of 2D linear GARCH}

We now consider 2 assets, $A$ and $B$, with volatility $\sigma_A$ and $\sigma_B$ and returns $r_A$ and $r_B$. The MQARCH volatility of the asset $i\in \{A,B\}$ is driven by 

\begin{equation*}
    \sigma^2_{i,t} = \sigma^2_{i,\infty}+\sum_{j\in \{A,B\}}\sum_{k=1}^{+\infty}\phi^i_j(k)r^2_{j,t-k}\quad \text{ with returns defined as } r_{i,t}=\sigma_{i,t} \xi_{i,t}, \quad \text{with}\quad \xi_{i,t}\sim\mathcal{N}(0,1).
\end{equation*}

\begin{figure}[h]
    \centering
    \includegraphics[width=0.55\linewidth]{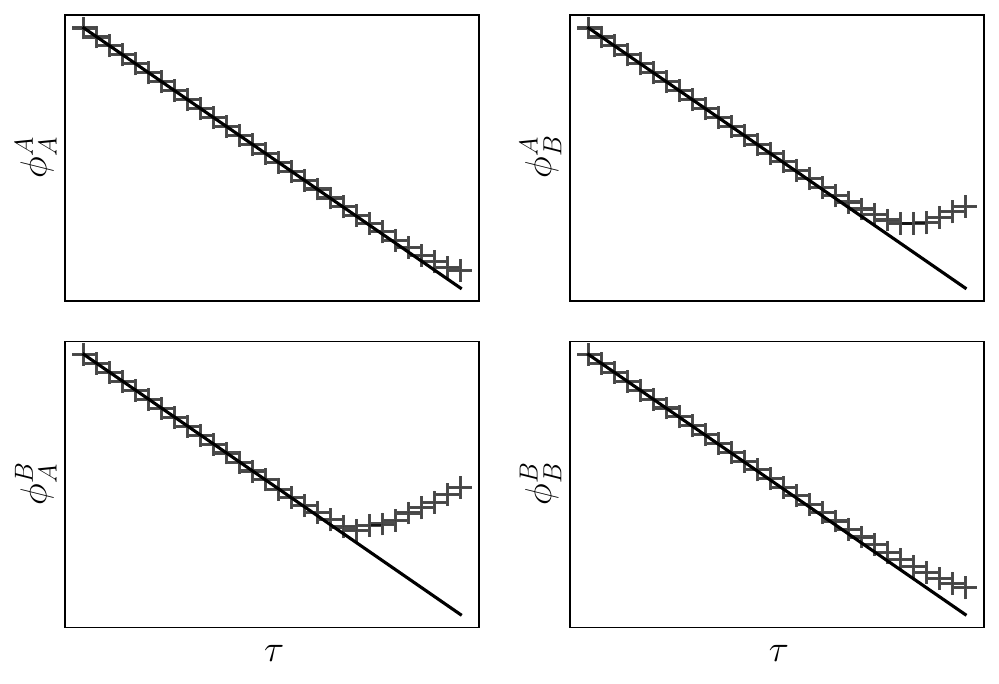}
    \caption{Proof of concept - Calibration of a 2D linear GARCH with exponential kernels. The scatter plots are the kernels obtained by the method of moments and the plain lines are the kernels used to generate the synthetic data and that we want to recover. For this simulation, the parameters are set as defined in Table~\ref{app_tab:parameters_GARCH2D}.}
    \label{app_fig:2DGARCH_linear}
\end{figure}

In the case of exponential kernels, for all $i,j\in \{A,B\}$, the kernel $\phi^i_j$ takes the form $\phi^i_j(k)={n_H}_{ij}\beta_{ij}\exp(-\beta_{ij}k)$. The goal of the calibration is then to determine the eight parameters $({n_H}_{ij},\beta_{ij})_{i,j\in \{A,B\}}$. The method of moments works as before. Figure~\ref{app_fig:2DGARCH_linear} shows our calibration results for the set of parameters presented in Table~\ref{app_tab:parameters_GARCH2D}. It clearly demonstrates that the method of moments enables to recover the input kernels in the case of the 2D linear GARCH model.  
\begin{table}[h]
    \centering
    \begin{tabular}{c|c|c}
        &$n_H$ & $\beta$ \\\hline
        $X_A^A$ & 0.8 &0.2\\
        $X_B^A$ & 0.2 &0.3\\
        $X_A^B$ & 0.3 &0.3\\
        $X_B^B$ & 0.7 &0.1\\
    \end{tabular}
    \caption{Input parameters to generate the synthetic 2D GARCH as defined above. Additionally, we set $\sigma^2_{A,\infty}=0.05$ and $\sigma^2_{B,\infty}=0.1$.}
    \label{app_tab:parameters_GARCH2D}
\end{table}

\subsubsection{Calibration of 2D quadratic GARCH (without cross trends feedback)}\label{app_sec:MQH_calib_simu}

\begin{figure}[h!]
    \centering
     \begin{subfigure}[b]{0.55\textwidth}
         \centering
         \includegraphics[width=\textwidth]{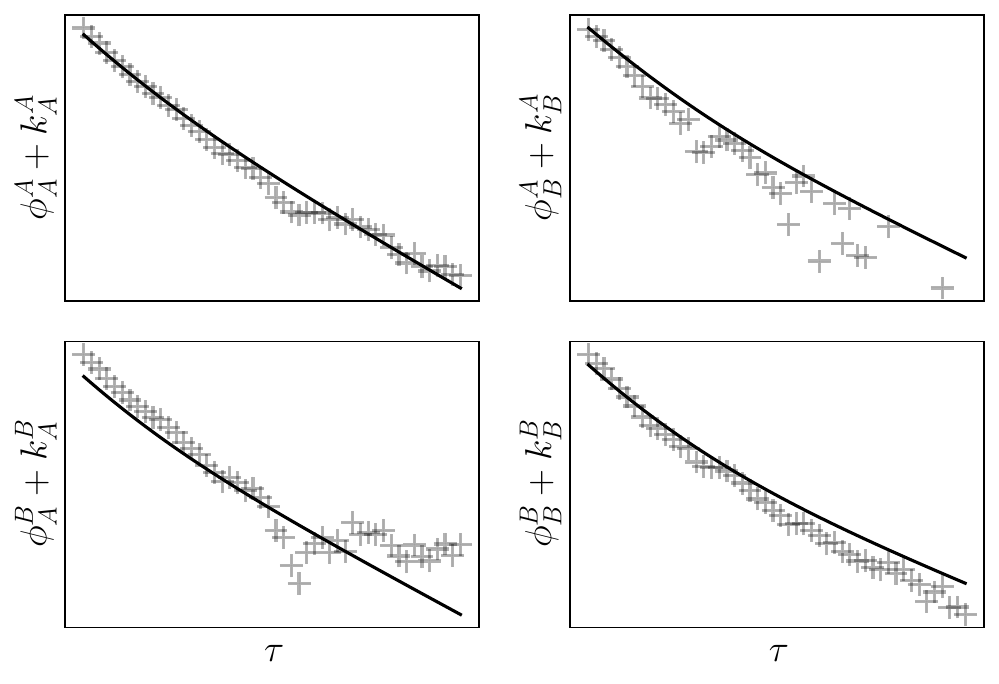}
         \caption{Calibration of the diagonal of the kernels, for $i,j\in \{A,B\}$, $K^i_j=\phi^i_j+(k^i_j)^2$ up to a lag $q=50$.}
         \label{app_subfig:MQARCH_phi}
     \end{subfigure}
     \hfill
     \begin{subfigure}[b]{0.55\textwidth}
         \centering
         \includegraphics[width=\textwidth]{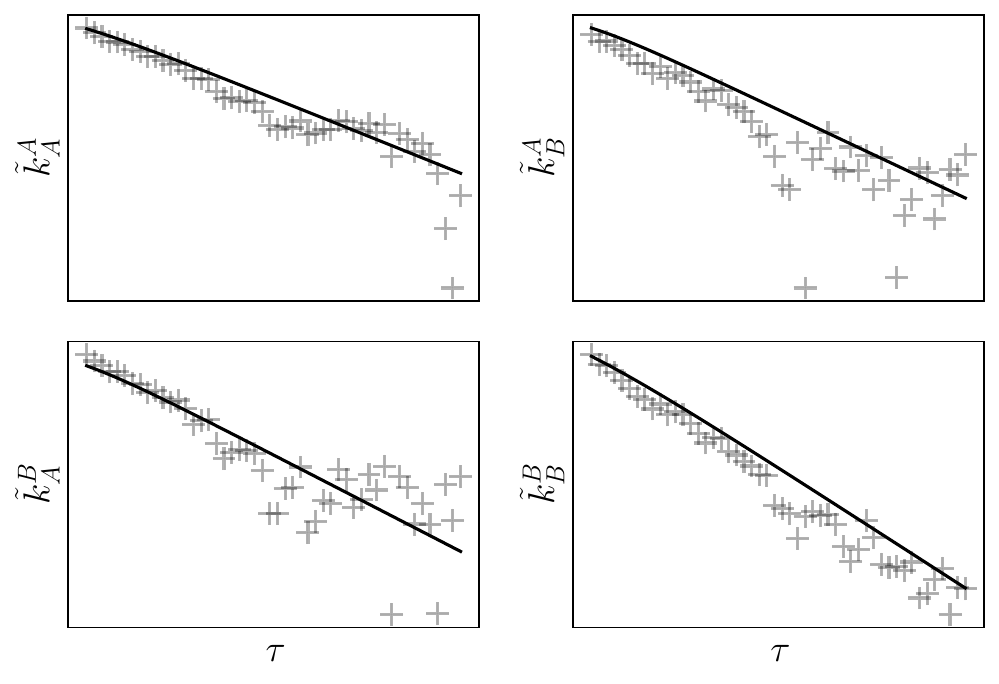}
         \caption{Calibration of the rank one decomposition of off-diagonal kernels, for $i,j\in \{A,B\}$, $(K^i_j(\tau_1,\tau_2))_{1\leq \tau_1,\tau_2\leq q,\tau_1\neq\tau_2} = \Tilde{k}^i_j(\tau_1)\Tilde{k}^i_j(\tau_2)$ up to a lag $q=50$.}
         \label{app_subfig:MQARCH_k}
     \end{subfigure}
    \caption{Proof of concept - Calibration of a 2D quadratic GARCH with exponential kernels, with no leverage or cross trend feedback. The scatter plots are the kernels obtained by the method of moments and the plain lines are the kernels used to generate the synthetic data and that we want to recover. For this simulation, the parameters are set as defined in Table~\ref{app_tab:parameters_QGARCH2D}.}
    \label{app_fig:2DQGARCH}
\end{figure}

This last section focuses on our model of interest the multivariate quadratic GARCH. The goal is to provide, in a simple case, a proof of concept that the method of moments can be extended to this case and give consistent results. 

Here, we consider two assets $A$ and $B$ whose volatility $(\sigma_i)_{i\in\{A,B\}}$ is driven by 

\begin{equation*}
    \sigma^2_{i,t} = \sigma^2_{i,\infty}+\sum_{j\in \{A,B\}}\sum_{k_1=1}^{+\infty}\sum_{k_2=1}^{+\infty}K^i_j(k_1,k_2)r_{j,t-k_1}r_{j,t-k_2},
\end{equation*}
with returns defined as $r_{i,t}=\sigma_{i,t} \xi_{i,t}$,  with $\xi_{i,t}\sim\mathcal{N}(0,1)$.

Note that, for the sake of simplicity, the leverage and cross-trends feedback (characterised by $L$ and $K_\times$ in Equation~\eqref{eq:2D_QGARCH_def_sigma}) are considered null in this framework. As for the univariate quadratic case above (Section~\ref{app_sec:univariate_quadratic_calib_sur_simu}), we can adopt the ZHawkes framework \cite{blanc2017quadratic}. Therefore, each kernel $K^i_j$ for $i,j\in \{A,B\}$ can be decomposed as the sum of a diagonal contribution and a rank-one component, meaning that for all $i,j\in \{A,B\}$, $K^i_j$ writes $K^i_j(\tau_1,\tau_2) = \phi^i_j(\tau_1)\delta_{\tau_1-\tau_2} + k^i_j(\tau_1)k^i_j(\tau_2)$. %To generate synthetic data, exponential kernels are used, and thus, $\phi^i_j$ and $k^i_j$ take the form defined above. 

For better representation, as for the univariate case, each kernel $K^i_j$ is considered as a matrix defined by its lags, up to a certain lag $q$. Thus, $K^i_j$ is represented by the symmetric matrix $(\mathbb{K}^i_j(\tau_1,\tau_2))_{1\leq \tau_1,\tau_2\leq q}$, where, in the ZHawkes framework, the diagonal is $(\phi^i_j(\tau) + (k^i_j)^2(\tau))_{1\leq \tau\leq q}$ and the upper triangle entries are $(k^i_j(\tau_1)k^i_j(\tau_2))_{1\leq \tau_1<\tau_2\leq q}$. As before, the method of moments (see Appendix~\ref{app:YW_calibration_matrix}) returns, for all $i,j\in\{A,B\}$, the estimated diagonal and upper triangle values of $\mathbb{K}^i_j$ up to a certain lag $q$. Subsequently, the off-diagonal of each kernel $\mathbb{K}^i_j$ can be approximated with a rank-one representation, i.e. a vector $\Tilde{k}^i_j$ such that $(\mathbb{K}^i_j(\tau_1,\tau_2))_{1\leq \tau_1,\tau_2\leq q,\tau_1\neq\tau_2} = \Tilde{k}(\tau_1)\Tilde{k}(\tau_2)$. 

% \newpage
To provide a proof of concept of the calibration of the 2D QGARCH, we generate synthetic time series of the QGARCH according to this framework, meaning that we generated two time series of volatility along with two time series of returns such that

\begin{equation*}
    \sigma^2_{i,t} = \sigma^2_{i,\infty}+\sum_{j\in \{A,B\}}\sum_{k=1}^{+\infty}\phi^i_j(k)r^2_{j,t-k} + \sum_{k_1=1}^{+\infty}\sum_{k_2=1}^{+\infty}k^i_j(k_1)k^i_j(k_2)r_{j,t-k_1}r_{j,t-k_2}, 
\end{equation*}

with returns defined as $r_{i,t}=\sigma_{i,t} \xi_{i,t}$ and $\xi_{i,t}\sim\mathcal{N}(0,1)$, and where kernels take the form $\phi^i_j(\tau) = {n_H}_{ij}\beta_{ij}\exp(-\beta_{ij}\tau)$ and $k^i_j(\tau) = \sqrt{{2n_Z}_{ij}\omega_{ij}}\exp(-\omega_{ij}\tau)$. The value of the parameters for the simulation are described in Table~\ref{app_tab:parameters_QGARCH2D}. Subsequently, the method of moments was applied as described in Appendix~\ref{app:YW_calibration_matrix} considering $L=\phi_\times=k_\times=0$. The results of such calibration are shown Figure~\ref{app_fig:2DQGARCH}.
% \newpage

\begin{table}[h]
    \centering
    \begin{tabular}{c|c|c|c|c}
        &$n_H$ & $\beta$ (min$^{-1}$) & $n_Z$ & $\omega$ (min$^{-1}$) \\\hline
        $X_A^A$ & 0.6 & 0.06 & 0.2 & 0.07  \\
        $X_B^A$ & 0.1 & 0.1  & 0.15& 0.1 \\
        $X_A^B$ & 0.2 & 0.08 & 0.1 & 0.09 \\
        $X_B^B$ & 0.4 & 0.04 & 0.21& 0.06 \\
    \end{tabular}
    \caption{Input parameters to generate the synthetic 2D QGARCH as defined above. Additionally, we set $\sigma^2_{A,\infty}=1.2$ and $\sigma^2_{B,\infty}=0.8$.}
    \label{app_tab:parameters_QGARCH2D}
\end{table}

Several observations can be made. First, the calibration appears to successfully replicate the exponential shape and amplitude of the input kernels. However, it is evident that the calibration results do not perfectly align with the input kernels, indicating room for further improvement. Enhancements could involve increasing the number of lags $q$ set in the method of moments, as larger values tend to yield more stable results but also require more computational time. Additionally, another approach to improve the calibration results is to utilize the method of moments' outcome as the initial point for optimization by Maximum Likelihood, as detailed in Appendix~\ref{app:calib_ML}, similar to the procedure used in univariate quadratic Hawkes calibration. 
\newpage
\section{``Surprise price''} \label{app:surprisePrice}
We know, for Gaussian variables: 

\begin{align*}
    \mathbb{E}(X \vert Y) = \mathbb{E}(X) + \Sigma_{XY}\Sigma_Y^{-1}(Y-\mathbb{E}(Y))
\end{align*}

So, if we consider returns as Gaussian, and we consider only the lag-1 has influence on the next return:

\begin{align*}
    \mathbb{E}({\rm d}P_t \vert {\rm d}P_{t-1}) =  \mathbb{E}({\rm d}P_t) + \Sigma_{t,t-1}\Sigma_{t-1}^{-1}({\rm d}P_{t-1}-\mathbb{E}({\rm d}P_{t-1}))
\end{align*}

Since we centered/normalized the returns $\mathbb{E}({\rm d}P_t)=\mathbb{E}({\rm d}P_{t-1})=0$ and $\Sigma_{t-1}=1$

and $\Sigma_{t,t-1}=\text{cov}({\rm d}P_t,{\rm d}P_{t-1})=\Gamma(1)$. Now in multidimension:
\begin{align*}
    \mathbb{E}\left(\begin{pmatrix}{\rm d}P^1_t\\{\rm d}P^2_t\end{pmatrix} \vert \begin{pmatrix}{\rm d}P^1_{t-1}\\{\rm d}P^2_{t-1}\end{pmatrix}\right) =  \mathbb{E}\left(\begin{pmatrix}{\rm d}P^1_t\\{\rm d}P^2_t\end{pmatrix}\right) + \Sigma_{t,t-1}\Sigma_{t-1}^{-1}\left[\begin{pmatrix}{\rm d}P^1_{t-1}\\{\rm d}P^2_{t-1}\end{pmatrix}-\mathbb{E}\left(\begin{pmatrix}{\rm d}P^1_{t-1}\\{\rm d}P^2_{t-1}\end{pmatrix}\right)\right]
\end{align*}

Similarly, thanks to the normalization and centralisation we have: 
\begin{align*}
    \Sigma_{t-1} = \begin{pmatrix}1&\text{cov}({\rm d}P^1,{\rm d}P^2)\\\text{cov}({\rm d}P^1,{\rm d}P^2)&1\end{pmatrix}
\end{align*}

we note $\text{cov}(dP^1,dP^2)=\nu$

\begin{align*}
    \Sigma_{t,t-1} = \begin{pmatrix}\text{cov}({\rm d}P^1_t,{\rm d}P^1_{t-1})&\text{cov}({\rm d}P^1_t,{\rm d}P^2_{t-1})\\\text{cov}({\rm d}P^1_{t-1},{\rm d}P^2_{t})&\text{cov}({\rm d}P^2_t,{\rm d}P^2_{t-1})\end{pmatrix}
\end{align*}

So we should have: 

\begin{align*}\small
    \begin{cases}
    {\rm d}\Tilde{P}^1_t = {\rm d}P^1_t-\frac{1}{1-\nu^2}\Big(\big(\text{cov}({\rm d}P^1_t,{\rm d}P^1_{t-1})-\nu \text{cov}({\rm d}P^1_t,{\rm d}P^2_{t-1})\big){\rm d}P^1_{t-1}+\big(-\nu \text{cov}({\rm d}P^1_t,{\rm d}P^1_{t-1})+ \text{cov}({\rm d}P^1_t,{\rm d}P^2_{t-1})\big){\rm d}P^2_{t-1}\Big)\\
    {\rm d}\Tilde{P}^2_t = {\rm d}P^2_t-\frac{1}{1-\nu^2}\Big(\big(\text{cov}({\rm d}P^1_{t-1},{\rm d}P^2_{t})-\nu \text{cov}({\rm d}P^2_t,{\rm d}P^2_{t-1})\big){\rm d}P^1_{t-1}+\big(-\nu \text{cov}({\rm d}P^1_{t-1},{\rm d}P^2_{t})+ \text{cov}({\rm d}P^2_t,{\rm d}P^2_{t-1})\big){\rm d}P^2_{t-1}\Big)
    \end{cases}
\end{align*}
\newpage
\section{Yule-Walker equations -- MQARCH framework}\label{app:YW_system_2DQGARCH}
This appendix details the calibration procedure for a Multivariate Quadratic GARCH (MQARCH) process. 

\subsection{MQARCH framework}

Consistently with the notations therein, we consider a MQARCH process with $N=2$ assets, where the squared volatility of asset $j$, $\sigma^2_j$, is defined by

\begin{equation}\label{app_eq_MM:sigma^2_def}
    \sigma^2_{j,t} = \sigma^2_{j,\infty} + \sum_{i=1}^N \sum_{\tau=1}^{+\infty}L^j_i(\tau)r_{i,t-\tau}+\sum_{i,k=1}^N \sum_{\tau_1,\tau_2=1}^{+\infty}K^{j}_{ki}(\tau_1,\tau_2)
    r_{k,t-\tau_1}r_{i,t-\tau_2}. 
\end{equation}

To ease the development of the Yule-Walker system, we separate the linear, quadratic and cross kernels, then, $\sigma^2_j$ writes 
\begin{equation}\label{app_eq_MM:sigma^2_def_details}
    \begin{aligned}
    \sigma^2_{j,t} =& \sigma^2_{j,\infty}\\
     &+ \sum_{i=1}^N \sum_{\tau=1}^{+\infty}L^j_i(\tau)r_{i,t-\tau}\\
    &+ \sum_{i=1}^N \sum_{\tau=1}^{+\infty}\phi^j_i(\tau)r^2_{i,t-\tau}\\&+2\sum_{i=1}^2\sum_{\tau_1=1}^{+\infty}\sum_{\tau_2=\tau_1+1}^{+\infty}K^j_i(\tau_1,\tau_2)
    r_{i,t-\tau_1}r_{i,t-\tau_2} \\&+\sum_{\tau=1}^{+\infty} \phi^{j}_\times(\tau)(r_{j,t-\tau}r_{\overline{\jmath},t-\tau}-\overline{r_jr_{\overline{\jmath}}})
    \\&+\sum_{\tau_1=1}^{+\infty}\sum_{\tau_2=\tau_1+1}^{+\infty} k^{j}_\times(\tau_1,\tau_2)r_{j,t-\tau_1}r_{\overline{\jmath},t-\tau_2}
    \\&+\sum_{\tau_1=2}^{+\infty}\sum_{\tau_2=1}^{\tau_1-1} k^{j}_\times(\tau_1,\tau_2)r_{j,t-\tau_1}r_{\overline{\jmath},t-\tau_2},
\end{aligned}
\end{equation}

% where $A$ and $B$ denote the first and second assets. In the following, we swap ($A, B$) and (1, 2) whichever makes the writing clearer. 

\begin{proof}

The transformation of the quadratic term (for $k=i$ in Eq.~\eqref{app_eq_MM:sigma^2_def}) goes as follows
    \begin{align*} \sum_{\tau_1,\tau_2=1}^{+\infty}K^j_i(\tau_1,\tau_2)
    r_{i,t-\tau_1}r_{i,t-\tau_2} 
    =&  \sum_{\tau=1}^{+\infty}\phi^j_i(\tau)r^2_{i,t-\tau}\\&+ \sum_{\tau_1=2}^{+\infty}\sum_{\tau_2=1}^{\tau_1-1}K^j_i(\tau_1,\tau_2)
    r_{i,t-\tau_1}r_{i,t-\tau_2}
    + \sum_{\tau_1=1}^{+\infty}\sum_{\tau_2=\tau_1+1}^{+\infty}K^j_i(\tau_1,\tau_2)
    r_{i,t-\tau_1}r_{i,t-\tau_2}\\ =&  \sum_{\tau=1}^{+\infty}\phi^j_i(\tau)r^2_{i,t-\tau}+ \sum_{\tau_1=2}^{+\infty}\sum_{\tau_2=1}^{\tau_1-1}K^j_i(\tau_1,\tau_2)
    r_{i,t-\tau_1}r_{i,t-\tau_2}
    + \sum_{\tau_2=2}^{+\infty}\sum_{\tau_1=1}^{\tau_2-1}K^j_i(\tau_1,\tau_2)
    r_{i,t-\tau_1}r_{i,t-\tau_2}\\
    &\text{thanks to the properties of symmetry,}\\
   =&  \sum_{\tau=1}^{+\infty}\phi^j_i(\tau)r^2_{i,t-\tau}+2 \sum_{\tau_1=2}^{+\infty}\sum_{\tau_2=1}^{\tau_1-1}K^j_i(\tau_1,\tau_2)
    r_{i,t-\tau_1}r_{i,t-\tau_2}
\end{align*}

\end{proof}

Additionally, if we consider the formulation of Equation~\eqref{app_eq_MM:sigma^2_def_details}, then the mean volatility meets the following condition

\begin{align*}
    \overline{\bm{\sigma}^2} = \bm{\sigma}_\infty^2 +\left(\sum_{\tau=1}^{+\infty}\begin{pmatrix}
        \phi^1_1&\phi^1_2\\
        \phi^2_1&\phi^2_2
    \end{pmatrix}(\tau)\right)\overline{\bm{r}^2}
    \iff\overline{\bm{\sigma}^2} = \bm{\sigma}_\infty^2 +\left(\sum_{\tau=1}^{+\infty}\begin{pmatrix}
        \phi^1_1&\phi^1_2\\
        \phi^2_1&\phi^2_2
    \end{pmatrix}(\tau)\right)\overline{\bm{\sigma}^2}
\end{align*}

\subsection{Yule-Walker equations for 
 a 4-step calibration}

For more stability, we undertake a calibration of the model Eq.~\eqref{app_eq_MM:sigma^2_def_details} in 4 steps.

During the first three steps, we consider the leverage kernel to be null and aim to determine the other kernels ($(\phi^j_i)_{i,j \in \{A,B\}}$, $(K^j_i)_{i,j \in \{A,B\}}$, $(\phi^j_\times)_{j \in \{A,B\}}$ and $(k^j_\times)_{j \in \{1,2\}}$). Therefore, to exclude the leverage effect from real data, we calibrate the linear, quadratic and cross kernels on ``symmetrised'' data, meaning that we use the datasets $[\sigma,\sigma]$ and $[r,-r]$ where $\sigma$ and $r$ represent our original datasets. Once step 3 is completed, we use the original datasets, in which the leverage effect is present, to determine the leverage kernels. The calibration is thus decomposed as follows:
\begin{enumerate}
    \item Determine the QGARCH 1D of the 2 processes ($(\phi^i_i)_{i\{1,2\}}$ and $(K^i_{i})_{i\{1,2\}}$)
    \item Determine the QGARCH 2D cross linear and quadratic kernels ($(\phi^i_j)_{i,j\{1,2\},i\neq j}$ and $(K^i_{j})_{i,j\{1,2\},i\neq j}$)  
    \item Determine the cross kernels of the QGARCH 2D ($(\phi^i_\times)_{i\{1,2\}}$ and $(k^i_\times)_{i\{1,2\}}$)
    \item Determine the QGARCH 2D Leverage kernels ($(L^i_i)_{i\{1,2\}}$.
\end{enumerate}

In order to stay agnostic on the shape of the kernels, for each step, we implement a general method of moments. We rely on a system of Yule-Walker type equations which we describe subsequently. Further details can be found in \cite{aubrun2024unraveling}.

\subsubsection{Covariance structure}\label{app:cov_structure_def}

The Yule-Walker system of equations and thus, the method of moments, rely on the following covariance structures:
\begin{equation}\label{app_eq_MM:Cdef}
    \Cb_{ij}(\tau) = \mathbb{E}(\sigma^2_{i,t}r^2_{j,t-\tau}) - \overline{\sigma^2_{i}}\;\overline{r^2_{j}}
\end{equation}

\begin{equation}\label{app_eq_MM:Crdef}
    \Cb^r_{ij}(\tau) = \mathbb{E}(r^2_{i,t}r^2_{j,s}) - \overline{r^2_{i}}\; 
 \overline{r^2_{j}}
\end{equation}

\begin{equation}
    \Db_{ijk}(\tau_1,\tau_2) = \mathbb{E}(\sigma_{i,t}^2r_{j,t-\tau_1}r_{k,t-\tau_2})
\end{equation}

\begin{equation}
    \Db_{p,(ij)kl}(\tau) = \mathbb{E}((r_{i,t}r_{j,t})r_{k,t-\tau_1}r_{l,t-\tau_2})
\end{equation}

\begin{equation}
    \Db_{\times,jAB}(\tau_1,\tau_2)=\mathbb{E}\left(\sigma^2_{j,t}r_{A,t-\tau_1}r_{B,t-\tau_2}\right)
\end{equation}

\begin{equation}\label{app_eq_MM:Vdef}
    \Vb_{ij}(\tau) = \mathbb{E}(\sigma^2_{i,t}r_{j,t-\tau})
\end{equation}

\begin{equation}\label{app_eq_MM:Vrdef}
    \Vb^r_{ij}(\tau) = \mathbb{E}(r^2_{i,t}r_{j,t-\tau})
\end{equation}

% % \begin{remark}
    By definition of the model and in simulation, Eq.~\eqref{app_eq_MM:Cdef} and Eq.~\eqref{app_eq_MM:Crdef} should result in the same values. However in practice, there is a slight difference due to noise, which influence the calibration. Hence, in order to give reproducible guidelines we note the difference in what follows.
% \end{remark}

\subsubsection{Yule-Walker system for QGARCH 2D without cross terms (steps 1 and 2)}

For the first two steps of the calibration, we need equations linking $\Cb$ and $\Db$ to the kernels $\phi$ and $K$. The goal is to characterise $(\phi^j_i)_{i,j\in\{1,2\}}$ and $(K^j_i)_{i,j\in\{1,2\}}$ up to a certain lag $q$. Hence, we want to find the values of, for $i,j\in\{1,2\}$, $\phi^j_i(\tau)_{\tau \in \llbracket 1,q\rrbracket}$ and $K^j_i(\tau_1,\tau_2)_{1\leq\tau_1<\tau_2\leq q}$ (since $K$ is symmetric, it is enough to find only its upper triangle entries). The Yule Walker equations on $(\Cb_{ij}(\tau))_{i,j\in\{1,2\},\tau \in \llbracket 1,q\rrbracket}$ will give $4\times q$ equations and the Yule Walker equations on $(\Db_{ij}(\tau_1,\tau_2))_{i,j\in\{1,2\},1\leq\tau_1<\tau_2\leq }$ will give $4\times q\frac{q-1}{2}$ equations, allowing to fully characterise $(\phi^j_i)_{i,j\in\{1,2\}}$ and $(K^j_i)_{i,j\in\{1,2\}}$ up to a certain lag $q$.

Without complicating things, we develop the general equations for $N$ assets.

\begin{lemma}[Yule-Walker equation for $\Cb$]\label{app_lemma_MM:YW_Cb}
    for $\tau>0$, in the framework described by Eq.~\eqref{app_eq_MM:sigma^2_def_details}, with $\phi^\times=k^\times=0$, the covariance $\Cb$ writes
    \begin{equation*}
    \Cb_{jl}(\tau)
    = \sum_{i=1}^N \sum_{k=1}^{+\infty}\phi^j_i(k)\Cb^r_{il}(\tau-k)+2\sum_{i=1}^N \sum_{k_2=\tau+1}^{+\infty}\sum_{k_1=k_2+1}^{+\infty}K^j_i(k_1,k_2)\Db_{p,lii}(k_1-\tau,k_2-\tau)
\end{equation*}
\end{lemma}

\begin{lemma}[Yule-Walker equation for $\Db$]\label{app_lemma_MM:YW_Db}
    For $\tau_1<\tau_2$, in the framework described by Eq.~\eqref{app_eq_MM:sigma^2_def_details}, with $\phi^\times=k^\times=0$, the covariance $\Db$ writes
    \begin{equation*}
   \Db_{jll}(\tau_1,\tau_2) = \sum_{i=1}^N \sum_{k=1}^{\tau_1-1}\phi^j_i(k)\Db_{p,ill}(\tau_1-k,\tau_2-k)+2\sum_{i=1}^N \sum_{k_2=\tau_1+1}^{+\infty}K^j_i(\tau_1,k_2)
   \Db_{p,(il)il}(-\tau_1+k_2,\tau_2-\tau_1)
\end{equation*}
\end{lemma}

\subsubsection{Yule-Walker system for QGARCH 2D with cross terms (step 3)}

The goal of the third step is to characterise $(\phi^j_\times)_{j\in \{1,2\}}$ and $(K^j_\times)_{j\in \{1,2\}}$ up to a certain lag $q$. In this case, we need $2\times q$ equations to characterise $(\phi^j_\times)_{j\in \{1,2\}}$ and $2\times q(q-1)$ equations to characterise $(K^j_\times)_{j\in \{1,2\}}$, as $K^j_\times$ is not symmetric so we need both the upper and lower triangle entries. The equations on $(\Db^j_\times(\tau_1,\tau_2))_{j\in \{1,2\}, 1\leq \tau_1,\tau_2\leq q}$, developed subsequently, allow to fully characterise the cross kernels $(\phi^j_\times)_{j\in \{1,2\}}$ and $(K^j_\times)_{j\in \{1,2\}}$.

\begin{lemma}[Yule-Walker equation for $\Db_\times$ for $\tau_1=\tau_2=\tau$]\label{app_lemma_MM:YW_Dx_diag}
for $\tau>0$, in the framework described by Eq.~\eqref{app_eq_MM:sigma^2_def_details}, the covariance $ \Db_\times$ writes
\begin{equation*}
        \begin{aligned}
    \Db_{\times,j}(\tau,\tau)=& \sigma^2_{j,\infty}\overline{r_{j}r_{{\overline{\jmath}}}}\\
    &+ \sum_{i=1}^N \sum_{k=1}^{+\infty}\phi^j_i(k)\Db_{p,(ii)j{\overline{\jmath}}}(\tau-k,\tau-k)\\
    &+2\sum_{i=1}^N \sum_{k_1=\tau+1}^{+\infty}\sum_{k_2=k_1+1}^{+\infty}K^j_i(k_1,k_2)\Db_{p,(j{\overline{\jmath}})ii}(k_1-\tau,k_2-\tau) \\&+\sum_{k=1}^{+\infty} \phi^j_{\times}(k)\left(\Db_{p(j{\overline{\jmath}})j{\overline{\jmath}}}(\tau-k,\tau-k) - \overline{r_jr_{\overline{\jmath}}}^2\right)
    \\&+\sum_{k_2=\tau+1}^{+\infty}\sum_{k_1=k_2+1}^{+\infty} k^j_{\times}(k_1,k_2)\Db_{p,(j{\overline{\jmath}})j{\overline{\jmath}}}(k_1-\tau,k_2-\tau)
    \\&+\sum_{k_1=\tau+1}^{+\infty}\sum_{k_2=k_1+1}^{+\infty} k^j_{\times}(k_1,k_2)\Db_{p,(j{\overline{\jmath}})j{\overline{\jmath}}}(k_1-\tau,k_2-\tau)    
    \end{aligned}
\end{equation*}

\end{lemma}

One of the difficulty in including the cross terms is the non-symmetry of the covariance structure $\Db_\times$ and of the kernels $k_\times$. Hence, we need to develop equations for $\Db_{\times,j}(\tau_1,\tau_2)$ for $\tau_1<\tau_2$ and $\tau_1>\tau_2$.

\begin{lemma}[Yule-Walker equation for $\Db_\times$ for $\tau_1<\tau_2$]\label{app_lemma_MM:YW_Dx_1}
for  $\tau_1<\tau_2$, in the framework described by Eq.~\eqref{app_eq_MM:sigma^2_def_details}, the covariance $ \Db_\times$ writes
\begin{equation*}
    \begin{aligned}
    \Db_{\times,j}(\tau_1,\tau_2)=&\sigma^2_{j,\infty}\overline{r_jr_{\overline{\jmath}}}\\
    &+ \sum_{i=1}^N \sum_{k=1}^{\tau_1-1}\phi^j_i(k)\Db_{p,(ii)j\overline{\jmath}}(\tau_1-k,\tau_2-k)\\
    &+2\sum_{i=1}^N \sum_{k_2=\tau_1+1}^{+\infty}K^j_i(\tau_1,k_2)
    \Db_{p,(ij)i\overline{\jmath}}(k_2-\tau_1,\tau_2-\tau_1)\\&+\sum_{k=1}^{\tau_1-1} \phi^j_{\times}(k)\left(\Db_{(j\overline{\jmath})j\overline{\jmath}}(\tau_1-k,\tau_2-k)-\overline{r_jr_{\overline{\jmath}}}^2)\right)
    \\&+\sum_{k_1=\tau_1+1}^{+\infty} k^j_{\times}(k_1,\tau_1)\Db_{p,(\overline{\jmath}j)j\overline{\jmath}}(k_1-\tau_1,\tau_2-\tau_1)
\\&+\sum_{k_2=\tau_1+1}^{+\infty} k^j_{\times}(\tau_1,k_2)\Db_{p,(jj)\overline{\jmath}\overline{\jmath}}(k_2-\tau_1,\tau_2-\tau_1)
\end{aligned}
\end{equation*}

\end{lemma}

\begin{lemma}[Yule-Walker equation for $\Db_\times$ for $\tau_1>\tau_2$]\label{app_lemma_MM:YW_Dx_2}
for  $\tau_1>\tau_2$, in the framework described by Eq.~\eqref{app_eq_MM:sigma^2_def_details}, the covariance $ \Db_\times$ writes
\begin{equation*}
        \begin{aligned}
    \Db_{\times,j}(\tau_1,\tau_2)=&\sigma^2_{j,\infty}\overline{r_jr_{\overline{\jmath}}}\\
    &+ \sum_{i=1}^N \sum_{k=1}^{\tau_1-1}\phi^j_i(k)\Db_{p,(ii)j\overline{\jmath}}(\tau_1-k,\tau_2-k)\\
    &+2\sum_{i=1}^N \sum_{k_2=\tau_1+1}^{+\infty}K^j_i(\tau_1,k_2)
    \Db_{p,(i\overline{\jmath})ij}(k_2-\tau_1,\tau_1-\tau_2)\\&+\sum_{k=1}^{\tau_1-1} \phi^j_{\times}(k)\left(\Db_{(j\overline{\jmath})j\overline{\jmath}}(\tau_1-k,\tau_2-k)-\overline{r_jr_{\overline{\jmath}}}^2)\right)
    \\&+\sum_{k_1=\tau_2+1}^{+\infty} k^j_{\times}(k_1,\tau_2)\Db_{p,(\overline{\jmath}\overline{\jmath})jj}(k_1-\tau_2,\tau_1-\tau_2)
    \\&+\sum_{k_2=\tau_1+1}^{+\infty} k^j_{\times}(\tau_1,k_2)\Db_{p,(j\overline{\jmath})\overline{\jmath}j}(k_2-\tau_1,\tau_1-\tau_2)\\
\end{aligned}
\end{equation*}

\end{lemma}

Appendix~\ref{app:YW_calibration_matrix} gives more details on who to concretely implement those systems of linear equations between kernels and covariance structures.

\subsubsection{Yule-Walker system for QGARCH 2D for Leverage kernel (step 4)}

Finally, once the linear, quadratic and cross kernels are determined, we can work on the original dataset (not symmetrised) and characterised the leverage kernels. To achieve this, additional Yule-Walker equations are required relying on $\Vb$ (defined in Equations~\eqref{app_eq_MM:Vdef} and~\eqref{app_eq_MM:Vrdef}), we derive them subsequently. The goal is to characterise $(L^j_i)_{i,j\in\{1,2\}}$ up to a certain lag $q$. Hence, we need $4\times q$ equations which are obtained using $(\Vb_{jl}(\tau))_{j,l\in\{1,2\},1\leq \tau\leq q}$.

\begin{lemma}[Yule-Walker equation for $\Vb$]\label{app_lemma_MM:YW_Vb}
    for $\tau>0$, in the framework described by Eq.~\eqref{app_eq_MM:sigma^2_def_details}, the covariance $\Vb$ writes
    \begin{equation*}
    \begin{aligned}
         \Vb_{jl}(\tau) = &L^j_i(\tau)\overline{r_{i}r_{l}}\\
    &+ \sum_{i=1}^N \sum_{k=1}^{\tau}\phi^j_i(k)\Vb^r_{il}(\tau-k)\\&+2\sum_{k_2=\tau+1}^{+\infty}K^j_i(\tau,k_2)\Vb^r_{(il)j}(k_2-\tau)\\&+\sum_{k=1}^{\tau} \phi^{j}_\times(k)\Vb^r_{(j\overline{\jmath})l}(\tau-k)\\&+\sum_{k_2=\tau+1}^{+\infty} k^{j}_\times(\tau,k_2)\Vb^r_{(jl)\overline{\jmath}}(k_2-\tau)
    \\&+\sum_{k_1=\tau+1}^{+\infty} k^{j}_\times(k_1,\tau)\Vb^r_{(\overline{\jmath}l)j}(k_1-\tau)
    \end{aligned}
    \end{equation*}
\end{lemma}

\newpage

\section{Calibration Matrices - QGARCH}\label{app:YW_calibration_matrix}

The goal of the calibration up to step 3 is to deduce the kernels $(\phi^i_j)_{i,j\in\{1,2\}}$, $(K^i_{j})_{i,j\in\{1,2\}}$, $(\phi^i_\times)_{i\in\{1,2\}}$ and $(k^i_\times)_{i\in\{1,2\}}$ that weight the contributions of the past on the squared volatility as defined in Eq.~\eqref{app_eq_MM:sigma^2_def_details}, from the observation of the covariance structures defined in Appendix~\ref{app:cov_structure_def}. The link between the kernels and the covariance structures is established by the Yule-Walker equations developed in Appendix~\ref{app:YW_system_2DQGARCH}. To implement this Yule-Walker system of equations, we propose here a matricel environment. The calibration matrices to deduce the leverage kernels are described in a subsequent section.

\subsection{2D QGARCH -- framework}

Consistently, we present a matricel system for the 2D QGARCH, considering the two assets 1 and 2. The goal is to link the kernels

\begin{align*}
    \bbphi=\begin{pmatrix}
        \phi_1^1&\phi_2^1\\
        \phi_1^2&\phi_2^2
    \end{pmatrix}, \quad 
    \mathbb{K}=\begin{pmatrix}
        K_1^1&K_2^1\\
        K_1^2&K_2^2
    \end{pmatrix}, \quad 
    \boldsymbol{\phi}_\times=\begin{pmatrix}
        \phi_\times^1\\
        \phi_\times^2
    \end{pmatrix}, \quad \text{ and }\quad
    \boldsymbol{k}_\times=\begin{pmatrix}
        k_\times^1\\
        k_\times^2
    \end{pmatrix}
\end{align*}

to the observable covariance structures

\begin{align*}
    \mathbb{C}=\begin{pmatrix}
        \Cb_1^1&\Cb_2^1\\
        \Cb_1^2&\Cb_2^2
    \end{pmatrix}, \quad 
    \mathbb{D}_\text{d}=\begin{pmatrix}
        \Db_{111}&\Db_{122}\\
        \Db_{211}&\Db_{222}
    \end{pmatrix} \quad \text{ and }\quad
    \boldsymbol{\Db}_\times=\begin{pmatrix}
        \Db_{112}\\
        \Db_{212}
    \end{pmatrix}.
\end{align*}

The goal of this appendix is to describe a system of tensors which describes the relationship between the kernels and the covariance structures up to a certain lag $q$. We construct the following system:

\begin{equation}\label{app_eq:YW_matrix_sys}
    \begin{aligned}
     \begin{pmatrix}
        \textcolor{violet}{\mathbb{A}_{1\text{d}}} & \textcolor{violet}{\mathbb{A}_{2\text{d}}} &0 &0\\
        \textcolor{violet}{\mathbb{A}_{3\text{d}}}&\textcolor{violet}{\mathbb{A}_{4\text{d}}}&0&0\\
        \textcolor{olive}{\mathbb{A}_{1\text{d}\times}}&\textcolor{olive}{\mathbb{A}_{2\text{d}\times}}&\textcolor{teal}{\mathbb{A}_{1\times}}&\textcolor{teal}{\mathbb{A}_{2\times}}\\
        \textcolor{olive}{\mathbb{A}_{3\text{d}\times}}&\textcolor{olive}{\mathbb{A}_{4\text{d}\times}}&\textcolor{teal}{\mathbb{A}_{3\times}}&\textcolor{teal}{\mathbb{A}_{4\times}}\\
     \end{pmatrix}
     \begin{pmatrix}
        \textcolor{blue}{\bbphi(\tau)}\\
        \textcolor{blue}{\mathbb{K}(\tau_1,\tau_2)_{\tau_1<\tau_2}}\\
        \textcolor{cyan}{\bm{\phi}_{\times}(\tau)}\\
        \textcolor{cyan}{\boldsymbol{k_{\times}}(\tau_1,\tau_2)_{\tau_1<\tau_2}}
        \\\textcolor{cyan}{\boldsymbol{k_{\times}}(\tau_1,\tau_2)_{\tau_1>\tau_2}}
    \end{pmatrix}=&\begin{pmatrix}
        \textcolor{red}{\mathbb{C}(\tau)}\\
        \textcolor{red}{\mathbb{D}_{\text{d}}(\tau_1,\tau_1)_{\tau_1<\tau_2}}\\
        \textcolor{magenta}{\boldsymbol{\Db_{\times}}(\tau,\tau)}\\
       \textcolor{magenta}{\boldsymbol{\Db_{\times}}(\tau_1,\tau_2)_{\tau_1<\tau_2}}
        \\\textcolor{magenta}{\boldsymbol{\Db_{\times}}(\tau_1,\tau_2)_{\tau_1>\tau_2}}
    \end{pmatrix}.
\end{aligned}
\end{equation}

Another way to comprehend the system is to see that $\bbphi$ is the time-diagonal of $K$ and thus, one can separate the cross components from the linear and quadratic components, as follows:

\begin{align*}
     \begin{pmatrix}
        \textcolor{violet}{\mathbb{A}_\text{d}} & {0}\\
        \textcolor{olive}{\mathbb{A}_{\text{d}\times}}&\textcolor{teal}{\mathbb{A}_{\times}}
     \end{pmatrix}
     \begin{pmatrix}
        \textcolor{blue}{\mathbb{K}_{\text{d}}}\\
        \textcolor{cyan}{\boldsymbol{K_{\times}}}
    \end{pmatrix}=&\begin{pmatrix}
        \textcolor{red}{\mathbb{D}_{\text{d}}}\\
        \textcolor{magenta}{\boldsymbol{\Db_{\times}}}
    \end{pmatrix}.
\end{align*}

Let us describe the tensors system above:
\begin{itemize}
    \item \textbf{Covariance structures}: Represented by the right hand in Eq.~\eqref{app_eq:YW_matrix_sys}, the covariance tensors are composed of observable elements.
    \begin{itemize}
        \item The \textcolor{red}{red matrices} are correlation tensors of size $q\times 2\times 2$ (or more generally $q\times N\times N$ if $N$ is the number of assets). They represent the covariance of the squared volatility and the covariance between the volatility and the past trends. 
        \item The \textcolor{magenta}{magenta tensors} are also covariance structures, capturing the influence of the past \textit{cross} trends on future volatility. The time-diagonal $\Db_\times(\tau,\tau)$, with $\tau \in \llbracket1,q\rrbracket$, is a tensor of 2D vectors with shape $q\times 2 \times 1$. Conversely to $\mathbb{D}_\text{d}$, $\Db_\times$ is non-time-symmetric, i.e, $\Db_\times(\tau_1,\tau_2)\neq\Db_\times(\tau_2,\tau_1)$, thus, we need to consider $\Db_\times(\tau_1,\tau_2)$ for both  $\tau_1<\tau_2$ and $\tau_2<\tau_1$, both of which shape $q\frac{q-1}{2}\times 2\times 1$. 
    \end{itemize}
    \item \textbf{Kernels}: The kernels weight the feedback of past realisations on the future square volatility and characterise the process. These kernels are the unknowns we want to recover.
    \begin{itemize}
        \item The \textcolor{blue}{blue tensors} represent the linear and quadratic we want to calibrate (steps 1 and 2 of Appendix~\ref{app:YW_system_2DQGARCH}). The shape of $\mathbb{K}_\text{d}$ is, consistently with $\mathbb{D}_\text{d}$, $q+\frac{q(q-1)}{2}\times 2\times 2$ ($\bbphi$ has shape  $q\times 2\times 2$ and $\mathbb{K}$ has shape  $\frac{q(q-1)}{2}\times 2\times 2$).
        \item The \textcolor{cyan}{cyan tensors} are the lags of the kernels weighting the feedback of the cross trend. Consistently with $\boldsymbol{\Db}_\times$, their shape is $q+q(q-1)\times 2\times 1$.  
    \end{itemize}
    \item \textbf{Yule-Walker matrices}: These matrices are built upon observables according to the Yule-Walker system of equations developed in Appendix~\ref{app:YW_system_2DQGARCH}. We explain briefly here their role, but the subsequent sections are dedicated to describe them precisely. 
    \begin{itemize}
        \item The \textcolor{violet}{violet calibration matrices} relate the linear and quadratic kernels represented by $\mathbb{K}_\text{d}$ to the correlations structure $\mathbb{D}_\text{d}$.
        \item The \textcolor{teal}{teal calibration matrices} relate the kernel of the cross trends $\boldsymbol{K}_\times$ to the cross correlations $\boldsymbol{\Db}_\times$.
        \item the \textcolor{olive}{olive components} relate the linear and quadratic kernel $\mathbb{K}_\text{d}$ to the cross correlations $\boldsymbol{\Db}_\times$.
    \end{itemize}
\end{itemize}

The above system already considers a calibration in several steps, otherwise the upper right bloc would not be 0 as $\boldsymbol{K}_\times$ could contribute to $\mathbb{D}_{\text{d}}$. Thus, we first solve the upper bloc for $\mathbb{K}_{\text{d}}$ using $\mathbb{A}_{\text{d}}\mathbb{K}_{\text{d}} = \mathbb{D}_{\text{d}}$, and then we solve for $\boldsymbol{K}_\times$ using $\mathbb{A}_{\text{d}\times}\mathbb{K}_{\text{d}} + \mathbb{A}_{\times}\boldsymbol{K}_\times = \boldsymbol{\Db}_{\times}$. Let us note that the first calibration step to determine $\mathbb{K}_{\text{d}}$ is also decomposed in two, as we first characterise the self feedback (1D calibration) and then the cross-linear and quadratic feedback. The rest of this appendix is dedicated to the description of the the Yule-Walker matrices which define the system.

\subsection{Building the Yule-Walker Matrices}

The Yule-Walker matrices $\mathbb{A}_\text{d}$, $\mathbb{A}_{\text{d}\times}$ and $\mathbb{A}_\times$ are built according to the Yule-Walker lemmas of Appendix~\ref{app:YW_system_2DQGARCH}. In fact, each matrix ($\mathbb{A}_{1\text{d}}$, $\mathbb{A}_{2\text{d}}$, $\mathbb{A}_{3\text{d}}$, $\mathbb{A}_{4\text{d}}$, $\mathbb{A}_{1\text{d}\times}$, $\mathbb{A}_{2\text{d}\times}$, $\mathbb{A}_{3\text{d}\times}$, $\mathbb{A}_{4\text{d}\times}$, $\mathbb{A}_{1\times}$, $\mathbb{A}_{2\times}$, $\mathbb{A}_{3\times}$, $\mathbb{A}_{4\times}$) replicates one term of these equations. Upon scrutiny, it appears that some terms have the same structure, and only the components, i.e. the kernels and the covariance involved in the term, change. In particular, there are 4 different structures to implement: $A_1$, $A_2$, $A_3$ and $A_4$. 

\begin{table}
\centering
\begin{tabular}{|c|c|c|c|}
\hline
${A}_1$ & ${A}_2$ &${A}_3$ & ${A}_4$\\
 \hline\hline
$\mathbb{A}_{1\text{d}}$ & $\mathbb{A}_{2\text{d}}$ &$\mathbb{A}_{3\text{d}}$ & $\mathbb{A}_{4\text{d}}$\\%\hline
$\mathbb{A}_{1\text{d}\times}$ & $\mathbb{A}_{2\text{d}\times}$ &$\mathbb{A}_{3\text{d}\times}$ & $\mathbb{A}_{4\text{d}\times}$\\%\hline
$\mathbb{A}_{1\times}$ & $\mathbb{A}_{2\times}$ &$\mathbb{A}_{3\times}$ & $\mathbb{A}_{4\times}$\\\hline
\end{tabular}
\caption{Calibration matrices with similar structure}
\label{app_tab:YW_matrices_categories}
\end{table}

Specifically, restating the Equations of lemmas~\ref{app_lemma_MM:YW_Cb}, \ref{app_lemma_MM:YW_Db}, \ref{app_lemma_MM:YW_Dx_diag}, \ref{app_lemma_MM:YW_Dx_1} and \ref{app_lemma_MM:YW_Dx_2}, we make the link between the matrices and the term they represent.

From lemma~\ref{app_lemma_MM:YW_Cb}, we have
\begin{align*}
    \Cb_{jl}(\tau)
    =& \underbrace{\sum_{i=1}^N \sum_{k=1,k\neq \tau}^{+\infty}\phi^j_i(k)\Cb^r_{il}(\tau-k)}_{\mathbb{A}_{1\text{d}} \in \mathbb{A}_\text{d}}+\underbrace{2\sum_{i=1}^N \sum_{k_2=\tau+1}^{+\infty}\sum_{k_1=k_2+1}^{+\infty}K^j_i(k_1,k_2)\Db_{p,lii}(k_1-\tau,k_2-\tau)}_{\mathbb{A}_{2\text{d}}\in \mathbb{A}_\text{d}}.
\end{align*}

From lemma~\ref{app_lemma_MM:YW_Db}, we have, for $\tau_1<\tau_2$,

\begin{align*}
   \Db_{jll}(\tau_1,\tau_2) =& \underbrace{\sum_{i=1}^N \sum_{k=1}^{\tau_1-1}\phi^j_i(k)\Db_{p,ill}(\tau_1-k,\tau_2-k)}_{\mathbb{A}_{3\text{d}}\in \mathbb{A}_{\text{d}}}+\underbrace{2\sum_{i=1}^N \sum_{k_2=\tau_1+1}^{+\infty}K^j_i(\tau_1,k_2)
   \Db_{p,(il)il}(-\tau_1+k_2,\tau_2-\tau_1)}_{\mathbb{A}_{4\text{d}}\in \mathbb{A}_{\text{d}}}
\end{align*}

From lemma~\ref{app_lemma_MM:YW_Dx_diag}, we have, for, $\tau>0$
  \begin{align*}
    \Db_{\times,j}(\tau,\tau)-\sigma^2_{j,\infty}\overline{r_{j}r_{\overline{\jmath}}}=
    & \underbrace{\sum_{i=1}^N \sum_{k=1}^{\tau-1}\phi^j_i(k)\Db_{p,(ii)j\overline{\jmath}}(\tau-k,\tau-k)}_{\mathbb{A}_{1\text{d}\times} \in A_{\text{d}\times}}\\
    &+\underbrace{2\sum_{i=1}^N \sum_{k_1=\tau+1}^{+\infty}\sum_{k_2=k_1+1}^{+\infty}K^j_i(k_1,k_2)\Db_{p,(j\overline{\jmath})ii}(k_1-\tau,k_2-\tau) }_{\mathbb{A}_{2\text{d}\times} \in A_{\text{d}\times}}\\&+\underbrace{\sum_{k=1}^{\tau-1} \phi^j_{\times}(k)\left(\Db_{p(j\overline{\jmath})j\overline{\jmath}}(\tau-k,\tau-k) - \overline{r_jr_{\overline{\jmath}}}^2\right)}_{\mathbb{A}_{1\times}\in\mathbb{A}_\times}
    \\&+\underbrace{\sum_{k_2=\tau+1}^{+\infty}\sum_{k_1=k_2+1}^{+\infty} k^j_{\times}(k_1,k_2)\Db_{p,(j\overline{\jmath})j\overline{\jmath}}(k_1-\tau,k_2-\tau)}_{\mathbb{A}_{2\times}\in \mathbb{A}_{\times}}
    \\&+\underbrace{\sum_{k_1=\tau+1}^{+\infty}\sum_{k_2=k_1+1}^{+\infty} k^j_{\times}(k_1,k_2)\Db_{p,(j\overline{\jmath})j\overline{\jmath}}(k_1-\tau,k_2-\tau)}_{\mathbb{A}_{2\times}\in \mathbb{A}_{\times}}    
    \end{align*}

From lemma~\ref{app_lemma_MM:YW_Dx_1}, we have, for, $\tau_1<\tau_2$
\begin{align*}
    \Db_{\times,j}(\tau_1,\tau_2)-\sigma^2_{j,\infty}\overline{r_jr_{\overline{\jmath}}}=& \underbrace{\sum_{i=1}^N \sum_{k=1}^{\tau_1-1}\phi^j_i(k)\Db_{p,(ii)j\overline{\jmath}}(\tau_1-k,\tau_2-k)}_{\mathbb{A}_{3\text{d}\times}\in \mathbb{A}_{\text{d}\times}}\\
    &+\underbrace{2\sum_{i=1}^N \sum_{k_2=\tau_1+1}^{+\infty}K^j_i(\tau_1,k_2)
    \Db_{p,(ij)i\overline{\jmath}}(k_2-\tau_1,\tau_2-\tau_1)}_{\mathbb{A}_{4\text{d}\times}\in \mathbb{A}_{\text{d}\times}}\\&
    +\underbrace{\sum_{k=1}^{\tau_1-1} \phi^j_{\times}(k)\left(\Db_{(j\overline{\jmath})j\overline{\jmath}}(\tau_1-k,\tau_2-k)-\overline{r_jr_{\overline{\jmath}}}^2)\right)}_{\mathbb{A}_{3\times}\in \mathbb{A}_{\times}}
    \\&+\underbrace{\sum_{k_1=\tau_1+1}^{+\infty} k^j_{\times}(k_1,\tau_1)\Db_{p,(\overline{\jmath}j)j\overline{\jmath}}(k_1-\tau_1,\tau_2-\tau_1)}_{\mathbb{A}_{4\times}\in \mathbb{A}_{\times}}
    \\&+\underbrace{\sum_{k_2=\tau_1+1}^{+\infty} k^j_{\times}(\tau_1,k_2)\Db_{p,(jj)\overline{\jmath}\overline{\jmath}}(k_2-\tau_1,\tau_2-\tau_1)}_{\mathbb{A}_{4\times}\in \mathbb{A}_{\times}}
\end{align*}

From lemma~\ref{app_lemma_MM:YW_Dx_2}, we have, for, $\tau_1>\tau_2$
    \begin{align*}
    \Db_{\times,j}(\tau_1,\tau_2)-\sigma^2_{j,\infty}\overline{r_jr_{\overline{\jmath}}}=& \underbrace{\sum_{i=1}^N \sum_{k=1}^{\tau_1-1}\phi^j_i(k)\Db_{p,(ii)j\overline{\jmath}}(\tau_1-k,\tau_2-k)}_{\mathbb{A}_{3\text{d}\times}\in \mathbb{A}_{\text{d}\times}}\\
    &+\underbrace{2\sum_{i=1}^N \sum_{k_2=\tau_1+1}^{+\infty}K^j_i(\tau_1,k_2)
    \Db_{p,(i\overline{\jmath})ij}(k_2-\tau_1,\tau_1-\tau_2)}_{\mathbb{A}_{4\text{d}\times}\in \mathbb{A}_{\text{d}\times}}\\
    &+\underbrace{\sum_{k=1}^{\tau_1-1} \phi^j_{\times}(k)\left(\Db_{(j\overline{\jmath})j\overline{\jmath}}(\tau_1-k,\tau_2-k)-\overline{r_jr_{\overline{\jmath}}}^2)\right)}_{\mathbb{A}_{3\times}\in \mathbb{A}_{\times}}
    \\&+\underbrace{\sum_{k_1=\tau_2+1}^{+\infty} k^j_{\times}(k_1,\tau_2)\Db_{p,(\overline{\jmath}\overline{\jmath})jj}(k_1-\tau_2,\tau_1-\tau_2)}_{\mathbb{A}_{4\times}\in \mathbb{A}_{\times}}
    \\&+\underbrace{\sum_{k_2=\tau_1+1}^{+\infty} k^j_{\times}(\tau_1,k_2)\Db_{p,(j\overline{\jmath})\overline{\jmath}j}(k_2-\tau_1,\tau_1-\tau_2)}_{\mathbb{A}_{4\times}\in \mathbb{A}_{\times}}\\
\end{align*}

Table~\ref{app_tab:YW_matrices_categories} recaps the different categories. We now proceed to the description of the four structures.

\subsection{$A_1$ Structure}\label{app_sec:A1_structure}

The matrices we call ``$A_1$-like'' are the one making the calibration between the diagonal of the observable ($\mathbb{C}$ and $\Db_{\times}$) and the diagonal of a kernel ($\phi$ or $\phi_{\times}$). The characteristic equation is
\begin{align*}
    E_{A_1}(\tau)= \sum_{u=1}^{+\infty}K(s)\Db(\tau-s), \quad \text{where }K \text{ and }\Db \text{ are a random kernel and covariance structure}.
\end{align*}

\begin{example}[-- $A_1$ relation for $q=3$]
Up to the lag $q=3$, we then have the following system
\begin{align*}
\begin{cases}
    &E_{A_1}(1)=K(1)\Db(0)+K(2)\Db(-1)+K(3)\Db(-2)\\
 &E_{A_1}(2)=K(1)\Db(1)+K(2)\Db(0)+K(3)\Db(-1)\\
 &E_{A_1}(3)=K(1)\Db(2)+K(2)\Db(1)+K(3)\Db(0)  
\end{cases}
\end{align*}

so the Yule-Walker matrix is 
\begin{align*}
    A_1 = \begin{pmatrix}
          \Db(0)& \Db(-1) & \Db(-2)\\
          \Db(1)&\Db(0)& \Db(-1)\\
          \Db(2)& \Db(1) & \Db(0)\\
    \end{pmatrix} \text{and when $\Db$ is even, }A_1 = \begin{pmatrix}
          \Db(0)& \Db(1) & \Db(2)\\
          \Db(1)&\Db(0)& \Db(1)\\
          \Db(2)& \Db(1) & \Db(0)\\
    \end{pmatrix}
\end{align*}

\end{example}

Generalizing to any $q$ and for a any covariance $\Db$, we can define 
\begin{align*}
%     f_{A_1}(\Db)=&\begin{pmatrix}
% \boldsymbol{\Db}(0)&\boldsymbol{\Db}(-1)&\boldsymbol{\Db}(-2)&\boldsymbol{\Db}(-3)&...&\boldsymbol{\Db}(-(q-1))\\
% \boldsymbol{\Db}(1)&\boldsymbol{\Db}(0)&\boldsymbol{\Db}(-1)&\boldsymbol{\Db}(-2)&...&\boldsymbol{\Db}(-(q-2))\\
% \boldsymbol{\Db}(2)&\boldsymbol{\Db}(1)&\boldsymbol{\Db}(0)&\boldsymbol{\Db}(-1)&...&\boldsymbol{\Db}(-(q-3))\\
% &&...&&&\\
% \boldsymbol{\Db}(q-1)&\boldsymbol{\Db}(q-2)&...&&\boldsymbol{\Db}(1)&\boldsymbol{\Db}(0)\end{pmatrix}\\
 f_{A_1}(\Db_{up},\Db_{down},\Sigma)=&\begin{pmatrix}
\Sigma&\boldsymbol{\Db}_{up}(1)&\boldsymbol{\Db}_{up}(2)&\boldsymbol{\Db}_{up}(3)&...&\boldsymbol{\Db}_{up}((q-1))\\
\boldsymbol{\Db}_{down}(1)&\Sigma&\boldsymbol{\Db}_{up}(1)&\boldsymbol{\Db}_{up}(2)&...&\boldsymbol{\Db}_{up}((q-2))\\
\boldsymbol{\Db}_{down}(2)&\boldsymbol{\Db}_{down}(1)&\Sigma&\boldsymbol{\Db}_{up}(1)&...&\boldsymbol{\Db}_{up}((q-3))\\
&&...&&&\\
\boldsymbol{\Db}_{down}(q-1)&\boldsymbol{\Db}_{down}(q-2)&...&&\boldsymbol{\Db}_{down}(1)&\Sigma\end{pmatrix}
\end{align*}

We now just need to determine $(\Db_{up},\Db_{down},\Sigma)$ for each of the ``$A_1$-like'' matrices.

\subsubsection{$\mathbb{A}_{1\text{d}}$}
$\mathbb{A}_{1\text{d}}$ is the same as for the linear GARCH model and we have $$\mathbb{A}_{1\text{d}}=f_{A_1}(\mathbb{C}^\top,\mathbb{C},\mathbb{C}(0)).$$

\subsubsection{$\mathbb{A}_{1\text{d}\times}$}

$\mathbb{A}_{1\text{d}\times}$ makes the relation between $\Db_{\times}(\tau,\tau)$ and $\bbphi$. From the Yule-Walker equation in lemma~\ref{app_lemma_MM:YW_Dx_diag}, we have $$\mathbb{A}_{1\text{d}\times}=f_{A_1}(\boldsymbol{\Db_{\times}},\tau\xrightarrow{}\boldsymbol{\Db_{\times}}(-\tau),\boldsymbol{\Db_{\times}}(0)).$$

\subsubsection{$A_{1\times}$}
Similarly, $A_{1\times}$ encodes the calibration between $\Db_{\times}(\tau,\tau)$ and $\phi_{\times}$, from lemma~\ref{app_lemma_MM:YW_Dx_diag}, we have

$$A_{5\times}= f_{A_1}(\begin{pmatrix}
   \Db_{(i\overline{\imath})i\overline{\imath}} - \overline{r_ir_{\overline{i}}}^2
\end{pmatrix}),\tau\xrightarrow{}\begin{pmatrix}
   \Db_{(i\overline{\imath})i\overline{\imath}}(-\tau) - \overline{r_ir_{\overline{i}}}^2
\end{pmatrix},\begin{pmatrix}
   \Db_{(i\overline{\imath})i\overline{\imath}}(0) - \overline{r_ir_{\overline{i}}}^2
\end{pmatrix})$$

\subsection{$A_2$ Structure}

``$A_2$-like'' matrices encode the calibration between the diagonal of the observable ($\mathbb{C}$ and $\Db_{\times}$) and the non diagonal of the kernels $\mathbb{K}_{\text{d}}$ and $\boldsymbol{K}_{\times}$. The characteristic equation is
\begin{align*}
     E_{A_2}(\tau)=2\sum_{k_2=\tau+1}^{+\infty}\sum_{k_1=k_2+1}^{+\infty}K(k_1,k_2)\Db(k_1-\tau,k_2-\tau),
\end{align*}
{where }$K${ and }$\Db${ are a random kernel and covariance structure}.

\begin{example}[-- $A_2$ relation for $q=3$]
Up to the lag $q=3$, we then have the following system

\begin{align*}
\begin{cases}
    % E_{A_2}(1)&=2 \sum_{u=0}^{q}\sum_{r=u+1}^{q}K(1+u,1+r)\Db(u,r) \\
    E_{A_2}(1)&= K(1,2)\Db(0,1)+K(1,3)\Db(0,2)+K(2,3)\Db(1,2)\\
    E_{A_2}(2) &= K(1,2)\Db(-1,0)+K(1,3)\Db(-1,1)+K(2,3)\Db(0,1)\\
    E_{A_3}(3) &= K(1,2)\Db(-2,-1)+K(1,3)\Db(-2,0)+K(2,3)\Db(-1,0)    
\end{cases}.
\end{align*}

Hence, we have the following matrix 

\begin{align*}
    \begin{pmatrix}
          E(1)\\
          E(2)\\
          E(3)
    \end{pmatrix}=\begin{pmatrix}
          \Db(0,1)&\Db(0,2)&\Db(1,2)\\
          \Db(-1,0)&\Db(-1,1)&\Db(0,1)\\
          \Db(-2,-1)&\Db(-2,0)&\Db(-1,0)\\
    \end{pmatrix}\begin{pmatrix}
          K(1,2)\\
          K(1,3)\\
          K(2,3)
    \end{pmatrix}\\
    \text{and if $\Db$ is causal,}\\
    \begin{pmatrix}
          E(1)\\
          E(2)\\
          E(3)
    \end{pmatrix}=\begin{pmatrix}
          \Db(0,1)&\Db(0,2)&\Db(1,2)\\
          0&0&\Db(0,1)\\
          0&0&0\\
    \end{pmatrix}\begin{pmatrix}
          K(1,2)\\
          K(1,3)\\
          K(2,3)
    \end{pmatrix}
\end{align*}
    
\end{example}

Generalizing for any lag $q$ and for general structure $\Db$, we can define $f_{A_2}(\Db)$ as following: if we consider the matrix $(\Db(i,j))_{i,j \in \llbracket-(q-1),(q-1)\rrbracket}$ then, the $n^{\text{th}}$ row of $A_2$-like matrices is two times the upper triangle of  $(\Db(i,j))_{i,j \in \llbracket-(q-1),(q-1)\rrbracket}[q-n:2q-n,q-n:2q-n]$. Calling $f_{A_2}$ the function performing the transformation $A_2=f_{A_2}(\Db)$, we now proceed to the definition of the $A_2$-like matrices.

\subsubsection{$\mathbb{A}_{2\text{d}}$}

For $\mathbb{A}_{2\text{d}}$, we use the link between $\mathbb{C}$ and the quadratic kernel $\mathbb{K}$ so
$$\mathbb{A}_{2\text{d}}=f_{A_2}(\mathbb{D}^\top_{\text{d}}).$$

\subsubsection{$\mathbb{A}_{2\text{d}\times}$}

For $\mathbb{A}_{2\text{d}\times}$, we use the link between the time diagonal of $\Db_{\times}$ and the off time-diagonal kernel $\mathbb{K}_{\text{d}}^{\text{off-diagonal}}$ from lemma~\ref{app_lemma_MM:YW_Dx_diag}, and we obtain

$$\mathbb{A}_{2\text{d}\times}=f_{A_2}(\begin{pmatrix}
          \Db_{i\overline{\imath}ii}\\
           \Db_{i\overline{\imath}\overline{\imath}\overline{\imath}}
       \end{pmatrix}).$$

\subsubsection{$\mathbb{A}_{2\times}$}

$\mathbb{A}_{2\times}$ encodes the calibration between the time diagonal of $\Db_{\times}$ and the kernel $\boldsymbol{k}_{\times}$. Since $\boldsymbol{k}_{\times}$ is not time-symmetric (conversely to $\mathbb{K}$) we need to considered $\boldsymbol{k}_{\times}$ for both  $\tau_1<\tau_2$ and $\tau_2<\tau_1$. This phenomenon intervenes in several calibration matrices, as $\mathbb{A}_{2\times}$. We set the left bloc being the one encoding the calibration of $\boldsymbol{k}_{\times}(\tau_1,\tau_2)_{\tau_1<\tau_2}$ and the right bloc for $\boldsymbol{k}_{\times}(\tau_1,\tau_2)_{\tau_1>\tau_2}$ (consistently with the framework developped above).\newline

Then, from lemma~\ref{app_lemma_MM:YW_Dx_diag}, and we obtain
$$\mathbb{A}_{2\times}^{\text{left bloc}}=f_{A_2}(\Db_{p(i\overline{\imath})i\overline{\imath}})\quad \& \quad \mathbb{A}_{2\times}^{\text{right bloc}}=f_{A_2}(\Db_{p(i\overline{\imath})\overline{\imath}i}).$$

\subsection{$A_3$ structure}

``$A_3$-like'' matrices are the one linking the off-time-diagonal of observable covariances ($\mathbb{D}_{\text{d}}$ and $\boldsymbol{\Db}_{\times}$ with $\tau_1\neq\tau_2$) with the time-diagonal kernels ($\bbphi$ and $\boldsymbol{\phi}_\times$). The characteristic equation is \begin{align*}
    E_{A_3}(\tau_1,\tau_2)=\sum_{u=1}^{+\infty}K(u)\Db(\tau_1-u,\tau_2-u)
\end{align*}

{where }$K${ and }$\Db${ are a random kernel and covariance structure}.

\begin{example}[-- $A_3$ relation for $q=3$]
Up to the lag $q=3$, we then have the following system

\begin{align*}
\begin{cases}
    \Db_{A_3}(1,2)=&K(3)\Db(-2,-1) + K(2)\Db(-1,0) + K(1)\Db(0,1)\\
    \Db_{A_3}(1,3)=&K(3)\Db(-2,0) + K(2)\Db(-1,1) + K(1)\Db(0,2)\\
    \Db_{A_3}(2,3)=&K(3)\Db(-1,0) + K(2)\Db(0,1) + K(1)\Db(1,2)
\end{cases}.
\end{align*}

Hence, we obtain the system:
\begin{align*}
    &\begin{pmatrix}
          \Db_{A_3}(1,2)\\
          \Db_{A_3}(1,3)\\
          \Db_{A_3}(2,3)
    \end{pmatrix}=\begin{pmatrix}
          \Db(0,1)&\Db(-1,0)&\Db(-2,-1)\\
          \Db(0,2)&\Db(-1,1)&\Db(-2,0)\\
          \Db(1,2)&\Db(0,1)&\Db(-1,0)
    \end{pmatrix}\begin{pmatrix}
          K(1)\\
          K(2)\\
          K(3)
    \end{pmatrix}\\
    &\text{and if $\Db$ is causal, we have}\\
    &\begin{pmatrix}
          \Db_{A_3}(1,2)\\
          \Db_{A_3}(1,3)\\
          \Db_{A_3}(2,3)
    \end{pmatrix}=\begin{pmatrix}
          \Db(0,1)&0&0\\
          \Db(0,2)&0&0\\
          \Db(1,2)&\Db(0,1)&0
    \end{pmatrix}\begin{pmatrix}
          K(1,1)\\
          K(2,2)\\
          K(3,3)
    \end{pmatrix}.
\end{align*}

\end{example}

The function to create ``$A_3$-like'' matrices $f_{A_3}$, has the same structure as $A_2$, only transposed. Hence, we define $f_{A_3}(\Db)=f_{A_2}(\Db)^\top/2$. We now detailed the definition of each ``$A_3$-like'' matrix.

\subsubsection{$\mathbb{A}_{3\text{d}}$}

$\mathbb{A}_{3\text{d}}$ encodes the feedback of $\bbphi$ on the time-off-diagonal of $\mathbb{D}_{\text{d}}$. Using lemma~\ref{app_lemma_MM:YW_Db}, we have $$\mathbb{A}_{3\text{d}}=f_{A_3}(\mathbb{D}_{\text{d}})$$%=A_2^\top/2$$

% \begin{remark}

In python, one can use \texttt{(A3=A2.transpose(1,0,3,2)/2)}.
% \end{remark}

\subsubsection{$\mathbb{A}_{3\text{d}\times}$}

$\mathbb{A}_{3\text{d}\times}$ encodes the feedback of $\bbphi$ on the off-time-diagonal of $\boldsymbol{\Db}_{\times}$. $\mathbb{A}_{3\text{d}\times}$ is then composed of two blocs to capture the time asymmetry of $\boldsymbol{\Db}_{\times}$. The upper bloc characterises the feedback of $\boldsymbol{\phi}_\times$ on $\boldsymbol{\Db}_{\times}(\tau_1,\tau_2)$ for $\tau_1<\tau_2$ and the lower bloc for $\tau_1>\tau_2$.

Using lemmas~\ref{app_lemma_MM:YW_Dx_1} and \ref{app_lemma_MM:YW_Dx_2}, we can write

$$\mathbb{A}_{3\text{d}\times}^{\text{upper}}=f_{A_3}(\begin{pmatrix}
    \Db_{p,(ii)i\overline{\imath}}\\\Db_{p,(\overline{\imath}\overline{\imath})i\overline{\imath}}
\end{pmatrix})\quad \&\quad \mathbb{A}_{3\text{d}\times}^{\text{lower}}=f_{A_3}(\begin{pmatrix}
    \Db_{p,(ii)\overline{\imath}i}\\\Db_{p,(\overline{\imath}\overline{\imath})\overline{\imath}i}
\end{pmatrix}).$$

The second level is done with the symmetric of the first level.

\subsubsection{$\mathbb{A}_{3\times}$}

$\mathbb{A}_{3\times}$ encodes the feedback of $\boldsymbol{\phi}_{\times}$ on the off-time-diagonal of $\boldsymbol{\Db}_{\times}$. As for $\mathbb{A}_{3\text{d}\times}$, we account for the time asymmetry of $\boldsymbol{\Db}_{\times}$ by allocating the upper bloc of $\mathbb{A}_{3\times}$ to $\boldsymbol{\Db}_{\times}(\tau_1,\tau_2)$ for $\tau_1<\tau_2$ and the lower bloc for $\tau_1>\tau_2$. Using lemmas~\ref{app_lemma_MM:YW_Dx_1} and \ref{app_lemma_MM:YW_Dx_2}, we have
$${\mathbb{A}_{3\times}^{\text{upper}}}=f_{A_3}(\begin{pmatrix}\Db_{(i\overline{\imath})i\overline{\imath}}\end{pmatrix})\quad \&\quad {\mathbb{A}_{3\times}^{\text{lower}}}=f_{A_3}(\begin{pmatrix}\Db_{(i\overline{\imath})\overline{\imath}i}\end{pmatrix}).$$

\subsection{$A_4$ structure}

``$A_4$-like'' matrices are the one linking off-time-diagonal of observable ($\mathbb{D}_{\text{d}}$ and $\boldsymbol{\Db}_{\times}$ with $\tau_1\neq\tau_2$) and the off-time-diagonal of kernels ($\mathbb{K}$ and $\boldsymbol{k}_{\times}$). The characteristic equation is

\begin{align*}
    E_{A_4}(\tau_1,\tau_2)=2\sum_{u=\tau_1+1}^{+\infty}K(u,\tau_1)\Db(u-\tau_1,\tau_2-\tau_1)%+2\int_{0^+}^{\tau_2^+}K(\tau_2,u)\Db(u-\tau_2,\tau_1-\tau_2){\rm d}u
\end{align*}
{where }$K${ and }$\Db${ are a random kernel and covariance structure}.

\begin{example}[-- $A_4$ relation for $q=3$]
Up to the lag $q=3$, we then have the following system
for $q=3$,

\begin{align*}
\begin{cases}
     \Db_{A_4}(1,2)=&2(K(2,1)\Db(1,1)+K(3,1)\Db(2,1))\\
    \Db_{A_4}(1,3)=&2(K(2,1)\Db(1,2)+K(3,1)\Db(2,2))\\
    \Db_{A_4}(2,3)=&2K(3,2)\Db(1,1)
\end{cases}.
\end{align*}

Hence, we obtain the system

\begin{align*}
    \begin{pmatrix}
          \Db_{A_4}(1,2)\\
          \Db_{A_4}(1,3)\\
          \Db_{A_4}(2,3)
    \end{pmatrix}=\begin{pmatrix}
          \Db(1,1)&\Db(2,1)&0\\
          \Db(1,2)&\Db(2,2)&0\\
          0&0&\Db(1,1)
    \end{pmatrix}\begin{pmatrix}
          K(1,2)\\
          K(1,3)\\
          K(2,3)
    \end{pmatrix}.
\end{align*}

\end{example}

The function to create ``$A_4$-like'' matrices from the correlation $\Db$ can be deduced from the example. It creates blocs of the matrix $(\Db(i,j))_{i,j}$ along the diagonal. We now detail the structure of the $A_4$-like matrices.

\subsubsection{$\mathbb{A}_{4\text{d}}$}

$\mathbb{A}_{4\text{d}}$ encodes the link between  $\mathbb{K}$ and $\mathbb{D}_{\text{d}}$. From lemma~\ref{app_lemma_MM:YW_Db}, we have 

$$\mathbb{A}_{4\text{d}}=f_{A_4}((\Db_{(ij)ij})_{ij}).$$%$,(\Db_{(ij)ij})_{ij})$

\subsubsection{$\mathbb{A}_{4\text{d}\times}$}

$\mathbb{A}_{4\text{d}\times}$ encodes the feedback of $\mathbb{K}_{\text{d}}$ on the off-time-diagonal of $\boldsymbol{\Db}_{\times}$. To account for the asymmetry of $\boldsymbol{\Db}_{\times}$, we build $\mathbb{A}_{4\text{d}\times}$ with two blocs: the upper bloc being for $\boldsymbol{\Db}_{\times}(\tau_1,\tau_2)$ with  $\tau_1<\tau_2$ and the lower bloc for $\tau_2<\tau_1$. From lemmas~\ref{app_lemma_MM:YW_Dx_1} and \ref{app_lemma_MM:YW_Dx_2}, we have

$$\mathbb{A}_{4\text{d}\times}^{\text{upper}}=f_{A_4}(\begin{pmatrix}
    \Db_{p(ii)i\overline{\imath}}\\
    \Db_{p(i\overline{\imath})\overline{\imath}\overline{\imath}}
\end{pmatrix})\quad\&\quad\mathbb{A}_{4\text{d}\times}^{\text{lower}}=f_{A_4}(\begin{pmatrix}
    \Db_{p(\overline{\imath}i)ii}\\
    \Db_{p(\overline{\imath}\overline{\imath})\overline{\imath}i}
\end{pmatrix})$$

\subsubsection{$\mathbb{A}_{4\times}$}

$\mathbb{A}_{4\times}$ encodes the feedback of $\boldsymbol{k}_\times$ on the off-time-diagonal of $\boldsymbol{\Db}_{\times}$. For this particular matrix, one has to account for the asymmetry of $\boldsymbol{\Db}_{\times}$ and of $\boldsymbol{k}_\times$. As before, we thus build blocs for each case. 
From lemmas~\ref{app_lemma_MM:YW_Dx_1} and \ref{app_lemma_MM:YW_Dx_2}, we have

\begin{itemize}
    \item upper left, $\boldsymbol{\Db}_{\times}(\tau_1<\tau_2)$ VS $\boldsymbol{k}_\times (\tau_1<\tau_2)$\\
    \begin{align*}
    \mathbb{A}_{4\times}^{\text{upper left}} =f_{A_4}(\begin{pmatrix}
        \Db_{p(ii)\overline{\imath}\overline{\imath}}
    \end{pmatrix})
\end{align*}
    \item upper right, $\boldsymbol{\Db}_{\times}(\tau_1<\tau_2)$ VS $\boldsymbol{k}_\times (\tau_1>\tau_2)$\\
     \begin{align*}
    \mathbb{A}_{4\times}^{\text{upper right}} =f_{A_4}(\begin{pmatrix}
        \Db_{p(i\overline{\imath})i\overline{\imath}}
    \end{pmatrix})
\end{align*}
    \item lower left, $\boldsymbol{\Db}_{\times}(\tau_1>\tau_2)$ VS $\boldsymbol{k}_\times (\tau_1<\tau_2)$\\
     \begin{align*}
    \mathbb{A}_{4\times}^{\text{lower left}} =f_{A_4}(\begin{pmatrix}
        \Db_{p(i\overline{\imath})\overline{\imath}i}
    \end{pmatrix})
\end{align*}
    \item lower right, $\boldsymbol{\Db}_{\times}(\tau_1>\tau_2)$ VS $\boldsymbol{k}_\times (\tau_1>\tau_2)$\\
     \begin{align*}
    \mathbb{A}_{4\times}^{\text{lower left}} =f_{A_4}(\begin{pmatrix}
        \Db_{p(\overline{\imath}\overline{\imath})ii}
    \end{pmatrix})
\end{align*}
\end{itemize}

\subsection{Calibration matrices for the leverage kernels}\label{app_sec:leverage}

\subsubsection{Leverage calibration - framework}

We now explicate a method to determine the leverage kernels $(L^i_j)_{i,j\in \{1,2\}}$. For those last kernels, we aim at building a linear system such that: 

\begin{align*}
    \begin{pmatrix}
        \mathbb{V}(1)\\
        \mathbb{V}(2)\\
        \vdots\\
        \mathbb{V}(q)\\
    \end{pmatrix} = \mathbb{A_L} \begin{pmatrix}
        \mathbb{L}(1)\\
        \mathbb{L}(2)\\
        \vdots\\
        \mathbb{L}(q)\\
    \end{pmatrix}
\end{align*}

where 
\begin{align*}
    \mathbb{L}(\tau) = \begin{pmatrix}
        L^1_1(\tau)&L^1_2(\tau)\\
        L^2_1(\tau)&L^2_2(\tau)\\
    \end{pmatrix} \quad \text{ and } \quad  \mathbb{V}(\tau) = \begin{pmatrix}
        \Vb_{11}(\tau)&\Vb_{12}(\tau)\\
        \Vb_{21}(\tau)&\Vb_{22}(\tau)\\
    \end{pmatrix}.
\end{align*}

As before, we rely on the Yule-Walker type equation to construct $\mathbb{A_L}$. As a reminder, lemma~\ref{app_lemma_MM:YW_Vb} gives

\begin{equation*}
    \begin{aligned}
         \Vb_{jl}(\tau) = &L^j_i(\tau)\overline{r_{i}r_{l}}\\
    &+ \sum_{i=1}^N \sum_{k=1}^{\tau}\phi^j_i(k)\Vb^r_{il}(\tau-k)\\&+2\sum_{k_2=\tau+1}^{+\infty}K^j_i(\tau,k_2)\Vb^r_{(il)i}(k_2-\tau)\\&+\sum_{k=1}^{\tau} \phi^{j}_\times(k)\Vb^r_{(i\overline{\imath})l}(\tau-k)\\&+\sum_{k_2=\tau+1}^{+\infty} k^{j}_\times(\tau,k_2)\Vb^r_{(jl)\overline{\jmath}}(k_2-\tau)
    \\&+\sum_{k_1=\tau+1}^{+\infty} k^{j}_\times(k_1,\tau)\Vb^r_{(\overline{\jmath}l)j}(k_1-\tau).
    \end{aligned}
\end{equation*}

The equation above demonstrates that $\Vb_{jl}$ not only depends on the leverage kernels $(L^i_j)_{i,j\in \{1,2\}}$ but also on the linear kernels $(\phi^i_j)_{i,j\in \{1,2\}}$, the quadratic kernels $(K^i_j)_{i,j\in \{1,2\}}$ and the cross kernels $(\phi^i_\times)_{i\in \{1,2\}}$ and $(k^i_\times)_{i\in \{1,2\}}$ that we determined in steps 1, 2 and 3. 
Hence, the equation system becomes:

\begin{align*}
    \begin{pmatrix}
        \Tilde{\mathbb{V}}(1)\\
        \Tilde{\mathbb{V}}(2)\\
        \vdots\\
        \Tilde{\mathbb{V}}(q)\\
    \end{pmatrix} = \begin{pmatrix}
        \begin{pmatrix}
            \overline{r_1}^2& \overline{r_1r_2}\\
            \overline{r_1r_2}&\overline{r_2}^2
        \end{pmatrix}&&0\\
        &\ddots&\\
        0& &\begin{pmatrix}
            \overline{r_1}^2& \overline{r_1r_2}\\
            \overline{r_1r_2}&\overline{r_2}^2
        \end{pmatrix}
    \end{pmatrix} \begin{pmatrix}
         \mathbb{L}(1)\\
        \mathbb{L}(2)\\
        \vdots\\
        \mathbb{L}(q)\\
    \end{pmatrix}
\end{align*}

where 
\begin{align*}
     \begin{pmatrix}
        \Tilde{\mathbb{V}}(1)\\
        \tilde{\mathbb{V}}(2)\\
        \vdots\\
        \tilde{\mathbb{V}}(q)\\
    \end{pmatrix}=\begin{pmatrix}
        {\mathbb{V}}(1)\\
        \mathbb{V}(2)\\
        \vdots\\
        \mathbb{V}(q)\\
    \end{pmatrix} - \mathbb{A}_{1L}\begin{pmatrix}
        \bbphi(1)\\
        \bbphi(2)\\
        \vdots\\
        \bbphi(q)\\
    \end{pmatrix}- \mathbb{A}_{1\times L}\begin{pmatrix}
       \boldsymbol{\phi}_\times(1)\\
        \boldsymbol{\phi}_\times(2)\\
        \vdots\\
        \boldsymbol{\phi}_\times(q)\\ 
    \end{pmatrix}- \mathbb{A}_{v L}\begin{pmatrix}
       \mathbb{K}_\times(1,2)\\
        \mathbb{K}_\times(1,3)\\
        \vdots\\
        \mathbb{K}_\times(q-1,q)\\ 
    \end{pmatrix}- \mathbb{A}_{v\times L}\begin{pmatrix}
       \boldsymbol{K}_\times(1,2)\\
        \boldsymbol{K}_\times(1,3)\\
        \vdots\\
        \boldsymbol{K}_\times(q-1,q)\\ \end{pmatrix}, 
\end{align*}

and $\mathbb{A}_{1L}$, $\mathbb{A}_{1\times L}$, $\mathbb{A}_{v L}$ and $\mathbb{A}_{v\times L}$ are to be determined with the Yule-Walker equation of lemma~\ref{app_lemma_MM:YW_Vb}. Notably, we have 

\begin{equation*}
    \begin{aligned}
         \Vb_{jl}(\tau) = &L^j_i(\tau)\overline{r_{i}r_{l}}\\
    &+ \underbrace{\sum_{i=1}^N \sum_{k=1}^{\tau}\phi^j_i(k)\Vb^r_{il}(\tau-k)}_{\mathbb{A}_{1L}}\\&+\underbrace{2\sum_{k_2=\tau+1}^{+\infty}K^j_i(\tau,k_2)\Vb^r_{(il)i}(k_2-\tau)}_{\mathbb{A}_{v L}}\\&+\underbrace{\sum_{k=1}^{\tau} \phi^{j}_\times(k)\Vb^r_{(i\overline{\imath})j}(\tau-k)}_{\mathbb{A}_{1\times L}}\\&+\underbrace{\sum_{k_2=\tau+1}^{+\infty} k^{j}_\times(\tau,k_2)\Vb^r_{(jl)\overline{\jmath}}(k_2-\tau)}_{\mathbb{A}_{v\times L}\text{ left bloc}}
    \\&+\underbrace{\sum_{k_1=\tau+1}^{+\infty} k^{j}_\times(k_1,\tau)\Vb^r_{(\overline{\jmath}l)j}(k_1-\tau)}_{\mathbb{A}_{v\times L}\text{ right bloc}}.
    \end{aligned}
\end{equation*}

Using the work above, we can already determine $\mathbb{A}_{1L}$ and $\mathbb{A}_{1\times L}$ as they are ``$A_1$-like'' matrices. Hence, we have
\begin{align*}
    \mathbb{A}_{1L}=f_{A_1}(\boldsymbol{0}_{2\times 2},(\mathbb{V}^r),\begin{pmatrix}
        \overline{\sigma_1^2r_1}&0\\
        0&\overline{\sigma_2^2r_2}
    \end{pmatrix}) \quad \text{ and }\quad \mathbb{A}_{1\times L} = f_{A_1}(\boldsymbol{0}_{2\times 2},\begin{pmatrix}
        \Vb^r_{(i\overline{\imath})i}\\\Vb^r_{(i\overline{\imath})\overline{\imath}}
    \end{pmatrix},\begin{pmatrix}
        0\\0
    \end{pmatrix}).
\end{align*}

\subsubsection{$A_5$ structure}

Additionally, we need to define another type of matrices for $\mathbb{A}_{v L}$ and $\mathbb{A}_{v\times L}$. Both have the same structure which we call ``$A_5$-like''.  The characteristic equation is as follows
\begin{align*}
    E_{A_5}(\tau)= \sum_{k=\tau+1}^{+\infty}K(\tau,k)\Db(k-\tau).
\end{align*}
{where }$K${ and }$\Db${ are a random kernel and covariance structure}.

\begin{example}[-- $A_5$ relation for $q=3$]
Up to the lag $q=3$, we then have the following system
\begin{align*}
\begin{cases}
    &E_{A_5}(1)=K(1,2)\Db(1)+K(1,3)\Db(2)\\
 &E_{A_5}(2)=K(2,3)\Db(1)\\
 &E_{A_5}(3)=0
\end{cases}
\end{align*}

so the Yule-Walker matrix is 
\begin{align*}
    A_5 = \begin{pmatrix}
           \Db(1) & \Db(2)&0\\
          0&0& \Db(1)\\
          0& 0 & 0\\
    \end{pmatrix}
\end{align*}

\end{example}

Hence, the generalisation of ``$A_5$-like'' matrices comes down to writing each row $i$ of $A_5$ as $A_5[i,q-i:2q-2i-1]+=\Db[:q-i-1]$ and taking the other entries as zeros. Then, calling $f_{A_5}$ the function to build ``$A_5$-like'' matrices such like $A_5=f_{A_5}(\Db)$ we can define the last Yule-Walker matrices as

\begin{align*}
    \mathbb{A}_{v L}=f_{A_5}(\begin{pmatrix}
        \Vb^r_{(ii)i}&\Vb^r_{(i\overline{\imath})i}\\\Vb^r_{(i\overline{\imath})\overline{\imath}}&\Vb^r_{(\overline{\imath}\overline{\imath})\overline{\imath}}
    \end{pmatrix}) ,\quad \mathbb{A}_{v\times L}^{\text{left bloc}}=f_{A_5}(\begin{pmatrix}
        \Vb^r_{(ii)\overline{\imath}}\\\Vb^r_{(i\overline{\imath})\overline{\imath}}
    \end{pmatrix})\quad \text{ and }\quad  \mathbb{A}_{v\times L}^{\text{right bloc}}=f_{A_5}(\begin{pmatrix}
        \Vb^r_{(\overline{\imath}i)\overline{\imath}}(-)\\\Vb^r_{(\overline{\imath}\overline{\imath})\overline{\imath}}(-)
    \end{pmatrix}).
\end{align*}

Empirically, we observe that the contribution of these kernels is not significant, and so, in practice for the sake of simplicity, we only consider $\mathbb{A}_{1L}$ and $\mathbb{A}_{1\times L}$ in the calibration.

% \subsection{Generalisation to $N$ dimensions}

\newpage
\section{Kernels profiles from 2D-QGARCH Calibration on futures on indices}\label{app:future_on_indices_profiles}

\begin{figure}[h]
    \centering
    \includegraphics[width=\linewidth]{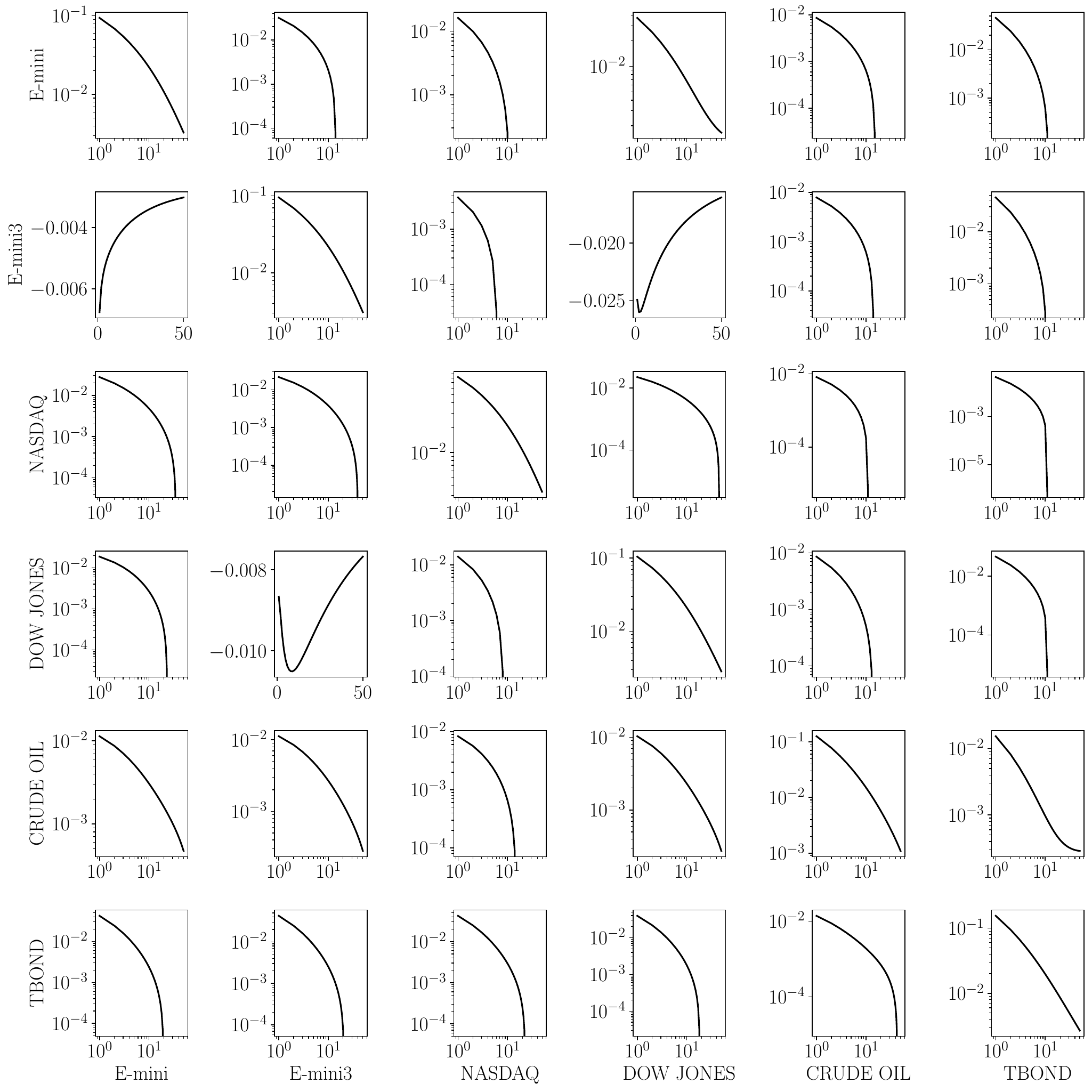}
    \caption{$(\phi^j_i)$ profiles -- the x-labels determine the index providing feedback and the y-labels determine the index receiving the feedback, i.e., for a kernel $\phi_i^j$, $i$ is labelled on the x-axis while $j$ is labelled on the y-axis.}
    \label{fig:phi_profile_futures}
\end{figure}

\begin{figure}[h]
    \centering
    \includegraphics[width=\linewidth]{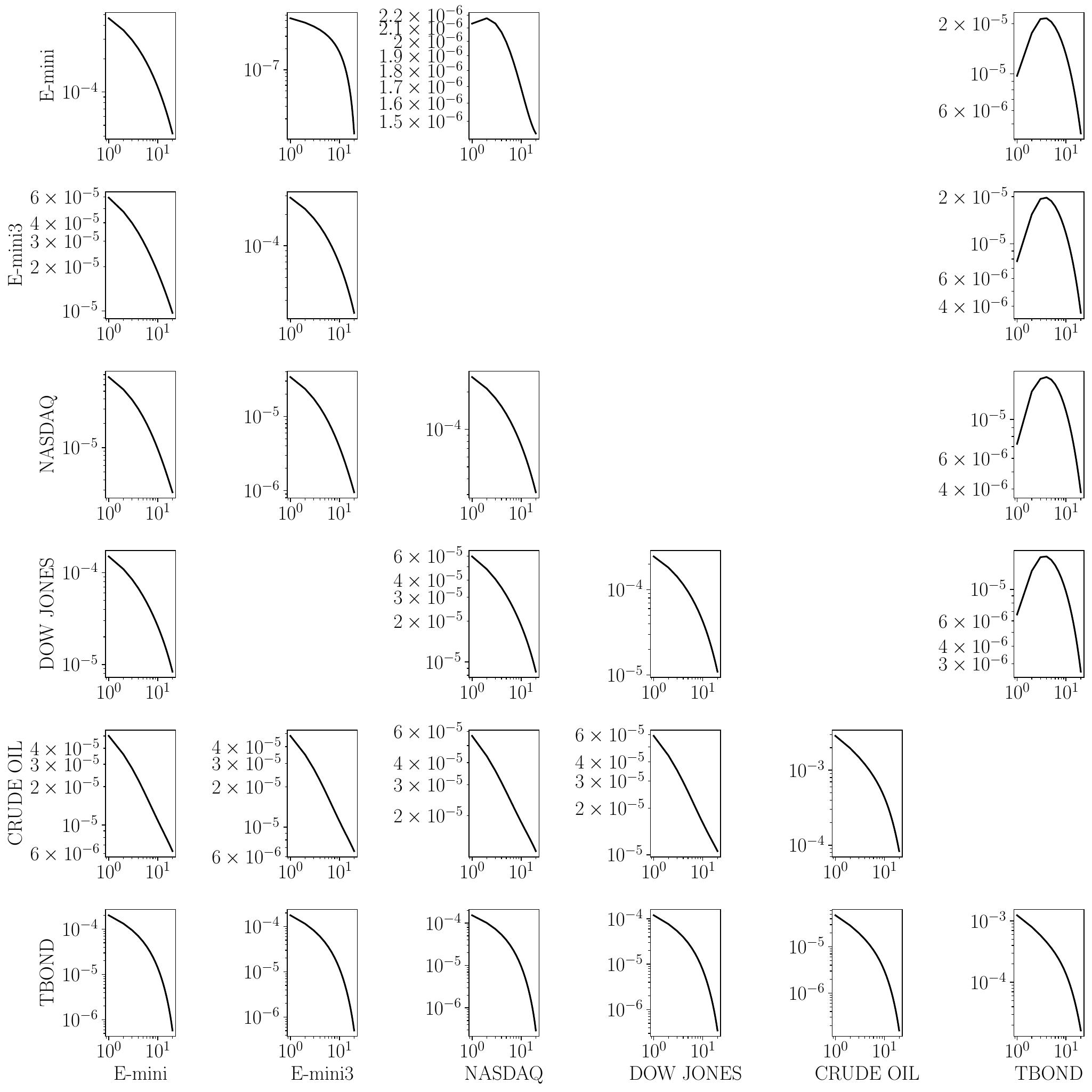}
    \caption{$(k_i^j)^2$ profiles -- the x-labels determine the index providing feedback and the y-labels determine the index receiving the feedback, i.e., for a kernel $(k_i^j)^2$, $i$ is labelled on the x-axis while $j$ is labelled on the y-axis.}
    \label{fig:ksq_profile_futures}
\end{figure}

\begin{figure}[h]
    \centering
    \includegraphics[width=\linewidth]{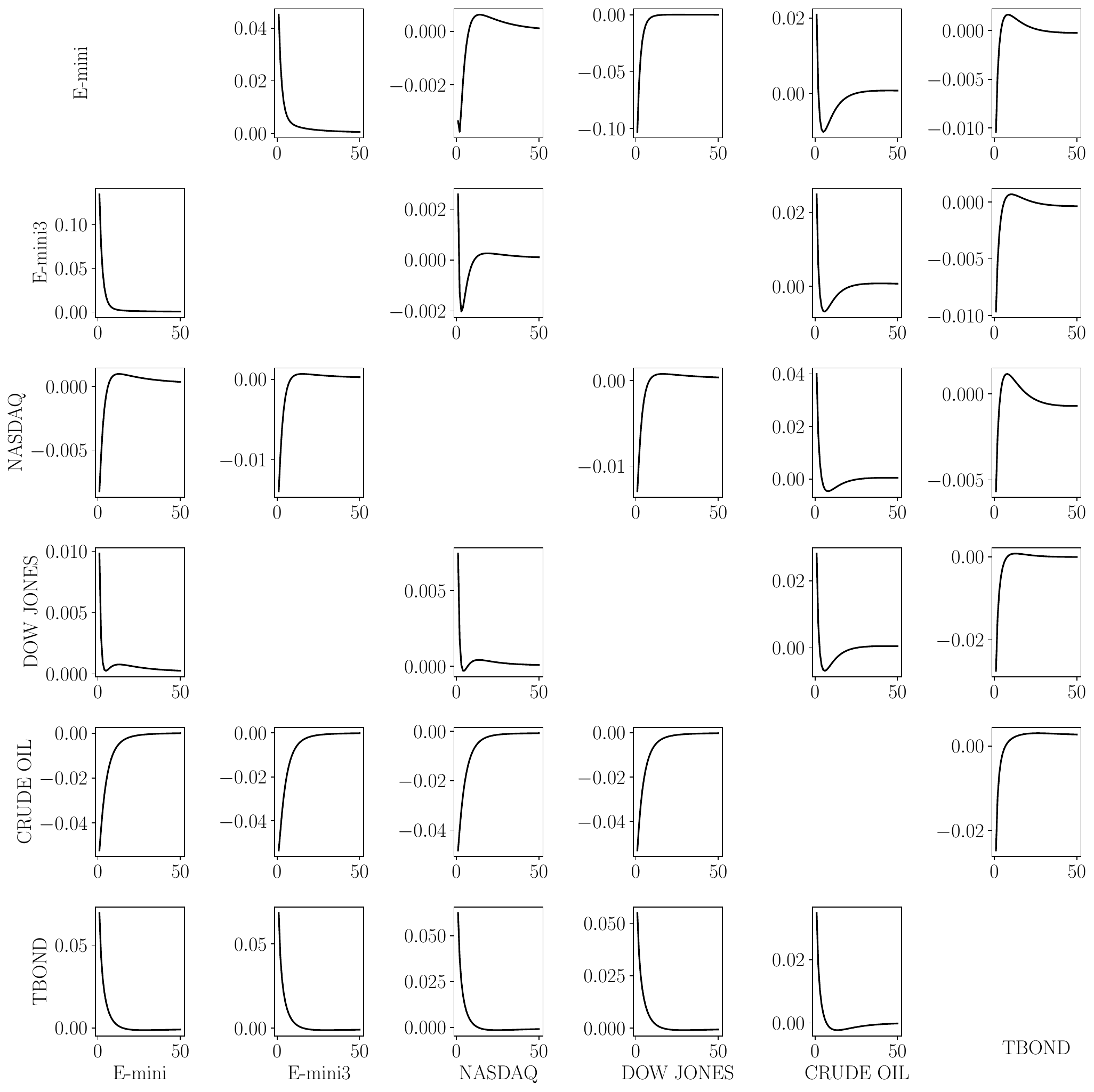}
    \caption{$\phi^j_\times$ profiles -- the y-labels determine the index receiving the feedback while the x-labels determine the index used for the correlation, i.e., the subplot $(i,j)$ show the profile of the kernel   $(\phi^i_\times)$ for the correlation between asset $i$ and asset $j$.}
    \label{fig:phix_profile_futures}
\end{figure}

\begin{figure}[h]
    \centering
    \includegraphics[width=\linewidth]{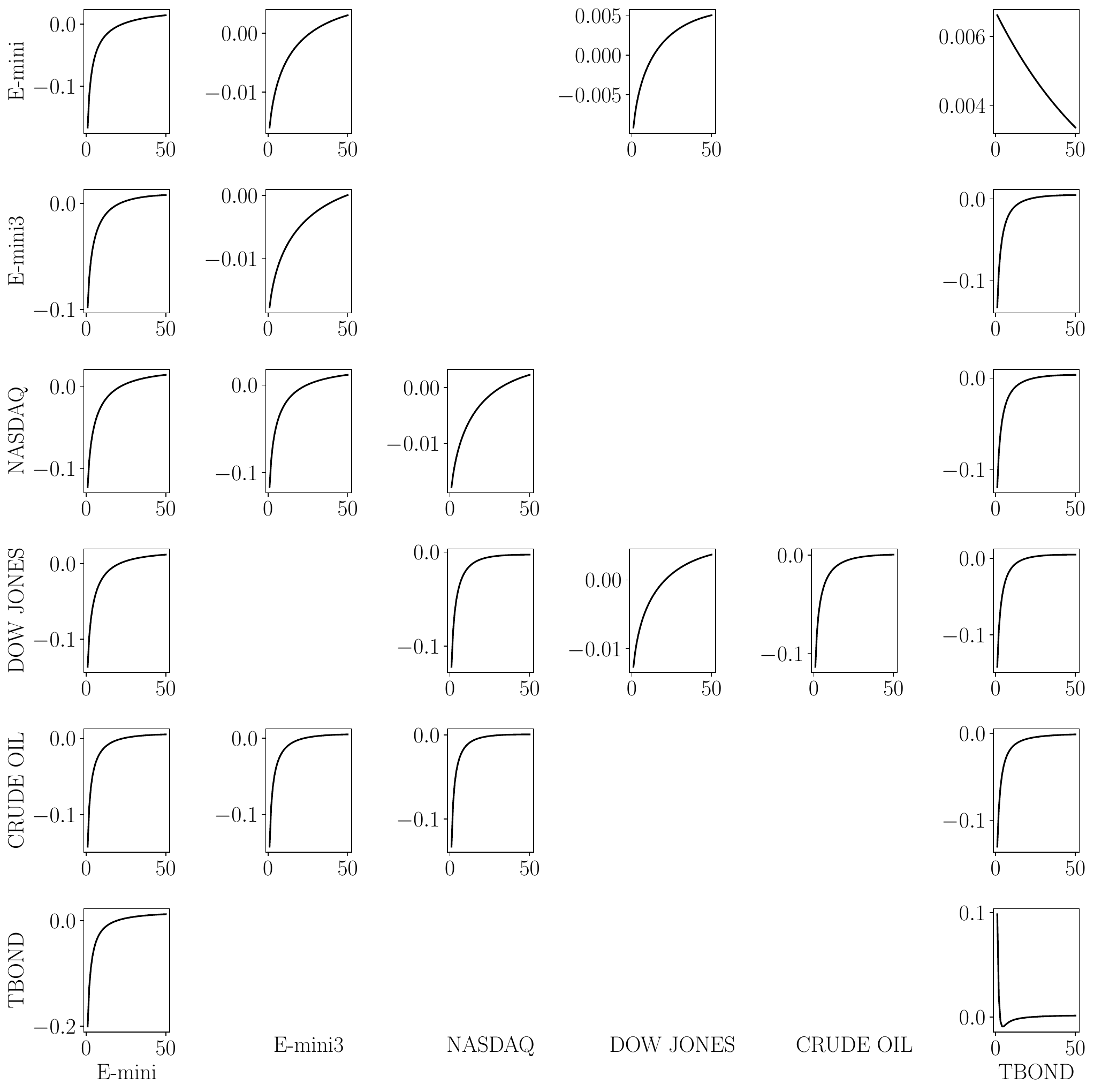}
    \caption{$L^i_j$ profiles -- the x-labels determine the index providing feedback and the y-labels determine the index receiving the feedback, i.e., for a kernel $L_i^j$, $i$ is labelled on the x-axis while $j$ is labelled on the y-axis.}
    \label{fig:L_profile_futures}
\end{figure}

\clearpage

\newpage

\section{Fits and noise reduction parameters}\label{app:fit_parameters}

\begin{table}[h!]
\begin{center}
\begin{tabularx}{0.9\textwidth}{|>{\centering\arraybackslash}X | >{\centering\arraybackslash}X>{\centering\arraybackslash}X>{\centering\arraybackslash}X>{\centering\arraybackslash}X|} 
 \hline
  &$\alpha$ & $\beta$ &$\gamma$ & $n$ \\ [0.5ex] 
 \hline\hline
 $\Cb_{E}$ & 0.740&1e-8 & 10.00&3.0845  \\ 
 \hline
 $\Cb_{E,T}$ & 0.1184 &1e-8 & 10.00&1.8810\\
 \hline
 $\Cb_{T,E}$ & 0.1151&1e-8 & 10.00& 2.0408 \\
 \hline
 $\Cb_{T}$ & 0.2088&1e-8 & 10.00& 3.5990 \\  
 \hline
\end{tabularx}
\caption{Fits parameters of the 2-points correlation $\Cb_{ij}$, as defined by Equation~\eqref{eq_MM:Cdef} with the fit of $\tau \xrightarrow{} {n\exp(-\beta \tau)}{(1+\gamma \tau)^{-\alpha}}$, for the pair {\sc e-mini vs tbond}}
\label{table:Cb_fit}
\end{center}
\end{table}

\begin{table}[h!]
\begin{center}
\begin{tabularx}{0.9\textwidth}{|>{\centering\arraybackslash}X | >{\centering\arraybackslash}X>{\centering\arraybackslash}X>{\centering\arraybackslash}X>{\centering\arraybackslash}X|} 
 \hline
  & $\alpha$ & $\beta$ &$\gamma$ & $n$ \\ [0.5ex] 
 \hline\hline
 $\Db_{\times E}$ & 0.0911&1e-8&10.00&-0.6743  \\ 
 \hline
 $\Db_{\times T}$ & 0.1062 &1e-8&10.00&-0.3118 \\
 \hline
\end{tabularx}
\caption{Fits parameters of the 3-points correlation $\Db_{\times j}$, as defined by Equation~\eqref{eq:Dx_def} with the fit of $\tau \xrightarrow{} {n\exp(-\beta \tau)}{(1+\gamma \tau)^{-\alpha}}$, for the pair {\sc e-mini vs tbond}}
\label{table:Dx_fit}
\end{center}
\end{table}

% \begin{table}[h!]
% \begin{center}
% \begin{tabularx}{0.9\textwidth}{|>{\centering\arraybackslash}X | >{\centering\arraybackslash}X>{\centering\arraybackslash}X|} 
%  \hline
%   & $a$ & $b$ \\ [0.5ex] 
%  \hline\hline
%  $\Db_{E}$ & 6&ff  \\ 
%  \hline
%  $\Db_{E,T}$ & 7 &ff \\
%  \hline
%  $\Db_{T,E}$ & 545&ff \\
%  \hline
%  $\Db_{T}$ & 545&dd \\  
%  \hline
% \end{tabularx}
% \caption{Fits parameters of the first eigenvector of the 3-points correlation $\Db_{i j}$, as defined in Equation~\eqref{eq:D_def} using the exponential fit $\tau \xrightarrow{} a \exp(-b \tau)$, for the pair {\sc e-mini vs tbond}}
% \label{table:Deig_fit}
% \end{center}
% \end{table}

% \begin{table}[h!]
% \begin{center}
% \begin{tabularx}{0.9\textwidth}{|>{\centering\arraybackslash}X | >{\centering\arraybackslash}X>{\centering\arraybackslash}X|} 
%  \hline
%   & $a$ & $b$ \\ [0.5ex] 
%  \hline\hline
%  $\Vb_{E}$ & 6&ff  \\ 
%  \hline
%  $\Vb_{E,T}$ & 7 &ff \\
%  \hline
%  $\Vb_{T,E}$ & 545&ff \\
%  \hline
%  $\Vb_{T}$ & 545&dd \\  
%  \hline
% \end{tabularx}
% \caption{Fits parameters of the 2-points correlation $\Vb_{i j}$, as defined in Equation~\eqref{eq:V_def} using the exponential fit $\tau \xrightarrow{} a \exp(-b \tau)$, for the pair {\sc e-mini vs tbond}}
% \label{table:Vb_fit}
% \end{center}
% \end{table}

\clearpage

\newpage

\section{Parametric Calibration--Maximum Likelihood}\label{app:calib_ML}

In this appendix, we describe a widely used parametric calibration method for (Q)Hawkes processes, with intensity $\lambda$. The method relies on maximising the likelihood of jump times. We present here two applications: 
\begin{itemize}
\setlength\itemsep{0.01em}
    \item maximum likelihood when exact jump times are available. Note, that it is the method used in the package \texttt{tick} of \texttt{Python} to calibrate (multidimensional) linear Hawkes.
    \item maximum likelihood with proxies, when event-by-event data is not available. 
\end{itemize}

\subsection{With exact time of events}

\subsubsection{Log likelihood for point process}

For this first method of likelihood maximisation, one needs to have access to the event times $(t_i)_i$. The likelihood is then defined as the joint distribution of those time of events. 
Considering a Hawkes process of intensity $\lambda$, we first give the probability of event occurring between time $t_i$ and $t_i+{\rm d}t$, without any event happening between time $t_{i-1}$ and $t_i$.

The probability of event between time $t_i$ and $t_i+{\rm d}t$ but no event between $t_{i-1}$ and $t_i$] is given by
    \begin{equation*}
        \mathbb{P}\left(\text{one event in }[t_i,t_i+{\rm d}t]\text{ but no event in }[t_{i-1},t_i]\right) = \exp{\left(-\int_{t_{i-1}}^{t_i}{\rm d}u \lambda(u|\theta)\right)}\lambda(t_i|\theta){\rm d}t
    \end{equation*}
% \end{definition}

One can then write the joint distribution of $N$ events happening in $[0,T)$ at arrival times $(t_i)_i$, and deduce the log-likelihood. 

% \begin{definition}[Log Likelihood of 1D point process]
For $N$ events in $[0,T)$ at arrival times $(t_i)_i$, then, the log-likelihood is:
    \begin{align}
\ln(\mathbf{L}(t_1,t_2,...,t_N,\theta)) = \sum_{i=1}^N \ln(\lambda(t_i/\theta)) - \int_{0}^T {\rm d}u \lambda(u/\theta)
    \end{align}
    where $\lambda$ is the intensity and $\theta$ the parameters we are looking to estimate.
% \end{definition}

\begin{proof}
From definition of the probability of event, we write the joint distribution of $N$ independent events happening in $[0,T)$ at arrival times $(t_i)_i$
\begin{equation*}
    \begin{aligned}
        \mathbf{L}(t_1,t_2,...,t_N,\theta) =& \left(\prod_{i=1}^N \exp{\left(-\int_{t_{i-1}}^{t_i}{\rm d}u \lambda(u|\theta)\right)}\lambda(t_i|\theta)\right)\exp{\left(-\int_{t_{N}}^{T}{\rm d}u \lambda(u|\theta)\right)}\\
        =& \left(\prod_{i=1}^N \lambda(t_i|\theta)\right)\exp{\left(-\int_{0}^{T}{\rm d}u \lambda(u|\theta)\right)}.
    \end{aligned}
\end{equation*}

Taking the log gives the result. 
    
\end{proof}

\subsubsection{Log likelihood for linear Hawkes process}

For linear Hawkes processes, the intensity is defined as
\begin{align*}
    \lambda(t/\theta) =& \lambda_\infty + \int_{-\infty}^t h_\theta(t-s){\rm d}N_s=\lambda_\infty + \sum_{t_i<t} h(t-t_i)
\end{align*}

where $(t_i)_i$ are the times of the jumps.

Particularly, with exponential kernels, we have $\theta=(\lambda_\infty,n_h,\beta)$, $h_\theta(t) = n_h\beta\exp(-\beta t)$. Then the expression of the log-likelihood evolves to:
\begin{align*}
    \ln(\mathbf{L}(t_1,t_2,...,t_N,\theta)) = -T\lambda_\infty + \sum_{i=1}^N\big(\ln(\phi_0+n_H\beta H_i)-n_H(\exp(-\beta(T-t_i))-1)\big)
\end{align*}
with 
\begin{align*}
    H_i = \sum_{k=1}^{i-1}\exp(-\beta(t_i-t_k))
\end{align*}

Indeed, we have $\lambda(t_i/\theta) = \lambda_\infty + n_H\beta H_i$. 
And 
\begin{align*}
    \int_0^T {\rm d}u\lambda(u/\theta) =& \lambda_\infty T + n_H\beta \int_0^T\int_{-\infty}^u \exp(-\beta(u-s)){\rm d}N_s {\rm d}u\\
    =& \lambda_\infty T + n_H\beta \int_{-\infty}^T\int_{s}^T \exp(-\beta(u-s)) {\rm d}u{\rm d}N_s\\
    =& \lambda_\infty T + n_H\beta \int_{-\infty}^T\frac{1}{\beta}(1-\exp(-\beta(T-s))){\rm d}N_s\\
    =& \lambda_\infty T + n_H \sum_{t_s<T}(1-\exp(-\beta(T-t_s)))
\end{align*}

\subsubsection{Log likelihood for ZHawkes process}

For ZHawkes processes, one needs to add the quadratic term to the intensity, which then becomes:
\begin{align*}
    \lambda(t/\theta) =& \lambda_\infty + \int_{-\infty}^t h_\theta(t-s){\rm d}N_s+ \big(\int_{-\infty}^t k_\theta(t-s){\rm d}P_s\big)^2\\
    =&\lambda_\infty + \sum_{t_i<t} h(t-t_i)+ \big(\sum_{t_i<t} k(t-t_i) m_{t_i}\big)^2
\end{align*}

where $(t_i)_i$ are the times of the jumps and $m(t_i)_t$ the marks of the jump at time $t_i$. Again, in the particular case of exponential kernels, we have $\theta=(\lambda_\infty,n_H,\beta,n_Z,\omega)$, $h_\theta(t) = n_H\beta\exp(-\beta t)$, $k_\theta(t) = \sqrt{2n_Z\omega}\exp(-\omega t)$. The log-likelihood can then entirely be defined by:
\begin{align*}
    \ln(\mathbf{L}(t_1,t_2,...,t_N,\theta)) = -T\lambda_\infty + \sum_{i=1}^N\big(\ln(\lambda_\infty+n_H\beta H_i + 2n_Z\omega Z_i^2)-n_H(\exp(-\beta(T-t_i))-1) - 2n_Z\omega Z_i^2\big)
\end{align*}
with 
\begin{align*}
    H_i = \sum_{k=1}^{i-1}\exp(-\beta(t_i-t_k)) \quad \text{ and }\quad 
    Z_i = \sum_{k=1}^{i-1}\exp(-\omega(t_i-t_k))m_{t_k}
\end{align*}

\subsubsection{Log likelihood for multivariate point process (2D)}

We consider a 2D process, with intensities $(\lambda_1,\lambda_2)$ and events arrival times $((t_i^1)_{i\in \llbracket 1, N \rrbracket},(t_i^2)_{i\in \llbracket 1, N \rrbracket})$, where $(t_i^1)_{i\in \llbracket 1, N \rrbracket}$, respectfully $(t_i^2)_{i\in \llbracket 1, N \rrbracket}$, are the times of events of process 1, respectfully process 2. The log likelihood consider the joint probability of the 2 jump processes, and thus results in:

% \begin{definition}
\textbf{[Log likelihood in 2D]}\\
    Log likelihood of two point processes  with intensities $(\lambda_1,\lambda_2)$, the log-likelihood of $N$ events from process 1 at times $(t_i^1)_{i\in \llbracket 1, N \rrbracket}$, and $N$ events of process 2 at times $(t_i^2)_{i\in \llbracket 1, N \rrbracket}$, between $0$ and time $T$ is given by:
    \begin{align*}
\ln(\mathbf{L}((t_i^1)_{i\in \llbracket 1, N \rrbracket},(t_i^2)_{i\in \llbracket 1, N \rrbracket},\theta)) = \sum_{i=1}^N \ln(\lambda_1(t^1_i/\theta))+\sum_{i=1}^N \ln(\lambda_2(t_i^2/\theta)) - \int_{0}^T {\rm d}u (\lambda_1(u/\theta)+\lambda_2(u/\theta))
    \end{align*}
    where $\theta$ is the vector of the parameters we are looking for.
% \end{definition}

\subsection{Without exact times of events - Using an intensity estimate}

\subsubsection{Defining the (log) likelihood}

It is sometimes impossible to access event times $(t_i)_i$ . However, we can estimate $\lambda_t$ for each time bin $t$. We can consider $\lambda_t$ constant over the bin $t$, then the probability to observe $n$ events in the bin $t$ is:
% \begin{definition}[Probability of number of events in a bin]
    \begin{align*}
    \mathbb{P}(\text{nb of events in bin }t = n) = \frac{\lambda_t^n}{n!}\exp(-\lambda_t)
\end{align*}
% \end{definition}

The number of events we observe in bin $t$ is our $\hat{\lambda}$ ($n=\hat{\lambda}$). 

In the QHawkes framework, $\lambda_t$ is defined by: 

\begin{align*}
    \lambda_t=\lambda_0 + \int_{-\infty}^{t}h(t-u){\rm d}N_u  + (\int_{-\infty}^{t}k(t-u){\rm d}P_u )^2
\end{align*}

% where we look for $h$,  $k$ with the form  $\gamma\frac{\exp(-\beta t)}{t^\alpha}$ (in the simulation, you can approximate a power law by a sum of exponential \blue{CHALLET REF} and thus end up with a sum of exponentials).
Using the approximation ${\rm d}N_u\equiv\mathbb{E}({\rm d}N_u)=\hat{\lambda}_u{\rm d}t$, we obtain:

\begin{align*}
    \lambda_t=\lambda_0 + \sum_{u=0}^{t}h(t-u)\hat{\lambda}_u{\rm d}u + (\sum_{u=0}^{t}k(t-u){\rm d}P_u )^2
\end{align*}
where $u=0$ represent the first bin of the day. Note that ${\rm d}P_t$ is the returns over the bin $t$, which is directly observable.

We can then define the log likelihood. We consider the joint density of observing the events ($(\hat{\lambda}^i)_{i\in \llbracket 1,N\rrbracket}$), where $N$ represent the number of bins we have, then the log likelihood is given by:
    \begin{align*}
\ln(\mathbf{L}((\hat{\lambda}^i)_{i\in \llbracket 1,N\rrbracket},\theta)) =& \sum_{i=1}^N \ln\left(\frac{\lambda_i(\theta)^{\hat{\lambda}_i}}{\hat{\lambda}_i!}\exp(-\lambda_i(\theta))\right) = \sum_{i=1}^N \hat{\lambda}_i\ln(\lambda_i(\theta)) - \ln(\hat{\lambda}_i!) -\lambda_i(\theta)\\
\sim&\sum_{i=1}^N \hat{\lambda}_i\ln(\lambda_i(\theta)) - \left(\hat{\lambda}_i\ln\left(\hat{\lambda}_i\right)-\hat{\lambda}_i\right) -\lambda_i(\theta)\\
\ln(\mathbf{L}((\hat{\lambda}^i)_{i\in \llbracket 1,N\rrbracket},\theta))=&\sum_{i=1}^N \hat{\lambda}_i\left(\ln(\lambda_i(\theta)) -\ln(\hat{\lambda}_i)\right)+\hat{\lambda}_i -\lambda_i(\theta)
    \end{align*}

% \end{definition}

\subsubsection{Defining the (log) likelihood for multidimensionnal process}

% \begin{definition}
\textbf{[likelihood of observing $(\hat{\lambda}^1,\hat{\lambda}^2)$]}\\
We consider 2 processes, with intensities $(\lambda^1,\lambda^2)$. We observe those two processes over $N$ bins, and the number of events from process $j$ in bin $i$ is $\hat{\lambda}_i^j$. Then, we can define the joint probability (likelihood), to observe  $(\hat{\lambda}_i^1,\hat{\lambda}_i^2)_{i\in\llbracket1,N\rrbracket}$ over the $N$ bins:
    
\begin{align*}
\mathbf{L}((\hat{\lambda}^1_i)_i,(\hat{\lambda}^2_i)_i,\theta)=& \prod_{i=1}^N \mathbb{P}(n_1 \text{ events from process 1 in bin }i)\mathbb{P}(n_2 \text{ events from process 2 in bin }i)\\
=& \prod_{i=1}^N \frac{(\lambda^1_i(\theta))^{n_1}}{n_1!}\exp(-\lambda^1_i(\theta)) \frac{(\lambda^2_i(\theta))^{n_2}}{n_2!}\exp(-\lambda^2_i(\theta))\\
=& \prod_{i=1}^N \frac{(\lambda^1_i(\theta))^{\hat{\lambda}^1_i}}{\hat{\lambda}^1_i!}\exp(-\lambda^1_i(\theta)) \frac{(\lambda^2_t(\theta))^{\hat{\lambda}^2_i}}{\hat{\lambda}^2_i!}\exp(-\lambda^2_i(\theta))
\end{align*}
where $\theta$ represent the vector of parameters we are trying to estimate.
% \end{definition}

Consequently, the log likelihood writes: 

\begin{align*}
\ln\big(\mathbf{L}((\hat{\lambda}^1_i)_i,(\hat{\lambda}^2_i)_i,\theta)\big)=& \sum_{i=1}^N\hat{\lambda}^1_i\ln(\lambda^1_i(\theta)) -\ln(\hat{\lambda}^1_i!)-\lambda^1_i(\theta)  +\hat{\lambda}^2_i\ln(\lambda^2_i(\theta)) -\ln(\hat{\lambda}^2_i!)-\lambda^2_i(\theta), \end{align*}
which we rewrite as:
\begin{equation}\label{app_eq:log_multi_proxy}
    \ln\big(\mathbf{L}((\hat{\lambda}^1_i)_i,(\hat{\lambda}^2_i)_i,\theta)\big)= \sum_{i=1}^N \hat{\lambda}^1_i\ln(\lambda^1_i(\theta)) -\big(\hat{\lambda}^1_i\ln(\hat{\lambda}^1_i)-\hat{\lambda}^1_i\big)-\lambda^1_i(\theta)  +\hat{\lambda}^2_i\ln(\lambda^2_i(\theta)) -\big(\hat{\lambda}^2_i\ln(\hat{\lambda}^2_i)-\hat{\lambda}^2_i\big)-\lambda^2_i(\theta) 
\end{equation}

where: 
\begin{itemize}
    \item $\hat{\lambda}^{1,2}_i$: the estimated intensities of processes 1,2 in bin $i$
    \item $\lambda^{1,2}_i$: the theoretical intensities of processes 1 \& 2 which depend on both the observation of $\hat{\lambda}$ and the parameters $\theta$ we are trying to estimate. In the case of a two dimensional ZHawkes, intensities are defined such that: 
    \begin{align*}
    \lambda^j_t=&\lambda^j_\infty + \int_{-\infty}^{t}h^j_1(t-u){\rm d}N^1_u+ \int_{-\infty}^{t}h^j_2(t-u){\rm d}N^2_u
     + (\int_{-\infty}^{t}k^j_{1}(t-u){\rm d}P^1_u )^2 + (\int_{-\infty}^{t}k^j_{2}(t-u){\rm d}P^2_u )^2.
\end{align*}
In our discrete environment, we have ordered bins $(t_i)_{i\in\llbracket1,N\rrbracket}$, and thus we can write the intensity of process $j$ in bin $M$ such that
\begin{align*}
    \lambda^j_{t_M}=&\lambda^j_\infty + \sum_{i=1}^{M-1}h^j_1(t_M-t_i)\hat{\lambda}^1_i{\rm d}t+ \sum_{i=1}^{M-1}h^j_2(t_M-t_i)\hat{\lambda}^2_i{\rm d}t  + (\sum_{i=1}^{M-1}k^j_{1}(t_M-t_i){\rm d}P^1_i)^2 + (\sum_{i=1}^{M-1}k^j_{2}(t_M-t_i){\rm d}P^2_i)^2.
\end{align*}
In the case of exponential kernels, we write $h^j_i(t)=n_i^j\exp(-\kappa^j_i t)$ and $k^j_i(t)=n_{iq}^j\exp(-\kappa^j_{iq} t)$, and the intensity becomes:
\begin{align*}
    \lambda^j_{t_M}=&\lambda^j_\infty \\&+ \sum_{i=1}^{M-1}{n}^j_1 \exp(-\kappa^j_1(t_M-t_i))\hat{\lambda}^1_i{\rm d}t+ \sum_{i=1}^{M-1}n^j_2\exp(-\kappa^j_2(t_M-t_i))\hat{\lambda}^2_i{\rm d}t\\&  + (\sum_{i=1}^{M-1}n^j_{1q}\exp(-\kappa^j_{1q}(t_M-t_i)){\rm d}P^1_i)^2 + (\sum_{i=1}^{M-1}n^j_{2q}\exp(-\kappa^j_{2q}(t_M-t_i)){\rm d}P^2_i)^2.
\end{align*}
The parameters we want to estimate in this case are $(n_i^j)_{i,j\in\llbracket 1,2 \rrbracket}$, $(n_{i,q}^j)_{i,j\in\llbracket 1,2 \rrbracket}$, $(\kappa_{i}^j)_{i,j\in\llbracket 1,2 \rrbracket}$ and $(\kappa_{i,q}^j)_{i,j\in\llbracket 1,2 \rrbracket}$. 
\end{itemize}

\subsubsection{Minimize the log-likelihood}

To minimise $-\ln\big(\mathbf{L}((\hat{\lambda}^1_i)_i,(\hat{\lambda}^2_i)_i,\theta)\big)$, we look at gradient descent methods. We need to calculate the gradient of $-\ln\big(\mathbf{L}((\hat{\lambda}^1_i)_i,(\hat{\lambda}^2_i)_i,\theta)\big)$ according to the parameters. 

% \begin{align}
%     -\ln\big(\mathbf{L}((\hat{\lambda}^1_i)_i,(\hat{\lambda}^2_i)_i,\theta)\big)= \sum_{i}- \hat{\lambda}^1_i\ln(\lambda^1_i) +\big(\hat{\lambda}^1_i\ln(\hat{\lambda}^1_i)-\hat{\lambda}^1_i\big)+\lambda^1_i  -\hat{\lambda}^2_i\ln(\lambda^2_i) +\big(\hat{\lambda}^2_i\ln(\hat{\lambda}^2_i)-\hat{\lambda}^2_i\big)+\lambda^2_i 
% \end{align}

Then, with expression~\eqref{app_eq:log_multi_proxy}, we can write the gradient: 
\begin{align*}
    \nabla\Big(-\ln\big(\mathbf{L}((\hat{\lambda}^1_i)_i,(\hat{\lambda}^2_i)_i,\theta)\big)\Big)=& \sum_{i=1}^N- \nabla\Big(\hat{\lambda}^1_i\ln(\lambda^1_i(\theta)) \Big)+\nabla\Big(\lambda^1_i(\theta)\Big)  -\nabla\Big(\hat{\lambda}^2_i\ln(\lambda^2_i(\theta))\Big) +\nabla\Big(\lambda^2_i(\theta)\Big) \\
    =& \sum_{i=1}^N -\hat{\lambda}^1_i\nabla\Big(\ln(\lambda^1_i(\theta)) \Big)+\nabla\Big(\lambda^1_i(\theta)\Big)  -\hat{\lambda}^2_i\nabla\Big(\ln(\lambda^2_i(\theta))\Big) +\nabla\Big(\lambda^2_i(\theta)\Big)\\
    =&\sum_{i=1}^N- \hat{\lambda}^1_i\frac{\nabla\Big(\lambda^1_i(\theta)\Big)}{\lambda^1_i(\theta)}+\nabla\Big(\lambda^1_i(\theta)\Big)  -\hat{\lambda}^2_i\frac{\nabla\Big(\lambda^2_i(\theta)\Big)}{\lambda^2_i(\theta)} +\nabla\Big(\lambda^2_i(\theta)\Big)\\
    =& \sum_{i=1}^N \nabla\lambda^1_i(\theta)\Big(-\frac{\hat{\lambda}^1_i}{\lambda^1_i(\theta)}+1\Big)  +\nabla\lambda^2_i(\theta)\Big(-\frac{\hat{\lambda}^2_i}{\lambda^2_i(\theta)}+1\Big).
\end{align*}

In practice, for financial time series, $\hat{\lambda}\equiv\frac{\sigma^2}{{\rm d}t}$ where $\sigma$ is the volatility, and ${\rm d}P$ are the returns.

\end{document}